\documentclass[]{aastex631}

\usepackage{graphicx,color}
\usepackage{subfigure}
\usepackage{mathrsfs,amsmath}
\usepackage{natbib}
\usepackage{ulem}
\usepackage{hyperref}
\usepackage{CJK}
\usepackage{comment}

\usepackage{float}
\begin{document}

\title{Anisotropy of Nanohertz Gravitational Wave Background and Source Clustering from Supermassive Binary Black Holes Based on Cosmological Simulation}

\author{Qing Yang}
\affiliation{College of Engineering Physics, Shenzhen Technology University, No.3002 Lantian Road, Shenzhen 518118, China}
\affiliation{Shenzhen Key Laboratory of Ultraintense Laser and Advanced Material Technology, Center for Intense Laser Application Technology, Shenzhen Technology University, No.3002 Lantian Road, Shenzhen 518118, China}

\author[0000-0001-5174-0760]{Xiao Guo$^*$}%
\affiliation{School of Fundamental Physics and Mathematical Sciences, Hangzhou Institute for Advanced Study, University of Chinese Academy of Sciences, No.1 Xiangshan Branch, Hangzhou 310024, China}
\author{Zhoujian Cao}
\affiliation{School of Fundamental Physics and Mathematical Sciences, Hangzhou Institute for Advanced Study, University of Chinese Academy of Sciences, No.1 Xiangshan Branch, Hangzhou 310024, China}
\affiliation{Institute of Applied Mathematics, Academy of Mathematics and Systems Science, Chinese Academy of Sciences, No. 55 Zhongguancun East Road, Beijing 100190, China}

\author{Xiaoyun Shao}
\affiliation{Observat\'orio Nacional, Rio de Janeiro, RJ, 20921-400, Brazil}

\author{Xi Yuan}
\affiliation{Department of Physics, Brown University, 182 Hope Street, Providence, 02912, RI, United States}

\email{$^*$\,guoxiao@nao.cas.cn}
\begin{abstract}
Several pulsar timing array (PTA) groups have recently claimed the detection of nanohertz gravitational wave background (GWB), but the origin of this gravitational wave (GW) signal remains unclear. Nanohertz GWs generated by supermassive binary black holes (SMBBHs) are one of the most important GW sources in the PTA band.
Utilizing data from cosmological simulation, we generate multiple realizations of a mock observable universe that self-consistently incorporates the cosmic large-scale structure, enabling a robust statistical analysis of SMBBH populations and their GW signatures. We systematically investigate the merger event distributions and both the isotropic and anisotropic properties of the resulting GWB signals under different hardening timescales. Specifically, we calculate the characteristic amplitude of the GWB signal, and the angular power spectrum for both the total energy density and energy density in different frequency bins accounting for cosmic variance through different realizations. We also study the clustering pattern of the positional distribution of SMBBHs to examine whether they can reproduce the large-scale structure of galaxies.
Furthermore, for the upcoming Chinese Pulsar Timing Array (CPTA) and Square Kilometre Array (SKA)-PTA, we predict the numbers and signal-to-noise ratio (SNR) distributions of resolvable individual GW sources that may be detected with SNR$>$8. We finally investigated the impact of weak lensing effects and found that their influence on the basic characteristics of the GWB signal and individual sources is rather limited.
\end{abstract}

\section{Introduction}
\label{sec:intro}

Observational data indicates that supermassive black holes (SMBHs) occupy the centers of nearly all massive galaxies \citep{1982MNRAS.200..115S,1995ARA&A..33..581K,1998AJ....115.2285M,2002MNRAS.335..965Y,2013ARA&A..51..511K}. While the precise evolutionary processes behind SMBHs remain unclear, there exist robust correlations between their masses and various observational properties of their host galaxies, encompassing factors such as velocity dispersion, star formation rate, and bulge stellar mass \citep{1996MNRAS.283.1388M,1998MNRAS.293L..49B,1998AJ....115.2285M,2000ApJ...539L...9F,2003ApJ...598..886U,2009ApJ...707.1566Z}. Furthermore, it is anticipated that during galaxy mergers, the SMBHs within them will coalesce into supermassive binary black holes (SMBBHs), generating gravitational waves (GW) during their orbital decay and ultimately merge into a single, more massive black hole \citep{2002MNRAS.335..965Y,2015ASSP...40..147S,2020ApJ...897...86C,2024MNRAS.533.1568S,2024arXiv240711669S}. The gravitational torques resulting from these galaxy-galaxy mergers promote the inflow of cold gas towards the galactic centers, igniting central starbursts and fuelling accretion onto the SMBHs \citep{1989Natur.340..687H,1991ApJ...370L..65B,1996ApJ...471..115B,1994ApJ...431L...9M,1996ApJ...464..641M,2005Natur.433..604D}. This process of galaxy/SMBH mergers has been proposed as a potential mechanism for activating central active galactic nuclei (AGN) and facilitating the growth of SMBHs \citep{2008ApJS..175..356H,1988ApJ...325...74S,2012ApJ...758L..39T}.

On the one hand, GWs emitted by numerous inspiraling SMBBHs are anticipated to constitute a GW background (GWB) spanning the frequency range of nHz-$\mu$Hz. The detection of this GWB holds profound and transformative implications for cosmology and galaxy evolution. On the other hand, if GWs emitted from some individual SMBBHs are so strong that they can be detected by GW detectors with significant signal-to-noise ratio (SNR), these GW signals can be resolved as individual GW sources and be separated from GWB.

Precision timing of an ensemble of millisecond pulsars, known as pulsar timing arrays (PTA), offers a unique approach to detect such low-frequency GW signals including GWB and individual sources \citep{1978SvA....22...36S,1979ApJ...234.1100D,1984JApA....5..369B,1990ApJ...361..300F,maggiore2008gravitational, 2009MNRAS.394.2255S, 2010CQGra..27h4016S, 2011gwpa.book.....C, 2011MNRAS.414.3251L, manchester2013pulsar, 2013CQGra..30x4009S, 2014gwdd.book.....V, 2014MNRAS.444.3709Z, 2015SCPMA..58.5748B,  mingarelli2015gravitational, 2016MNRAS.459.1737S, 2019BAAS...51c.336T}. 
Currently, a diverse array of PTA experiments is operational, including the Parkes PTA (PPTA\footnote{\url{http://www.atnf.csiro.au/research/pulsar/ppta/}}; \citealt{manchester2013pulsar,2023ApJ...951L...6R}), the European PTA (EPTA\footnote{\url{http://www.epta.eu.org/}}; \citealt{2013CQGra..30v4009K,2023A&A...678A..50E}), the North American Nanohertz Observatory for Gravitational Waves (NANOGrav\footnote{\url{http://nanograv.org/}}; \citealt{2013CQGra..30v4008M, 2019BAAS...51g.195R,2023ApJ...951L...8A}), the Indian Pulsar Timing Array (InPTA; \citealt{2018JApA...39...51J,2023A&A...678A..50E}), the Chinese PTA (CPTA; \citealt{2011IJMPD..20..989N, 2009A&A...505..919S,2023RAA....23g5024X}), and the MeerKAT PTA (MPTA; \citealt{2023MNRAS.519.3976M}). 
Among which, the first three PTA groups  have amassed data spanning over a decade in pursuit of potential low-frequency GW signals. They have also collaborated to form the International PTA (IPTA\footnote{\url{http://www.ipta4gw.org/}}; \citealt{2013CQGra..30v4010M,10.1093/mnras/stw347, 2019MNRAS.490.4666P}) by sharing data. Looking forward, the Square Kilometer Array (SKA) \citep[e.g.,][]{Lazio2013SKA, 2017PhRvL.118o1104W} is anticipated to discover numerous stable millisecond pulsars (MSPs) and establish a high-sensitivity PTA, henceforth referred to as SKA-PTA, to detect low-frequency GW sources.  Recently, CPTA, NANOGrav, EPTA (+InPTA), and PPTA have presented evidence for the presence of a nanohertz GWB with a confidence level of approximately 2-4\,$\sigma$\citep{2023RAA....23g5024X,2023ApJ...951L...8A, 2023ApJ...951L...6R, 2023A&A...678A..50E}(see also \citealt{2020ApJ...905L..34A}). Although 
the origin of this GWB is still unclear, it is likely that this signal can be attributed to, or at least partly attributed to GWs generated from cosmic inspiralling SMBBHs, indicating that the detection of individual SMBBHs may also soon become a reality \citep[e.g.,][]{2023ApJ...955..132C, 2023ApJ...952L..37A, 2023arXiv230616227A}.

In recent years several studies have also presented predictions on this GWB based on phenomenological models or simulations \citep{2003ApJ...583..616J,2008MNRAS.390..192S,2013MNRAS.433L...1S,2016MNRAS.463L...6S,2017MNRAS.464.3131K,2019MNRAS.483..503Y,2024ApJ...965..164G,2024PhRvD.110l3507A,2024MNRAS.533.1568S,2024arXiv241100532S,2025JCAP...03..011G}. The predicted GW strain amplitude ($A_{\rm yr^{-1}}$) lies approximately within the range of $1\times10^{-16}$ to $5\times10^{-15}$. The practical GW strain amplitude $A_{\rm yr^{-1}}$ from GWB detected by NANOGRav, CPTA, EPTA and PPTA is around $2\times10^{-15}$. The predictions regarding the characteristic amplitude of the GWB focus on its isotropic nature, which corresponds to an idealized case where the GWB sources are sufficiently abundant to approximate a continuous spatial distribution, and thus results in the characteristic Hellings and Downs curve that could be observed in PTA experiments \citep{1983ApJ...265L..39H}. However, in reality, the number of SMBBHs is finite, leading to a GWB signal comprising both a dominant isotropic component and also anisotropic contributions \citep{2024PhRvD.110l3507A}.
The impact of anisotropies on PTA measurements and the extension of the Hellings and Downs curve method to analyze anisotropies in the GWB have been explored in various studies, including \cite{2013CQGra..30v4005C,2013PhRvD..88f2005M,2013PhRvD..88h4001T,2014arXiv1406.4511C,mingarelli2015gravitational,2019OJAp....2E...8H,2024PhRvD.109l3544S}. Several PTAs have also provided constraints on the level of anisotropy in the GWB \citep{2015PhRvL.115d1101T}.
 
There has been some work exploring the anisotropy property of the GWB signal by utilizing random Poisson sampling of GW sources \citep{2019MNRAS.483..503Y, 2024PhRvD.109l3544S, 2024ApJ...965..164G}; by using cosmological simulations to estimate the annisotropic characteristics of GWB signal or its cross correlation with galaxies \citep{2024MNRAS.533.1568S,2024arXiv241100532S}; or by using theoretical framework to quantify the impact of GW source discreteness and clusterings on the Hellings and Downs curve \citep{2024PhRvD.110l3507A, 2025JCAP...03..011G}. In this work, we intend to employ a semi-analytic galaxy formation model (SAM) grounded on dark matter (DM) halo merger trees derived from the Millennium simulation \citep{2005Natur.435..629S,2011MNRAS.413..101G}. We will focus on the SAM variants proposed by \citet{2011MNRAS.413..101G,2013MNRAS.428.1351G} (see also \citealt{2015MNRAS.451.2663H}) to construct multiple realizations of a comprehensive dataset of SMBBHs within the mock observable universe, from which we forecast the statistical properties of SMBBHs regarding their redshifts, mass and mass ratios, their spatial and angular distributions, and the characteristic strain amplitude of the resulting GWB, along with its anisotropic characteristics. Our method allows us to incorporate the cosmic large-scale structure and hardening timescales in a self-consistent manner at the level of light-cone SMBBH events construction.
In addition, we study the two dimensional angular distributions of GW sources with the SMBBHs that we constructed in the observable universe, and present the resulting angular power spectrum coefficients and angular two-point correlation functions. We also predict resolvable individual GW sources for future CPTA and SKA-PTA, and show their statistical properties. Finally, we will also make an estimation on how gravitational lensing effect will affect all of our calculational results.

This paper is organized as follows. In Section~\ref{sec:models}, we introduce the galaxy evolution model and mock data we used, as well as the basic setups and methods for constructing light-cone SMBBHs in the observable universe. In Section~\ref{sec:results}, we present our basic results concerning the isotropic and anisotropic characteristics of the GWB signal, which we derived from the SMBBH data constructed, and also the clustering properties of the GW sources. We further discuss the detection capabilities of different PTAs for individual bright sources in Section~\ref{sec:detect}. In Section~\ref{sec:lensing}, we take into account the effects of gravitational lensing and discuss their impact on all of our calculational results derived above. Finally, Section~\ref{sec:concl} is devoted to conlusions and discussions. Throughout this paper, we adopt that dimensionless Hubble constant $h_0=0.704$.

\section{Models}
\label{sec:models}
\begin{figure}[ht!]
\centering
\includegraphics[width=0.45\textwidth]{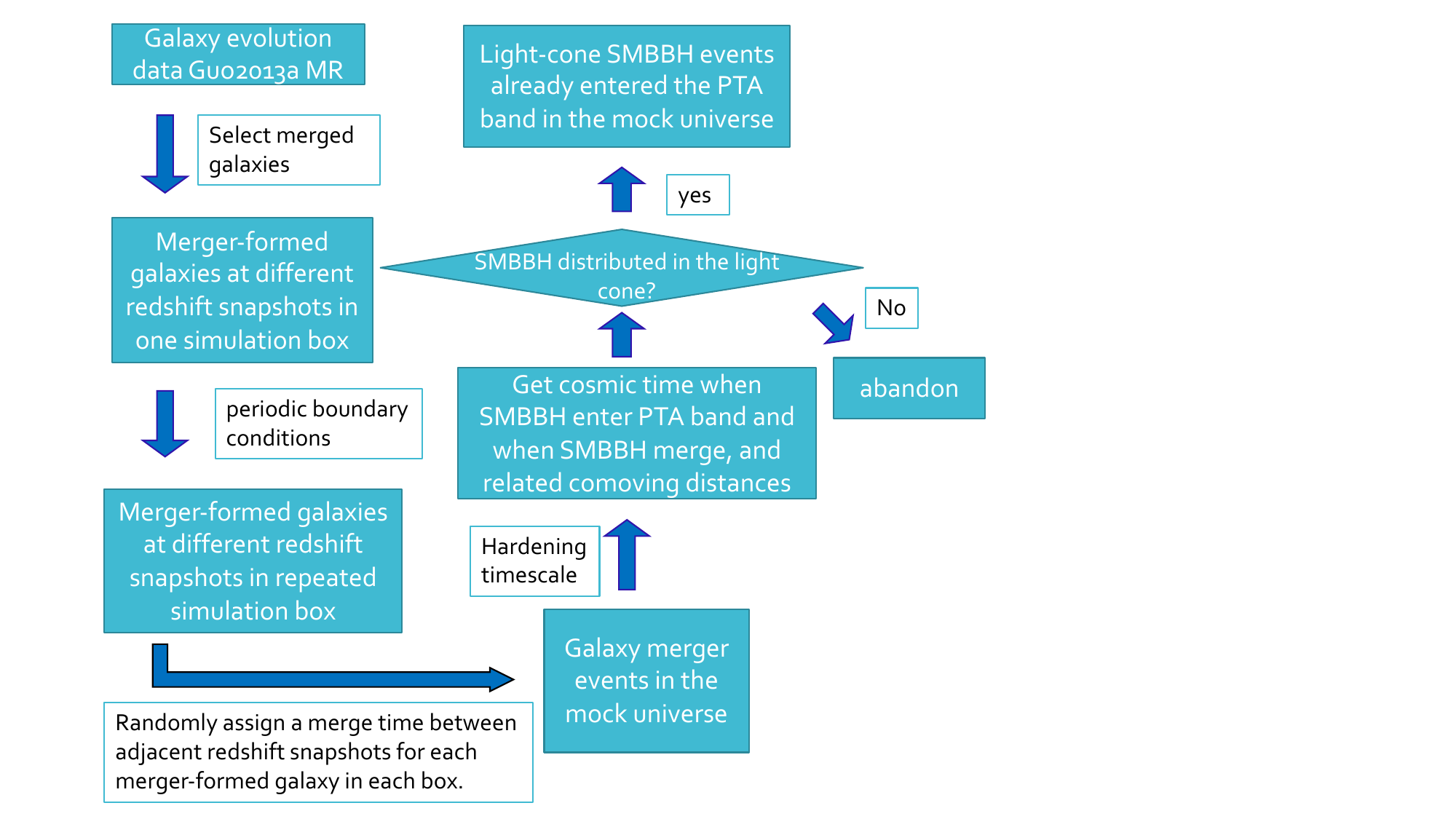}
\hspace{0.5cm}
\includegraphics[width=0.45\textwidth]{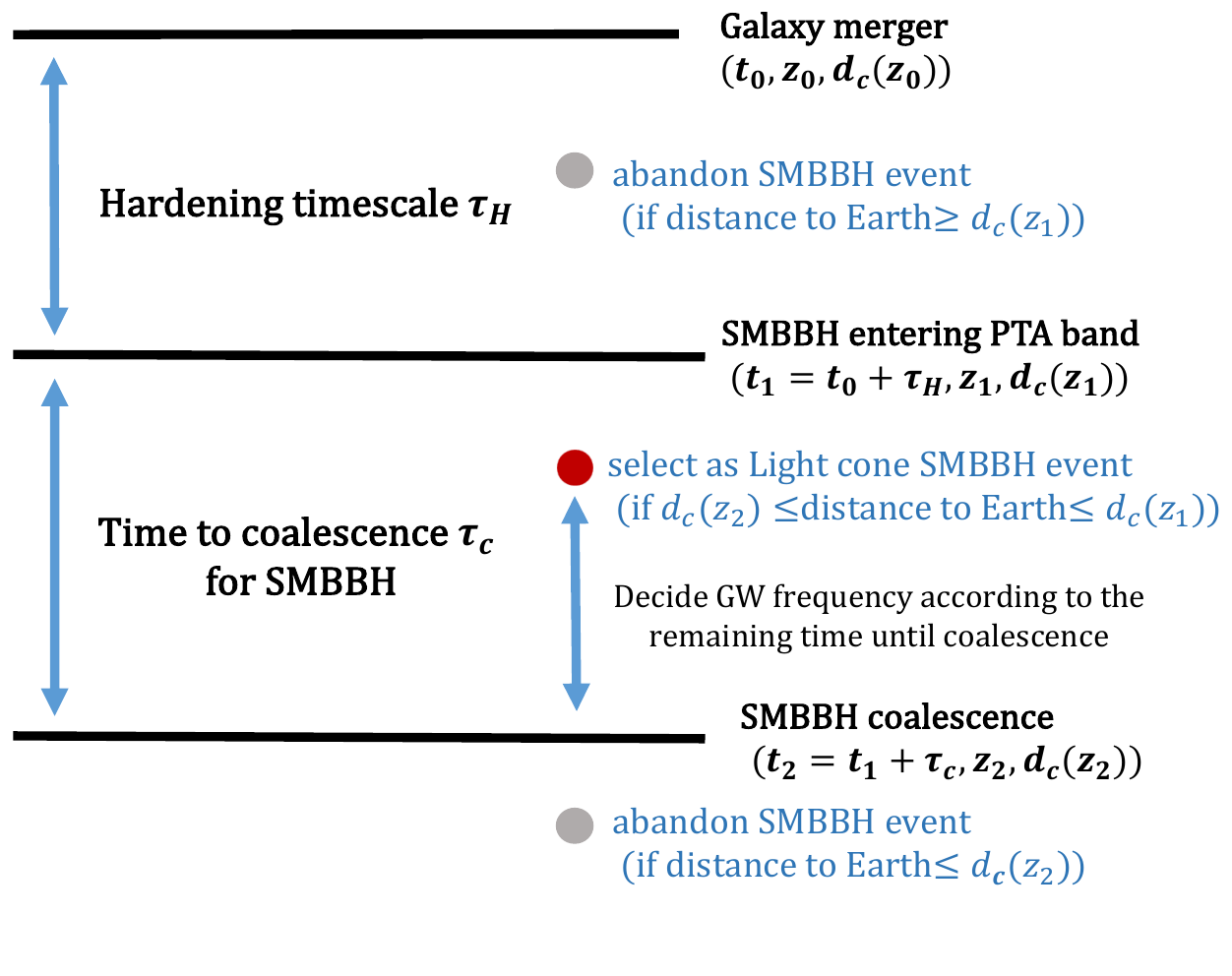}
\caption{Left panel: a flowchart for the entire process to deal with simulation data and construct light-cone GW events in the observable universe from Guo2013a mock galaxy data based on Millennium Run; Right panel: a schematic diagram illustrating the method for determining whether a SMBBH event lies within the light cone. Here, $t_0$, $t_1$ and $t_2$ are the cosmic times when the two progenitor galaxies merge, the SMBBH inside them entering PTA band and SMBBH coalescence respectively; $z_0$, $z_1$ and $z_2$ are the  corresponding redshifts of these events; $d_c(z)$ is the comoving distance related to the redshift $z$. We apply this selection process to each SMBBH in every merger-formed galaxy to obtain all light-cone events.}
\label{fig:flowchart}
\end{figure}

In this section, we introduce how we construct the light-cone SMBBH data in the observable universe. The flowchart and schematic of the entire process to deal with simulation data and generate light-cone GW sources are shown in Figure~\ref{fig:flowchart}. We will introduce our method related to each part of the flowchart in detail in the following subsections.

\subsection{Mock Data}

We utilize the results of L-Galaxies semi-analytical galaxy formation runs using the model described in \cite{2011MNRAS.413..101G,2013MNRAS.428.1351G} stored in the Millennium simulation database Guo2013a. This is a semi-analytical modelling applied to the DM halo merger trees of the Millennium simulations, which follows the formation, evolution and mergers of galaxies in a box with side length of 500 Mpc$/h_0$ from $z>14$ to $z=0$. The mock simulation data are output in 61 snapshots at different redshifts. From the output data, we can extract those galaxies formed by mergering in each redshift snapshot, each of these galaxies carries information about its location and its two progenitor galaxies. Thus the merger history of galaxies across the entire redshift range can be constructed. Since SMBH and galactic nuclear activity are closely related to galaxy formation, SAM always encompass the growth and evolution of SMBHs in their modeling. In Guo2013a, “quasar” mode and “radio” mode are considered to grow SMBHs from initial seeds, leading to a co-evolution of SMBHs and their host galaxies that is compatible with observations.
We assume that SMBHs will form inspiralling binaries following the merger of their host galaxies, so that the merger history of supermassive black holes can be correspondingly constructed following the merger history of galaxies.

\subsection{Observable Universe with different hardening timescale}
{We assume that supermassive black holes will reach the PTA frequency band ($10^{-9}{\rm Hz}-10^{-6}{\rm Hz}$) after undergoing a uniform hardening time following galaxy mergers. We consider three choices for the hardening timescale $\tau_H$, i.e., $\tau_H=0.1\,\mathrm{Gyr}$, $5\,\mathrm{Gyr}$, and $10\,\mathrm{Gyr}$. We randomly place Earth in 50 different galaxies, as is shown in Figure~\ref{fig:earthposition} (displacements at the order of $10$ kpc relative to the galaxy centers are randomly assigned to each Earth position to ensure they are not too close to any SMBBH merger events).  
\begin{figure}[ht!]
\centering 
\includegraphics[width=0.6\textwidth]{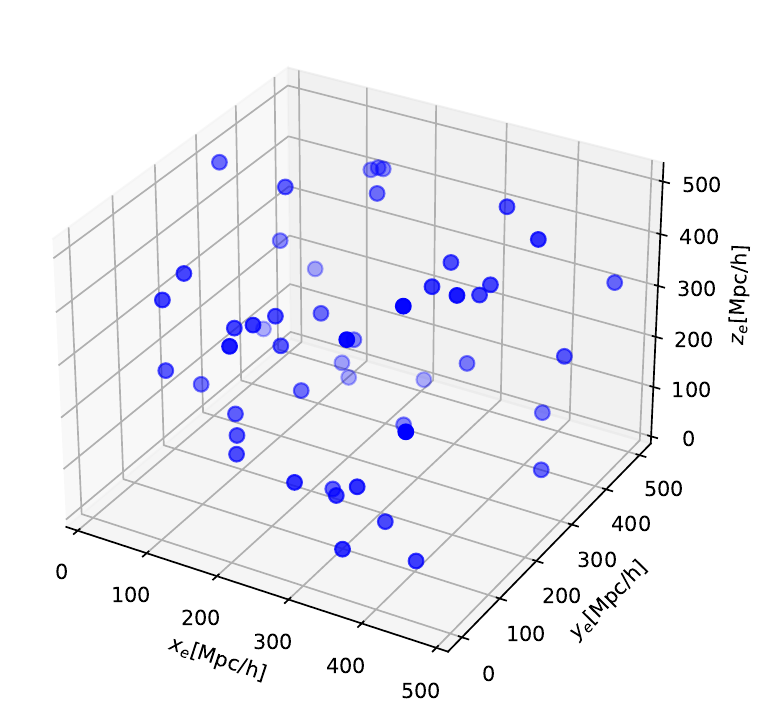} 
\caption{Earth's positions are randomly distributed across 50 different galaxies, allowing us to generate 50 distinct realizations of the light-cone SMBBH events for each of the three hardening timescales.}
\label{fig:earthposition}
\end{figure}
This allows us to generate 50 different realizations of the SMBBH merger events for each hardening timescale.}
However, instead of SMBBHs at specific redshifts in a box with limited size, signals observed by PTA originate from SMBBHs distributed along the light cone, with larger distance to us at higher redshifts. These SMBBHs constitutes the observable universe centered at the earth, and the distance of them may exceed the side length of the simulation box at relatively high redshifts. In order to construct this observable universe with mock data, and considering the universe is uniform and isotropic on large scale ($>100$\,Mpc), we adopt periodic boundary conditions, i.e., we piece together the simulation box in a repeating pattern to generate a universe that can theoretically stretch out indefinitely in all three directions.
For each simulation box and  every merger-formed galaxy recorded at certain redshift snapshots, we randomly assign to them a cosmic time between the current redshift snapshot and the previous one to represent the merger time of the two progenitor galaxies. The time when the SMBBH enters the PTA band is then determined by adding an extra hardening time to this merger time. For each SMBBH event, we calculate the evolution time spanning from its entry into the PTA band to final coalescence (i.e., from $(1+z)\times10^{-9}\rm Hz$ to coalescence, where $z$ is redshift of the SMBBH when it enters the PTA band, and can be determined by converting cosmic time to redshift) according to the following equation:
\begin{equation}
    \tau=\frac{5}{256}(\pi f_r)^{-\frac{8}{3}}\left(\frac{G M_c}{c^3}\right)^{-\frac{5}{3}}
    \label{eq:timetocoa}
\end{equation}
This evolutionary timescale spans a specific comoving distance interval, any SMBBH event whose distance to Earth falls within this interval is considered to be within the light cone and are saved as a potential observable GW source event. The distances of these GW events to Earth can determine a specific look-back time. By calculating how much time remains from this look-back time to the time of coalescence, the rest frame frequency of each binary black hole event can be determined by solving for $f_r$ with a certain $\tau$ through Equation~\eqref{eq:timetocoa}. Here we have assumed the orbits of SMBBHs are circular for simplicity. In the right panel of Figure \ref{fig:flowchart}, we also present a schematic diagram of our method for constructing the light-cone events. 


\subsection{GW Sources}
With methods described in the previous subsection, we ultimately obtained $N_{\rm total}\approx1\times10^{8},6\times10^6,1\times10^4$ SMBBH events in the observable universe for hardening time scales of $0.1,5,10\,\mathrm{Gyr}$ respectively. {Among them, we consider the SMBBHs whose $f_{\rm obs}$ falls between $f_{min}=1/T_{\rm obs}$ and  $f_{max}=1/\Delta t$ to serve as GW sources whose signals are detectable for the PTA configurations we will consider in subsequent sections, where $T_{\rm obs}$ is the total observation duration and $\Delta t$ is the observation cadence.}  Each binary system possesses a unique combination of spatial coordinates, redshift, black hole masses, and are emitting GWs at a certain frequency. We can determine the correponding GW strain amplitude for each SMBBH event with all this information.
Given that the pulsars utilized by PTA may be situated at various locations with diverse inclinations, the orientation of the PTA itself can be considered arbitrary. Furthermore, the SMBBHs under observation can also exhibit a wide range of inclinations. 
Hence we adopt the following formula to estimate the average GW
strain for each SMBBH:

\begin{equation}
    h=\sqrt{\frac{32}{5}}\frac{G^{5/3}\mathcal{M}_z^{5/3}(\pi f_{\rm obs})^{2/3}}{c^4d_{\rm L}},
\label{eq:h}
\end{equation}
where $d_{\rm L}$ is the luminosity distance of GW source, $\mathcal{M}_z=\mathcal{M}(1+z)$ is the redshifted mass, with chirp mass $\mathcal{M}=\frac{(m_{1}m_{2})^{3/5}}{M^{1/5}}$, and $m_1$, $m_2$ are the masses of the two components of this SMBBH, $M=m_1+m_2$ is the total mass of this SMBBH. Observational frequency $f_{\rm obs}=\frac{f_{\rm r}}{1+z}$, where $f_{\rm r}$ is the GW frequency in the rest frame of GW source.

{ It should be noted that, under conditions of extended hardening timescales, a SMBBH within a host galaxy may experience multiple subsequent merger processes before coalescence; we did not model the dynamical evolution of such multi-body systems. Instead, we adopt a simplified treatment assuming the new incoming black hole will form a new binary system with the remnant of the prior coalescence. The treatment of sequential mergers as independent GW sources, rather than interconnected dynamical systems, represents a limitation in capturing full self-consistency; however, it provides a tractable framework with enough computational efficiency. Due to the same reason, we also didn’t account for black hole mass evolution after the merger of galaxies. We evaluate the potential impact of this simplification in Appendix \ref{App:massevo} for $\tau_H=10\,\mathrm{Gyr}$, in which case mass evolution has the greatest impact.}

\section{Results}
\label{sec:results}
We superimposed the gravitational wave signals from all the PTA band light-cone SMBBH events to obtain the GWB signal and studied its properties. In this section, we will present the basic characteristics of the SMBBH mock data we have obtained, as well as our main results for the GWB signal.

\subsection{Statistical Properties of Light-Cone SMBBH Mock Data}
\begin{figure}[ht!]
\centering 
\includegraphics[width=0.45\textwidth]{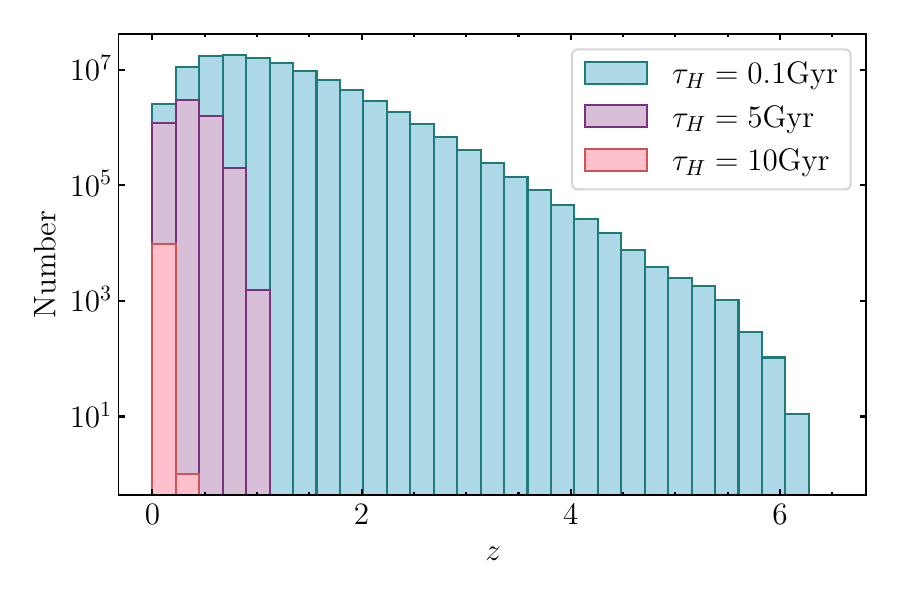} 
\includegraphics[width=0.45\textwidth]{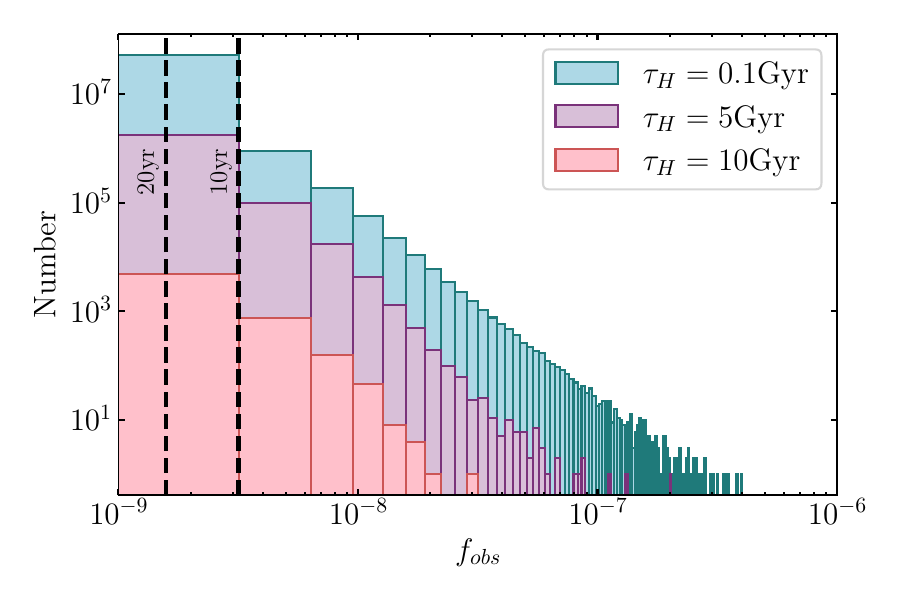}
\includegraphics[width=0.45\textwidth]{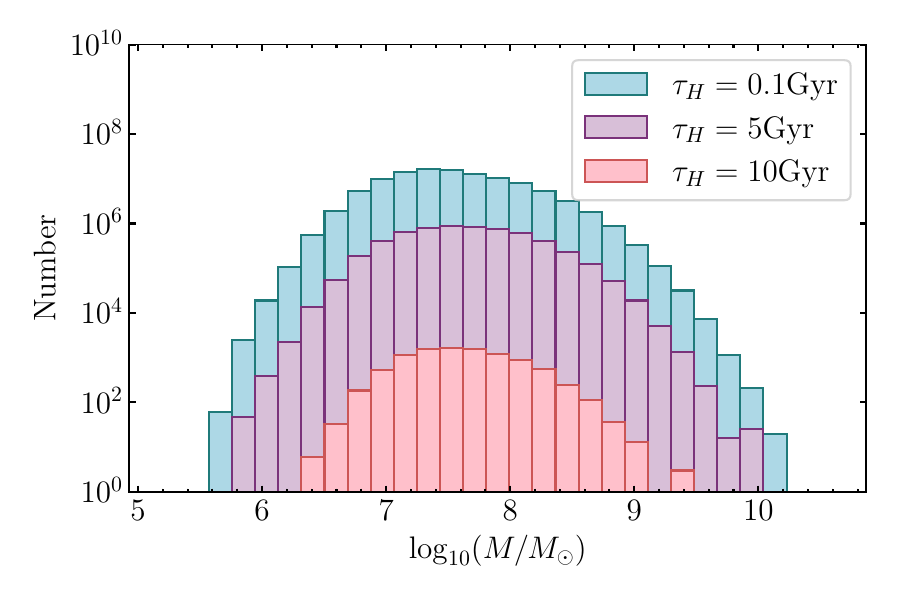} 
\includegraphics[width=0.45\textwidth]{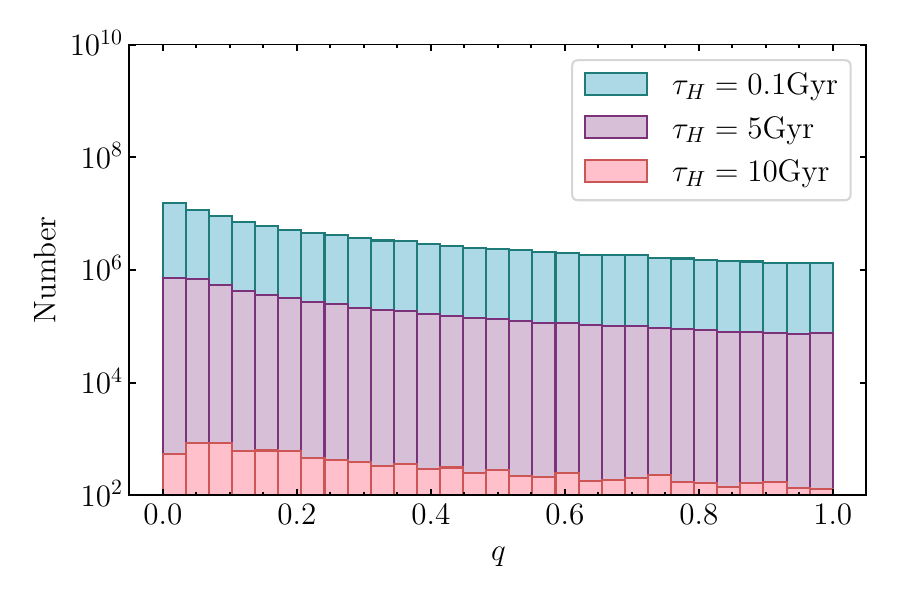} 
\caption{The redshift $z$ (upprt left), observed frequency $f_{\rm obs}$ (upper right),  total mass $M$ (lower left) and mass ratio $q$ (lower right) distribution of the light-cone SMBBHs from one representative realization for all the three hardening timescales. {The vertical black dashed lines in the frequency distribution plot indicate the lower frequency limits detectable with total observation durations of 10 years and 20 years respectively.}}
\label{fig:BHm}
\end{figure}

We present the basic statistical properties of our light-cone SMBBH mock data in this subsection.
{The redshift $z$, observed frequency $f_{obs}$, total mass $M$ and mass ratio $q$ distributions of the SMBBHs of all the three hardening timescales are demonstrated in Figure~\ref{fig:BHm}. The redshift distribution of the SMBBHs peaks around redshift $z\sim0.7$ and $z\sim0.3$, and the maximum redshift can reach $z=6$ and $z=1$ for hardening timescale of $0.1\,\mathrm{Gyr}$ and $5\,\mathrm{Gyr}$ respectively. While for the $\tau_h=10\,\mathrm{Gyr}$ case, the majority of SMBBH events are distributed in the lowest redshift bin ($z<0.21$)\footnote{In this extreme scenario, our data indicate that approximately 0.1\% of galaxies with stellar mass $M_{\star} > 10^{10}\,M_{\odot}$ below redshift z=0.21 host PTA-band SMBBHs. If it is possible for future experiments to conclusively identify nearby sub-pc SMBBHs, this occupation fraction may provide an additional observational test of our model.}, with the highest redshift not exceeding 0.5. The frequency distribution basically follow the power law distribution while subjecting to larger fluactuations in the higher frequency bins. The total masses of the binary systems peak around $3\times10^{7}M_\odot$ for all the three hardening timescales. 
On the other hand, the mass ratio distribution of the SMBBHs generally exhibits a monotonically decreasing trend with increasing mass ratio $q$, which indicates that in our data, the majority of the binary black hole systems exhibit a significant difference in mass between the two black holes.
{Note that our assumption of a uniform hardening timescale may bias the mass ratio distribution. Crucially, binaries with comparable masses (high mass ratios) may experience shorter dynamical friction timescales and transition more rapidly into the GW-driven regime. This will result in a higher population of such systems evolving within the PTA frequency band, thereby increasing the number of high mass ratio sources captured in our observable light cone. The modeling of more refined hardening process is beyond the scope of this study, we'll explore related issues in future work.}
}

\subsection{Isotropic Part of the Simulated GWB Signal}
\label{subsec:isotropic}

Our methodology incorporates a vast amount of individual sources scattered across the observable universe to construct GWB. For the sake of operational convenience, we considered the time interval for SMBBHs to evolve to coalescence when selecting light-cone events. 
{Note that we do not account for binaries with frequencies above $f_{\rm max}=1/\Delta t$}, as they lie beyond the observational window of PTA experiments under the parameter configuration we assumed. That is to say, we will impose a frequency window and only analyze the behavior of the resulting GWB signal between $f_{min}=1/T_{\rm obs}$ to $f_{max}=1/\Delta t$, where $T_{\rm obs}$ is the total observation duration and $\Delta t$ is the observation cadence. 

\begin{figure}[ht!]
\centering
\includegraphics[width=0.35\textwidth]{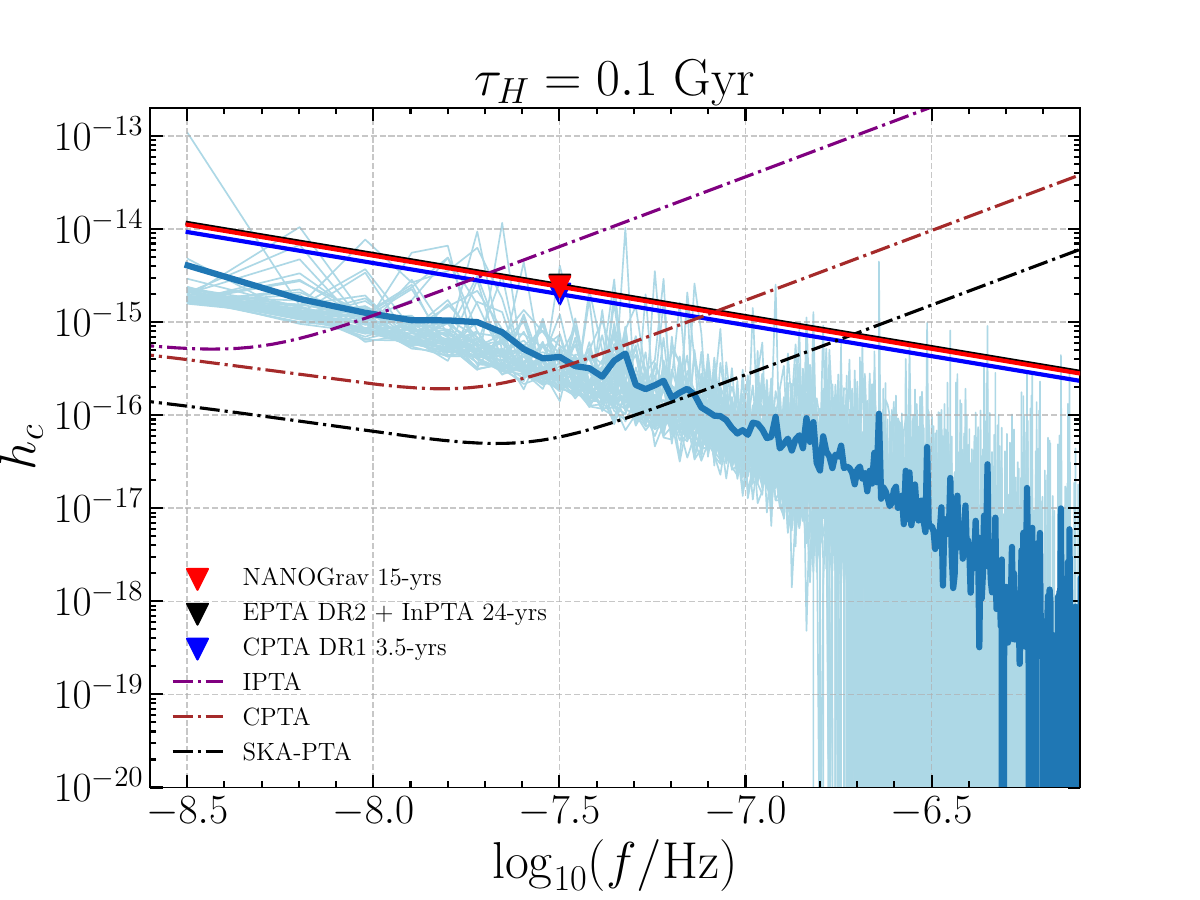}
\hspace{-0.83cm}
\includegraphics[width=0.35\textwidth]{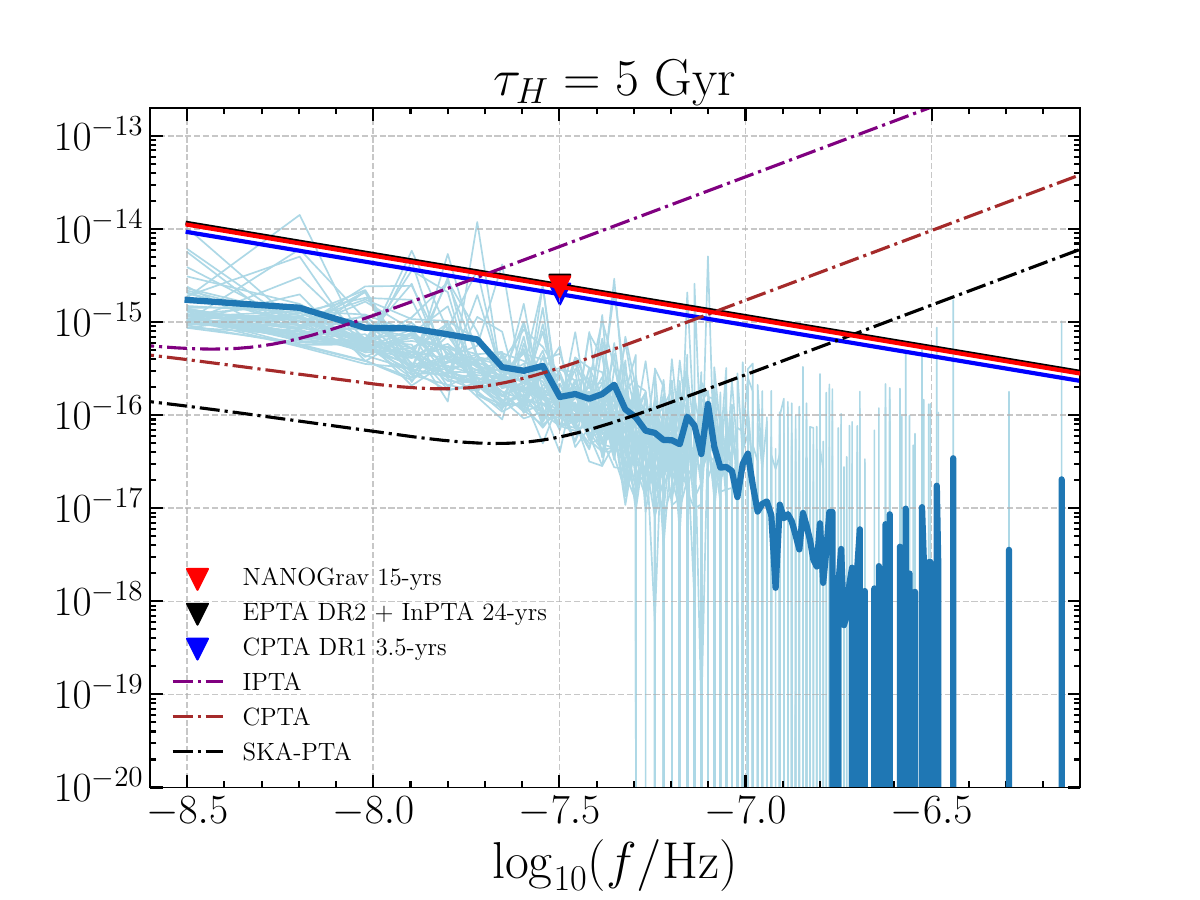}
\hspace{-0.83cm}
\includegraphics[width=0.35\textwidth]{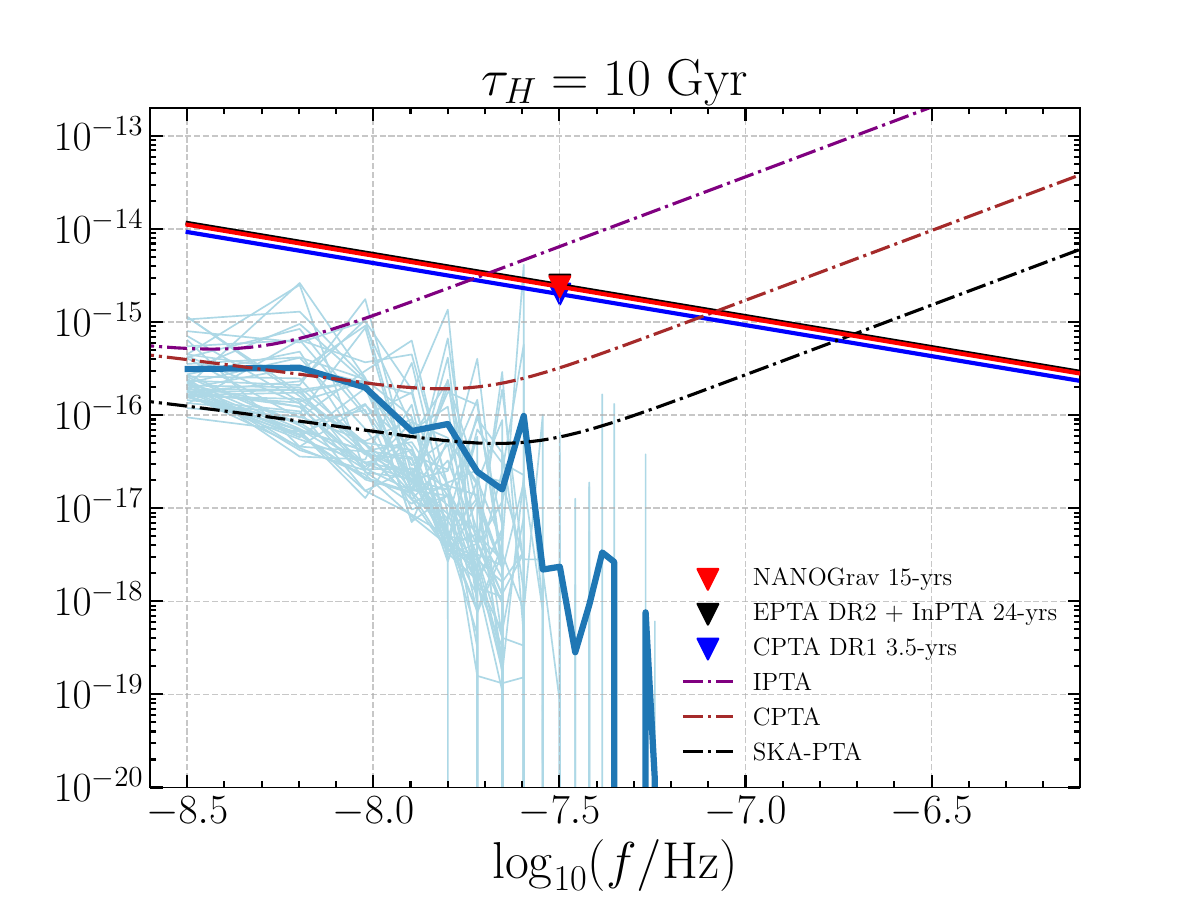}
\caption{The characteristic strain spectrum of GWB signal from our light-cone SMBBH events. { The three panels from left to right represent the results for hardening times of $0.1, 5$ and $10\,\mathrm{Gyr}$ respectively. The light blue lines represent the results of different realizations, while the dark blue line indicates the average of all realizations.} The latest results from NANOGrav \citep{2023ApJ...951L...8A}, EPTA \citep{2023A&A...678A..50E} and CPTA \citep{2023RAA....23g5024X} at the characteristic frequency of $\rm yr^{-1}$ are presented in the figure as a comparison. { The sensitivity curves of the three PTA configurations that we utilized for SNR calculations, including the current IPTA in this section and the future SKA-PTA and CPTA presented in Table \ref{tab:para} of Section 3.4, are also shown.}}
\label{fig:h_f}
\end{figure}

{ To calculate the resulting monopole part of the GWB signal and its characteristic strain spectrum, we divide the frequency range within the PTA frequency band spanning from {$f_{\rm min}$ to $f_{\rm max}$} into different frequency bins. The bin width is determined by $\Delta f = 1/T_{\rm obs}$, where $T_{\rm obs}$ is set to 10 years in this subsection. We select those SMBBHs whose observed frequencies are within the specified range in each bin. The resulting GW characteristic strain near a certain frequency can be calculated as a summation of strains from selected GW sources as:
{$$
h_{\rm c}(f_i)=\sqrt{\sum_{f_i<f_j\leq f_{i+1}} h_j^2f_j T_{\rm obs}},
$$
where $h_j$ can be computed according to Equation~\eqref{eq:h}, the summation over $j$ denote the superposition over all the GW events whose observed frequency $f_j$ falls within a specific frequency bin.}}
We present our results regarding the characteristic strain spectrum of GWB { for all the three hardening time scales} in Figure~\ref{fig:h_f}. In this figure, { we plot the results of all 50 realizations with light blue lines and show the average result of all realizations with a dark blue line.} We also indicated the data points at the characteristic frequency of ${\rm yr^{-1}}$ with the presupposed $-\frac{2}{3}$ power-law spectrum from the latest NANOGrav \citep{2023ApJ...951L...8A}, EPTA \citep{2023A&A...678A..50E} and CPTA \citep{2023RAA....23g5024X} results.
{ From the figure, it is evident that as the hardening time increases, the averaged amplitude of GWB is suppressed, particularly when the hardening time reaches a scale of $10\,\mathrm{Gyr}$, where a significant reduction in the amplitude is observed. {The characteristic amplitude $h_c$ at frequency $f = 1\,\mathrm{yr}^{-1}$ is $4.08 \times 10^{-16}$, $3.38 \times 10^{-16}$, and $2.19 \times 10^{-18}$ for $\tau_H = 0.1\,\mathrm{Gyr}$, $5\,\mathrm{Gyr}$, and $10\,\mathrm{Gyr}$, respectively; while at $f = 0.1\,\mathrm{yr}^{-1}$, $h_c=1.96 \times 10^{-15}$, $1.42 \times 10^{-15}$, and $3.13 \times 10^{-16}$ for the same $\tau_H$ values in corresponding order.} For the two cases with $\tau_H=$ 0.1\,Gyr and 5\,Gyr, where there are enough SMBBHs to demonstrate statistical trends,} our results generally exhibit the characteristic $-\frac{2}{3}$ power-law dependence, { although the higher probability of encountering bright sources in the low-frequency range has led to  slight deviations in the averaged curves.} On the other hand, in the high-frequency region the curve drops, which can be easily understood as the frequency upper limit $f_{\rm ISCO}$ is different from event to event, only events with small total mass can reach a relatively high signal frequency. This will lead to a decrease in the number of events in the high-frequency bins and a deviation from the $-2/3$ power-law behavior, { the same reason has also led to the relatively large fluctuation observed  in the higher frequency band of the characteristic strain spectrum. 
It can be noted that while the averaged curve lies below the observational results which have $A_{\rm yr^{-1}}\approx2\times10^{-15}$ at $f=1\,{\rm yr}^{-1}$, some random realizations can still reach observation levels at the characteristic frequency. Nevertheless, the projected averaged value resides well within the theoretical range [$1 \times 10^{-16}$, $5 \times 10^{-15}$], as supported by existing literature. We will discuss this point in the discussion section.}

\begin{figure}[ht!]
\centering
\includegraphics[width=0.32\textwidth]{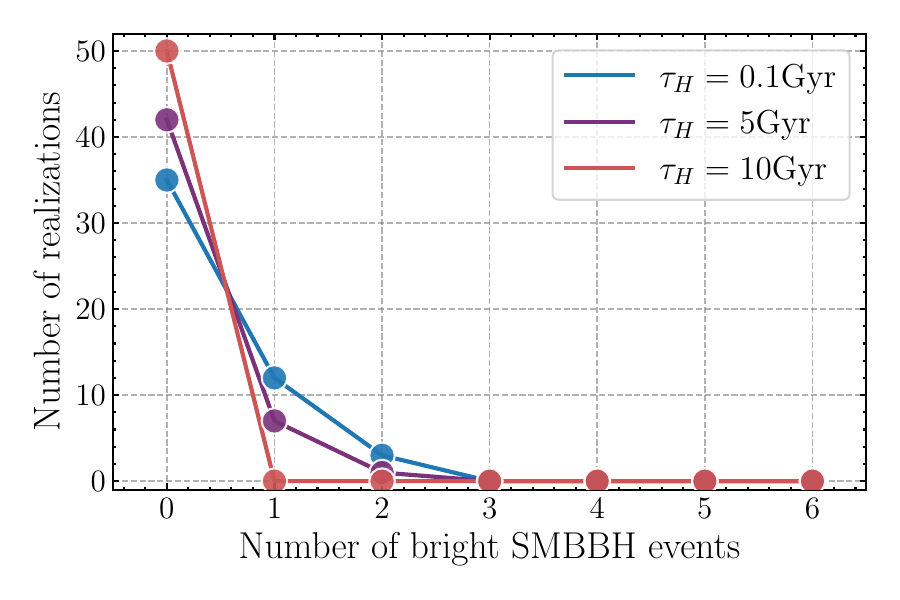}
\hspace{-0.3cm}
\includegraphics[width=0.32\textwidth]{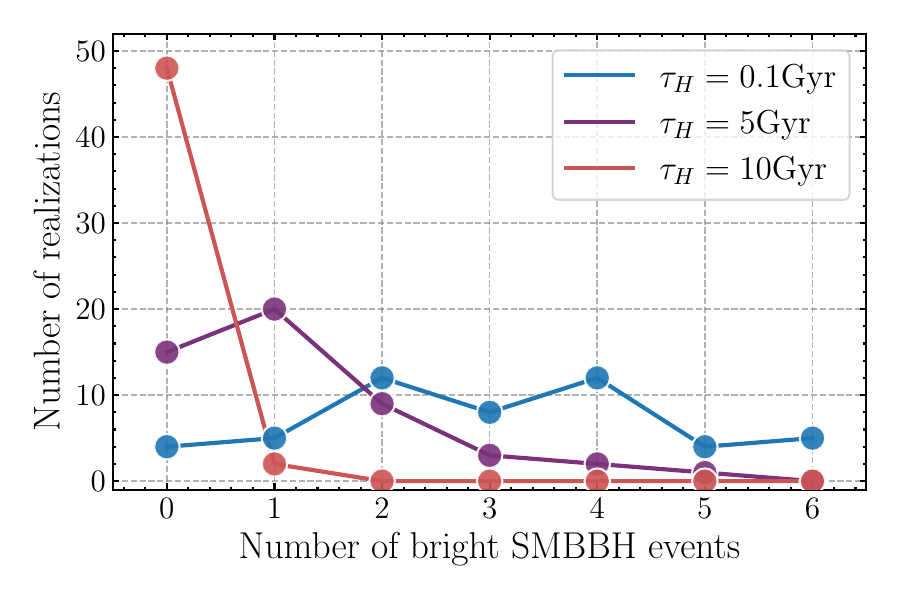}
\hspace{-0.3cm}
\includegraphics[width=0.32\textwidth]{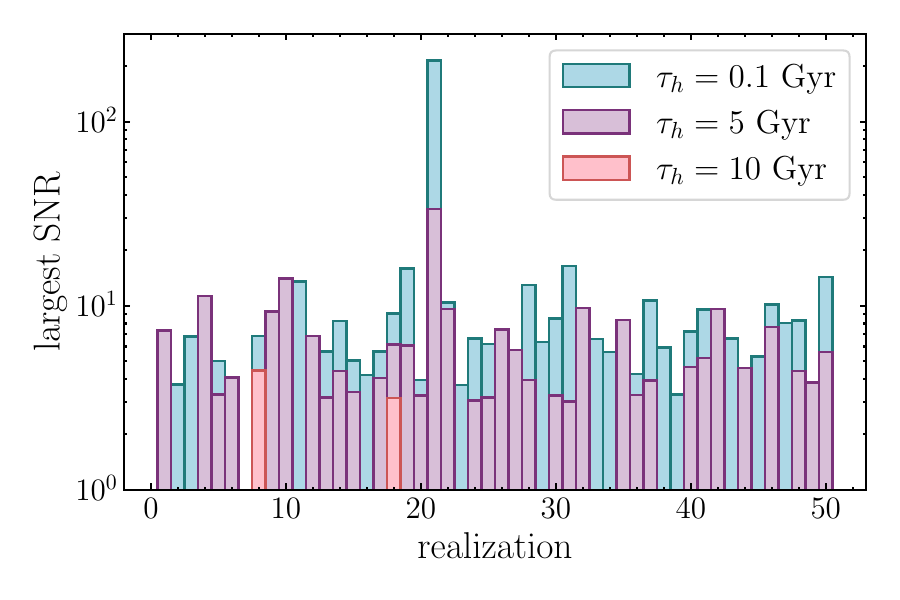}
\caption{ Left panel: The distribution of number of realizations versus different number counts of SMBBH events with SNR exceeding 8. Middle panel: The distribution of number of realizations versus different number counts of SMBBH events with SNR exceeding 3. Right panel: The maximum SNR in different realizations containing bright sources.}
\label{fig:currentSNR}
\end{figure}

{ We also investigated whether resolvable individual sources exist in SMBBH samples under different hardening times for current PTA detectors. We assumed 68 pulsars, a 10-year observation time, and a timing residuals of 350ns, which are typical for current IPTA. We calculated the SNR for all SMBBH events in the sample and counted the number of events with SNR exceeding 8 or 3. The results are shown in Figure \ref{fig:currentSNR}\footnote{The event with the largest SNR$>$200 in realization 21 has a distance of 0.67 Mpc, and the masses of the two SMBHs are $5.3\times10^{9}\,M_{\odot}$ and $6.7\times10^{7}\,M_{\odot}$ respectively. It coincidentally has both a relatively close distance and large black hole masses. We have checked that no SMBBH events come from the galaxy where Earth is placed.}. It can be seen from the figure that generally the number of bright sources shows an inverse correlation with the hardening timescale. Considering that current PTA experiments haven't  detected any individual sources yet, we will subtract these bright individual sources when investigating the anisotropic properties of background signals and {resovable individual sources under future PTAs} to ensure that our results better matches the actual observations. }

\subsection{Anisotropy}
Now we study the anisotropy properties of the GWB signal we get from our mock data, which is a result of GW sources and the GW strains come from them not being distributed isotropically in the virtual observable universe. The method we use allow us to estimate the anisotropy of GWB in a simulation environment more reflective to the real universe, as the cosmic large scale structure is incoporated. In this subsection, we will present the angular power spectrum of the GWB signal and the spatial distribution properties of the GW sources. {Given the observational challenges associated with detecting anisotropy, a 20-year total observation duration was adopted here for characterizing anisotropy properties.}

\subsubsection{Angular Power Spectrum}
\begin{figure}[ht!]
\centering
\includegraphics[width=0.6\textwidth]{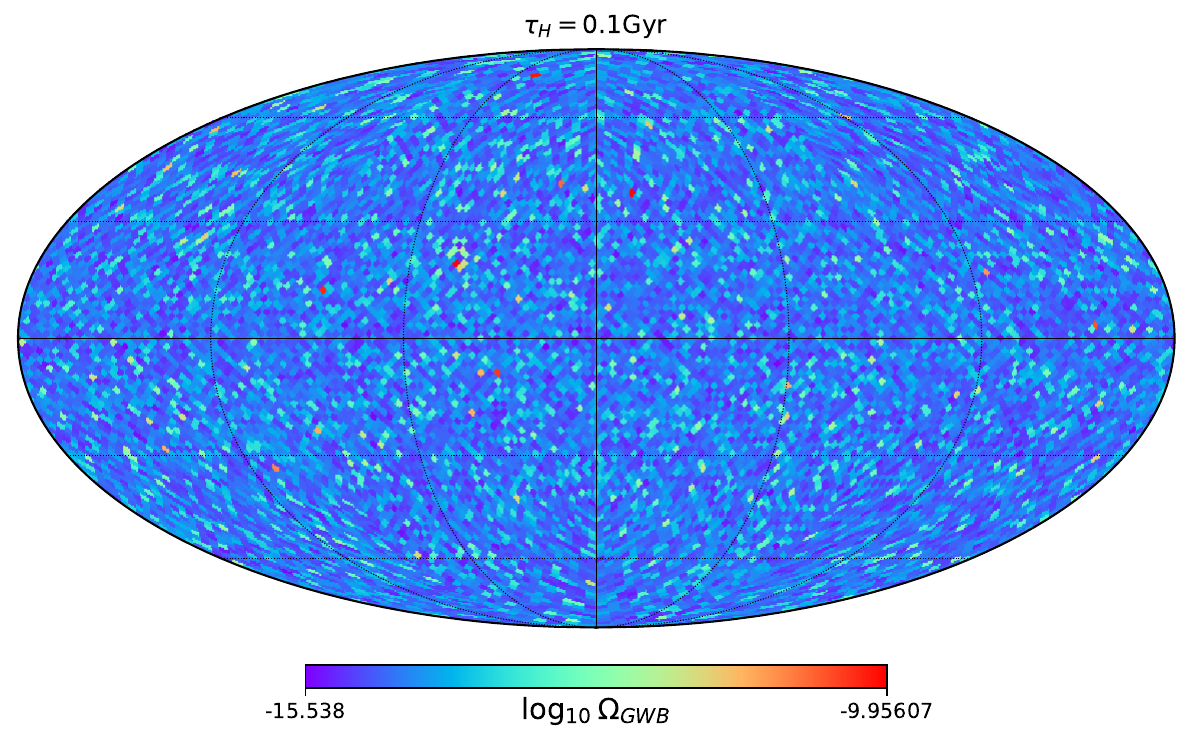}
\hspace{0cm}
\includegraphics[width=0.6\textwidth]{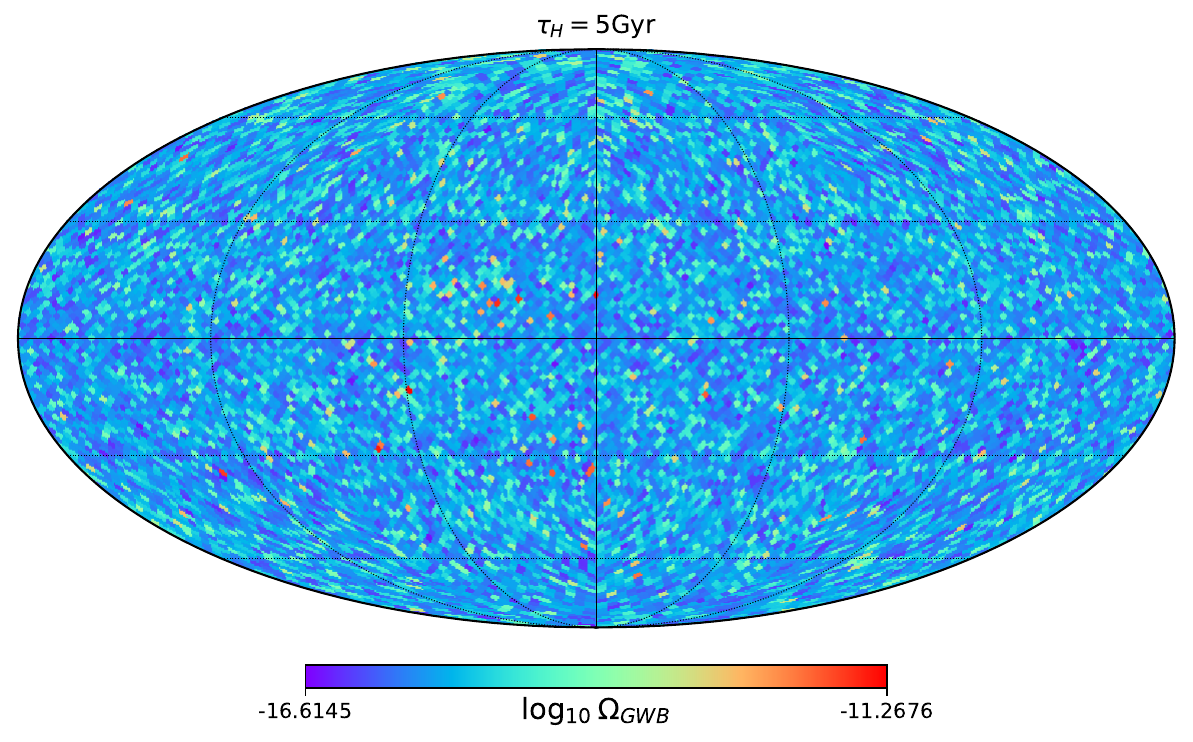}
\hspace{0cm}
\includegraphics[width=0.6\textwidth]{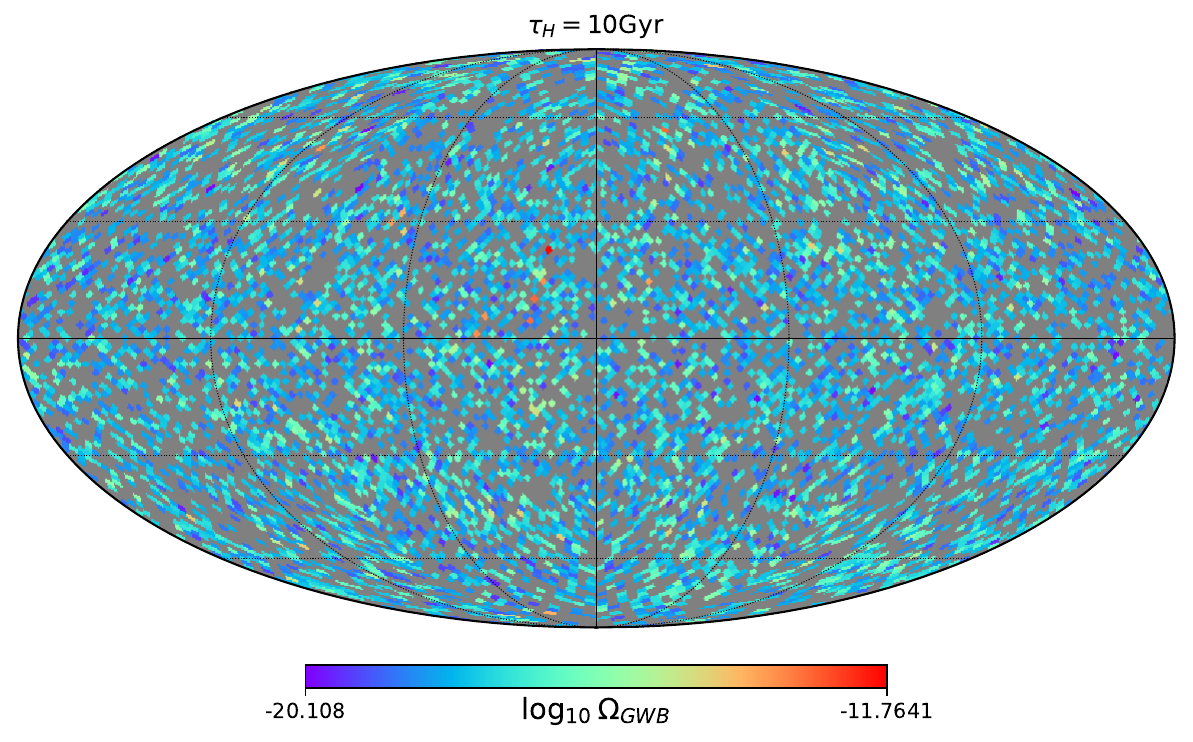}
\caption{ The skymap of GWB relative energy density from one representative realization, where $\Omega_{GWB}$ is the ratio of GWB energy density to the critical density. The three panels from up to down represent the results for hardening times of $0.1, 5$ and $10\,\mathrm{Gyr}$ respectively. The dark gray areas in the lowest panel denote the sky regions where no GW events are distributed.}
\label{fig:GWmap}
\end{figure}

We transform the three dimensional Cartesian coordinates $(x_s,y_s,z_s)$ of the mock SMBBH locations to spherical polar coordinates $(r,\theta,\phi)$, where $r=\sqrt{x_s^2+y_s^2+z_s^2}$, $\phi$ represents right ascension (the azimuthal angle respective to $x_s$-axis positive direction), and $\theta$ represents declination (elevating angle respective to $x_sOy_s$ plane). By superimposing the gravitational wave signals from all possible sources along a particular direction, sky maps of the background gravitational wave signal across the celestial sphere could be obtained. Similar to \cite{2019MNRAS.483..503Y,2013PhRvD..88h4001T},  spherical harmonic analysis could be used to quantitatively analyze the anisotropy properties of the GWB signal from mock SMBBHs.
The GW energy density $\rho(\hat{\Omega})$:
\begin{equation}
\rho(\hat{\Omega})=\frac{\pi}{4G}\int f^2h^2_c(f,\Omega)d\ln f
\label{eq:rho}
\end{equation}
is the superposition of the energy densities across all frequency bins in a specific direction, and can be expanded under the spherical harmonic basis:
\begin{equation}
\rho(\hat{\Omega})=\sum_{l=0}^{\infty}\sum_{m=-l}^{l}c_{lm}Y_{lm}(\hat{\Omega}),
\label{eq:rho_Y_lm}
\end{equation}
where the coefficient $c_{lm}$ can be calculated as:
\begin{equation}
c_{lm}=\int d\hat{\Omega} \rho(\hat{\Omega})Y_{lm}(\hat{\Omega} ).
\label{eq:c_lm}
\end{equation}
{  In practice}, the celestial sphere is divided into $N$ small regions to perform the superposition of GW signals in a specific direction, so the integral above should be replaced with a summation:
\begin{equation}
c_{lm}=\sum_{i=1}^{N} \rho(\hat{\Omega}_i)Y_{lm}(\hat{\Omega}_i )\frac{4\pi}{N}.
\label{eq:c_lmN}
\end{equation}
Then the angular power spectrum can be expressed as an average of all $|c_{lm}|^2$ with the same $l$ as:
$$
C_l=\sum_{m=-l}^{l} \frac{|c_{lm}|^2}{2l+1}.
$$

\begin{figure}[ht!]
\centering
\includegraphics[width=0.35\textwidth]{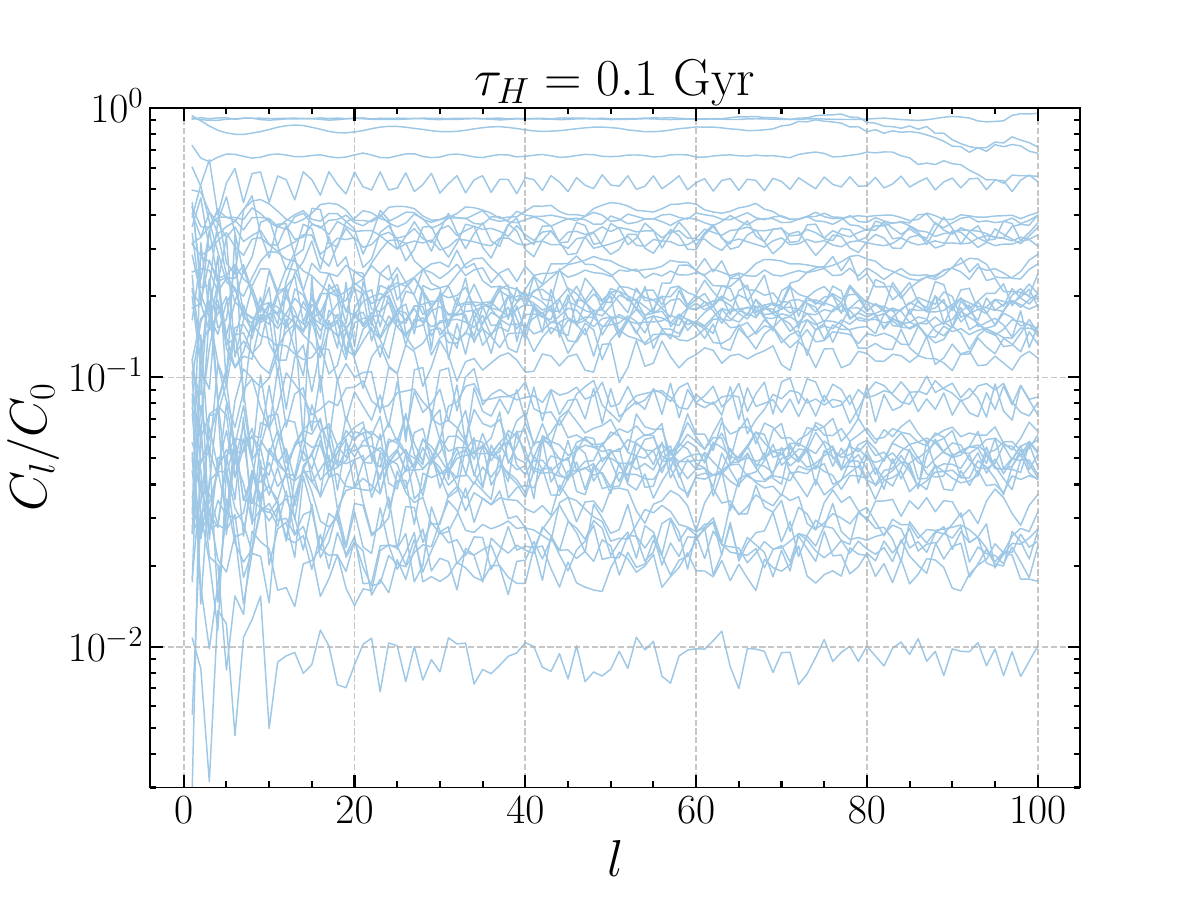}
\hspace{-0.83cm}
\includegraphics[width=0.35\textwidth]{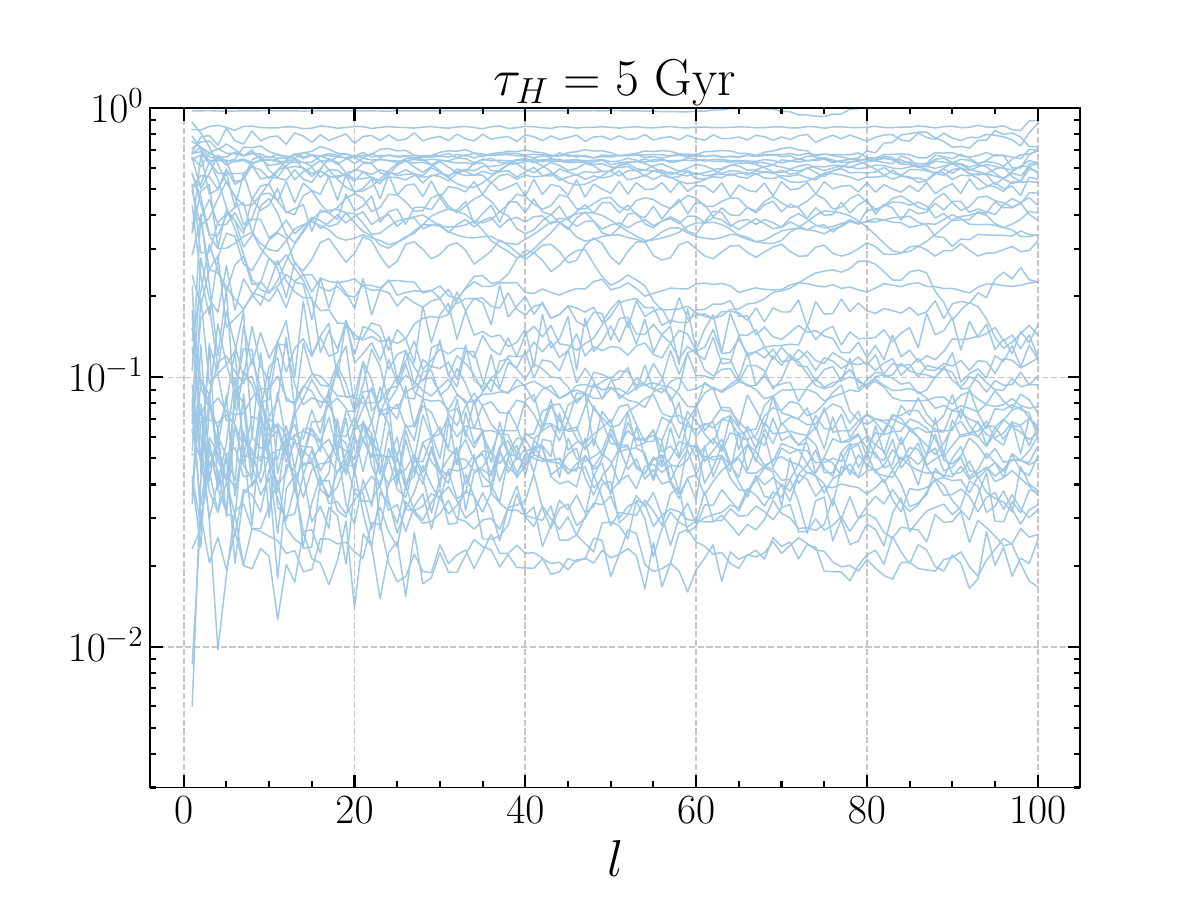}
\hspace{-0.83cm}
\includegraphics[width=0.35\textwidth]{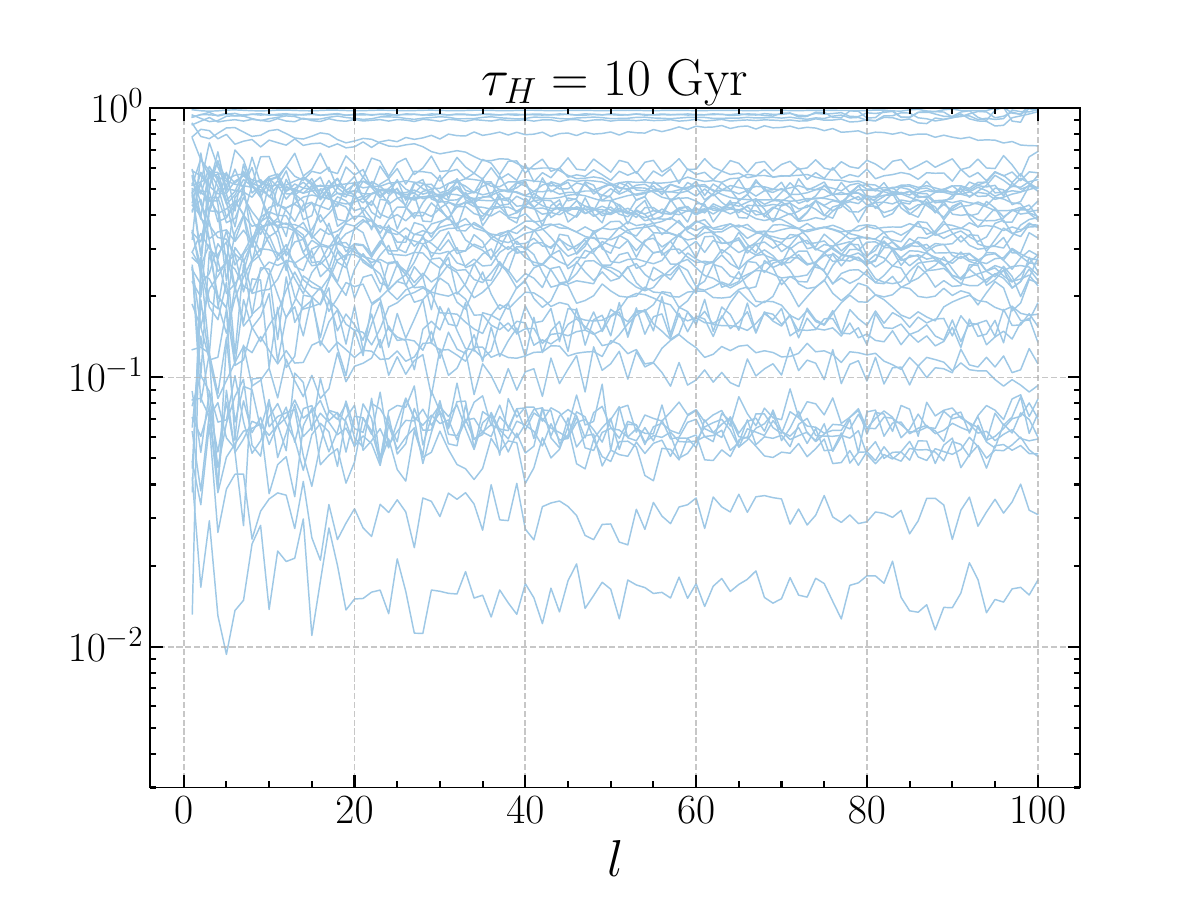}
\caption{The angular power spectrum coefficients $C_l$ as a function of $l$ for GWB energy density $\rho$ { after subtracting the events with SNR greater than 8 for all the 50 realizations of each hardening timescale. Panels from left to right represent the results for hardening times of $0.1, 5, 10\,\mathrm{Gyr}$ respectively.} }
\label{fig:c_lrho}
\end{figure}

\begin{figure}[ht!]
\centering 
\includegraphics[width=0.45\textwidth]{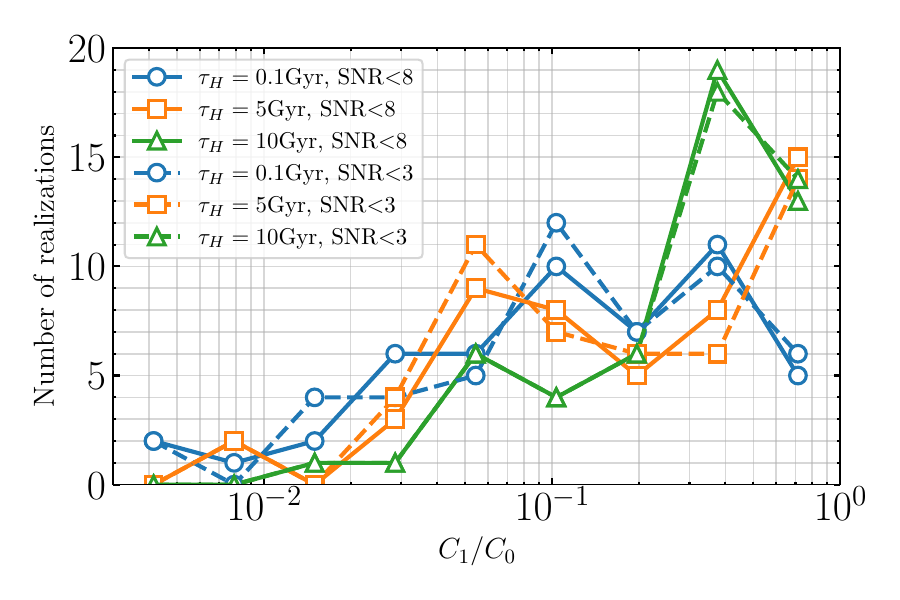}
\includegraphics[width=0.45\textwidth]{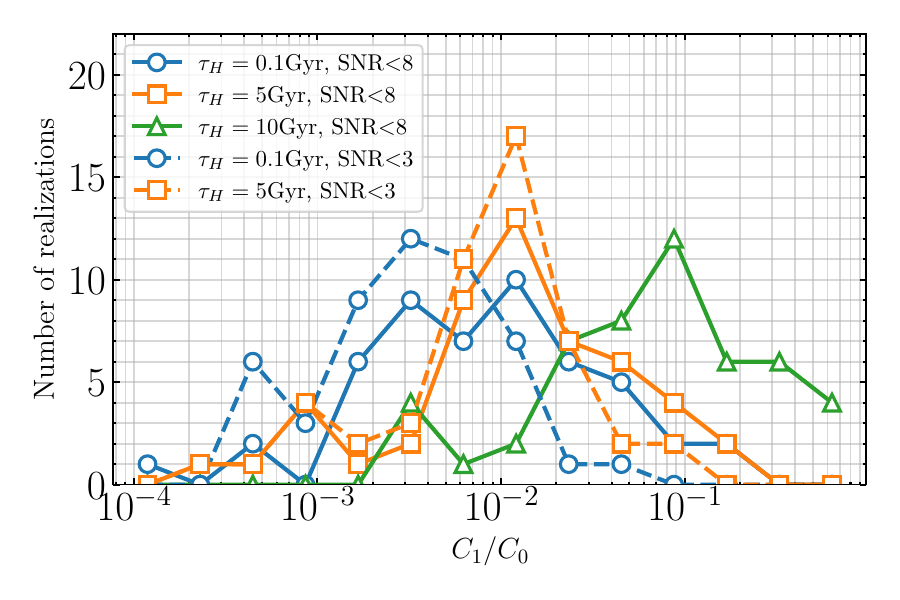}
\caption{ Left panel: the distribution of number of realizations versus the value of $C_1/C_0$ of total energy density for all the three hardening timescales.  Right panel: the distribution of number of realizations versus the value of $C_1/C_0$ for energy density in the lowest frequency bin for all the three hardening timescales. In both panels solid lines represent results after subtracting the events with SNR greater than 8, and dashed lines represent results after subtracting the events with SNR greater than 3. In the right panel, for $\tau_H=10$ Gyr case, since bright sources are absent in the lowest-frequency bin, the green line is completely unaffected by subtraction of additional bright sources.}
\label{fig:cldistri}
\end{figure}

\begin{figure}[ht!] 
\centering 
\includegraphics[width=0.35\textwidth]{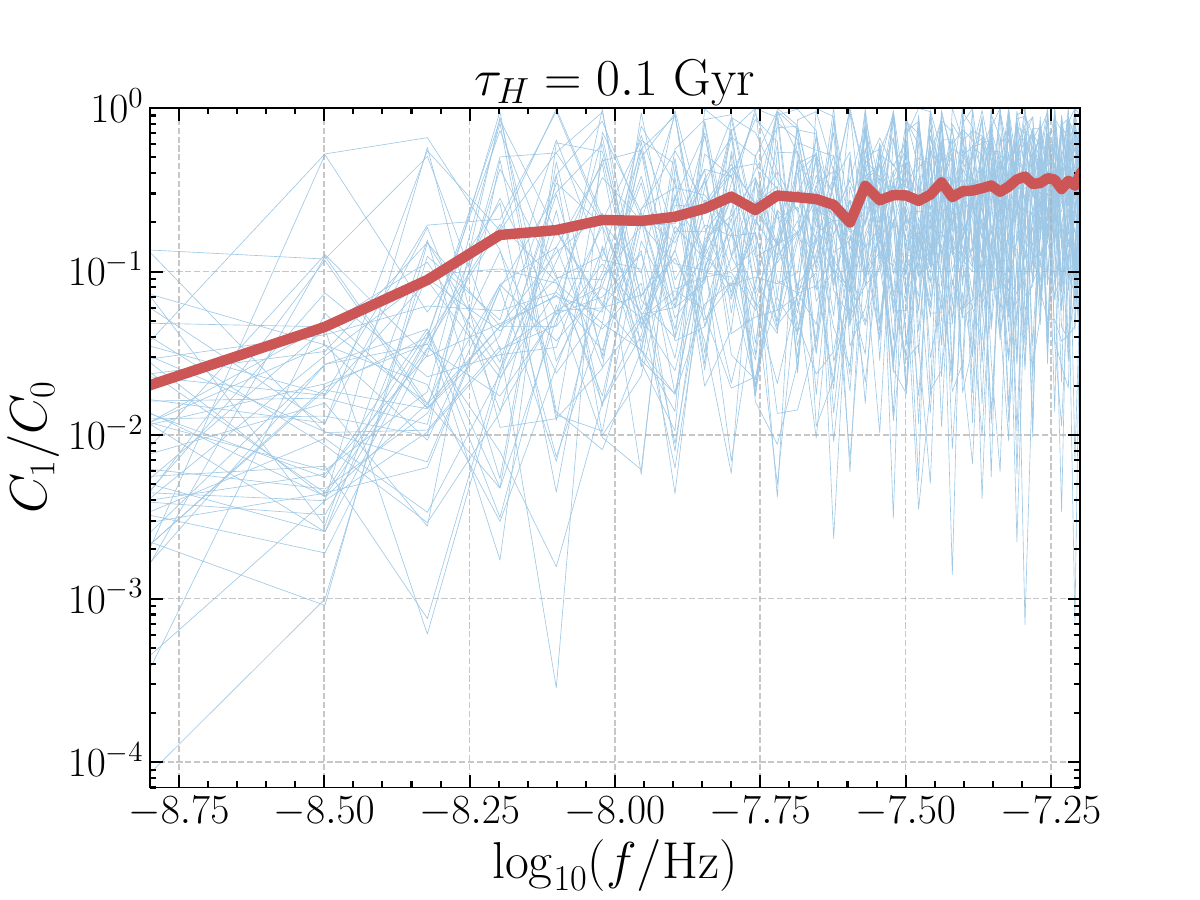}
\hspace{-0.83cm}
\includegraphics[width=0.35\textwidth]{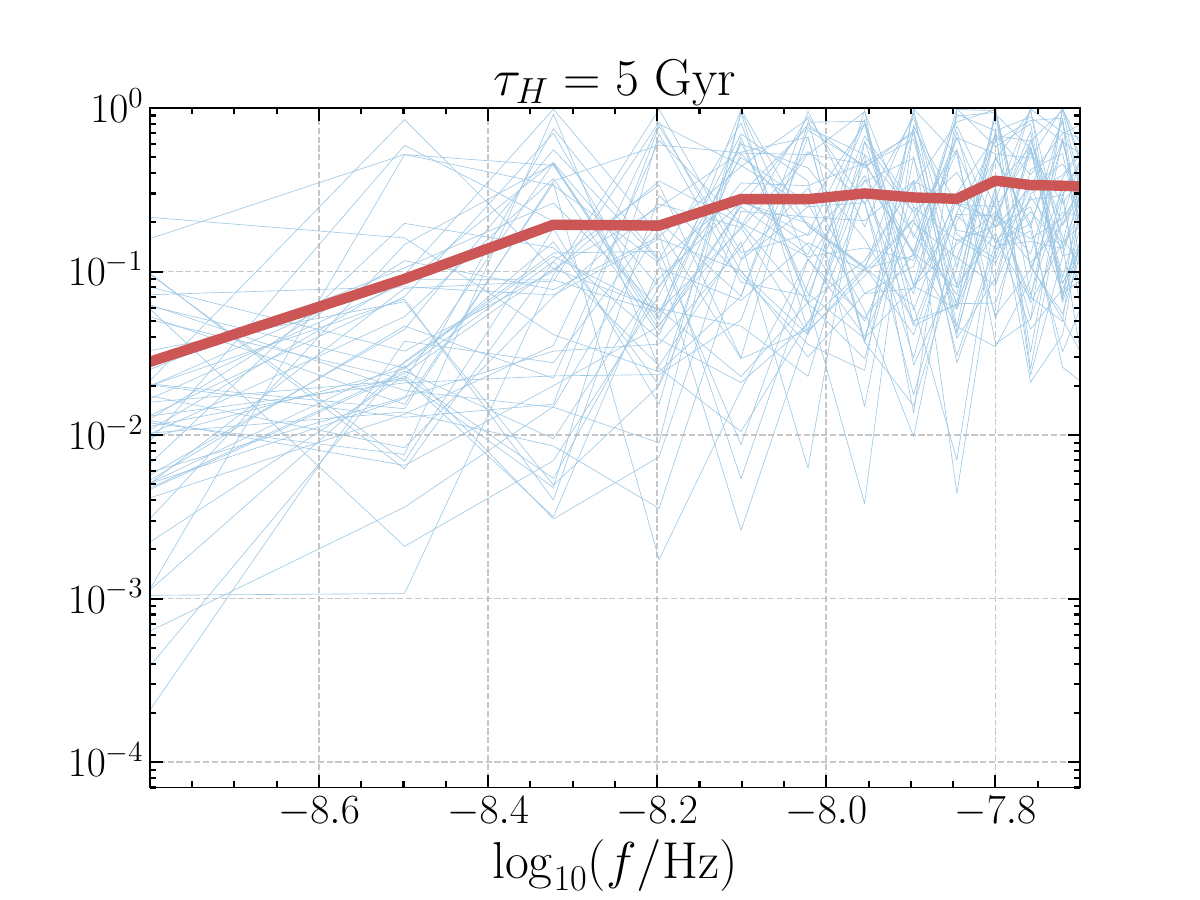}
\hspace{-0.83cm}
\includegraphics[width=0.35\textwidth]{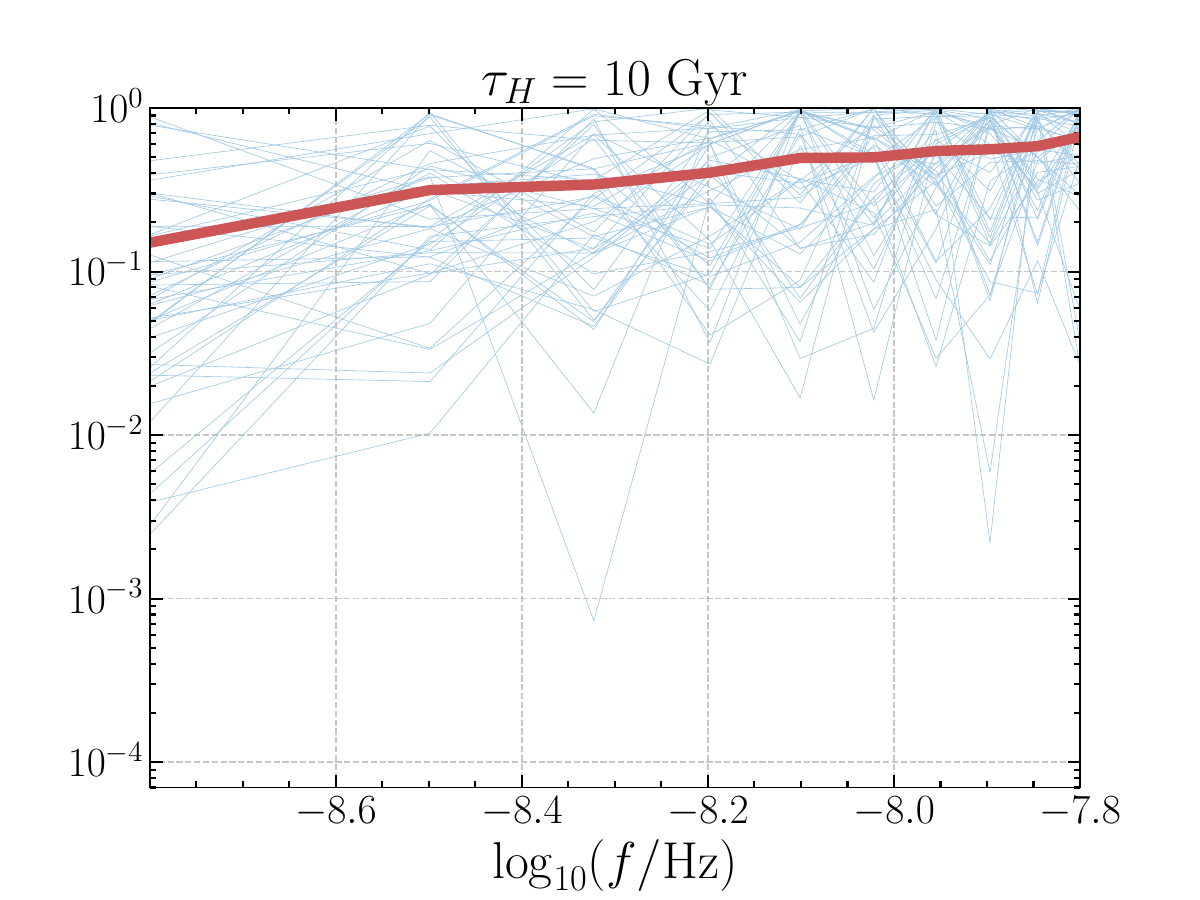}
\caption{ The angular power spectrum coefficients $C_1/C_0$ of GWB energy density $\rho$ in different frequency bins for all the three hardening time scales. The three panels from left to right represent the results for hardening times of $0.1, 5, 10\,\mathrm{Gyr}$ respectively. The light blue lines represent the results of different realizations, while the red lines indicate the average of all realizations. }
\label{fig:c_lrhof}
\end{figure}

{ We utilize \texttt{Healpix} to generate sky maps and calculate the corresponding angular power spectra of the GWB total energy density for all realizations. In Figure~\ref{fig:GWmap}, we show the resulted sky maps for one representative realization for each of the hardening time scale. For the angular power spectrum, considering individual sources hasn't been detected by current PTA experiments yet, we calculated the $C_l$ coefficients after bright sources (with SNR$>$8 or SNR$>$3) were subtracted and presented the SNR$<$8 case in Figure \ref{fig:c_lrho}. We can observe from all the panels that $C_l/C_0$ exhibits significant fluctuations in all cases. Overall, for most of the realizations of $0.1$, $5$ and $10\,\mathrm{Gyr}$ hardening timescales, the distribution of $C_1/C_0$ falls within the range of { $[3.0\times10^{-2}, 4.3\times10^{-1}]$, $[5.3\times10^{-2}, 6.8\times10^{-1}]$, $[7.7\times10^{-2}, 6.2\times10^{-1}]$ respectively (16\%-84\% distribution range).}
We also examined whether the level of anisotropy is sensitive to the substraction of additional bright sources by plotting in the left panel of Figure \ref{fig:cldistri} the distribution of number of realizations versus the value of $C_1/C_0$ after bright sources with SNR$>$8 or SNR$>$3 were subtracted. From this figure we can see that, although cases with shorter hardening time scales tend to exhibit higher $C_1/C_0$ ratios, the distribution ranges of $C_1/C_0$ values for all three hardening timescales show significant overlap in general. {While the removal of additional bright individual sources show negligible effect when $\tau_H = 10\,\mathrm{Gyr}$ (This is expected, as only one realization contains a single source with SNR$>$3 in this case),} it leads to a slight reduction in $C_1/C_0$ values for a small subset of realizations in the $\tau_H=0.1\,\mathrm{Gyr}$ and  $\tau_H=5\,\mathrm{Gyr}$ cases. Collectively, the $C_1/C_0$ ratio of total energy density is insensitivity to the subtraction of extra bright sources.
In Figure \ref{fig:c_lrhof}, we also plotted the $C_1/C_0$ value of energy density from all SNR$<$8 events across different frequency bins for all realizations. It can be seen from the figure that $C_1/C_0$ general increase with increasing frequency, this can be easily understood because as frequency increases, the number of SMBBHs decreases, leading to an increase in anisotropy. In the lowest frequency bin, {$C_1/C_0$ falls mostly within the range of $[2.2\times10^{-3}, 3.5\times10^{-2}$], $[4.0\times10^{-3}, 6.0\times10^{-2}]$, and  $[2.1\times10^{-2}, 2.9\times10^{-1}]$ for $\tau_H=0.1$, $5$, and $10\,\mathrm{Gyr}$ respectively (16\%-84\% distribution range). After substracting more bright sources with SNR$>$3, these distribution range further decreased to $[1.2\times10^{-3}, 9.1\times10^{-3}$], $[2.7\times10^{-3}, 2.0\times10^{-2}]$, and  $[2.1\times10^{-2}, 2.8\times10^{-1}]$ for the same $\tau_H$ values in corresponding order.} Compared to the total energy density distribution, all three hardening timescale cases exhibit lower anisotropy levels for energy density in the lowest frequency bin, and also demonstrating more divergent distribution ranges among themselves. We have checked that this arises because low-frequency events dominate the SMBBH population, while the $f^2$ factor in the energy density integral in  Eq.~\eqref{eq:rho} does not affect the $C_1/C_0$ ratio within individual frequency bins, it does modify this ratio for the total energy density by preferentially weighting higher-frequency contributions. Thus the directional dependence of GW frequency distributions become an important source of anisotropy in the total energy case, which introduces extra anisotropy in the distribution, and also mitigates the differences between results of different hardening timescales. As a comparison, we also plotted the distribution of number of realizations versus the value of $C_1/C_0$ for energy density in the lowest frequency bin after bright sources with SNR$>$8 or SNR$>$3 were subtracted in the right panel of Figure \ref{fig:cldistri}. As clearly demonstrated in the figure, the subtraction of additional bright sources yields greater reduction in anisotropy for energy density in the lowest frequency bin compared to the total energy density case, indicating that bright individual sources has more important contribution to anisotropy in this case. 

The distribution range of the power spectrum coefficients and their frequency-dependence are generally consistent with the results of \cite{2024ApJ...965..164G}, while our results demonstrate relatively lower anisotropy in the lowest frequency bin compared to this literature (which may be attributed to the fact that we have subtracted unrealistic bright sources that haven't been detected by current PTAs yet), and the fluctuations appear more pronounced in higher frequency bins (which may be attributed to the incorporation of large-scale structure in our sample distribution).
It also worth mention that in the lowest frequency bin, the behavior of the $C_l$ coefficients from most realizations resides well within the observational constraints, i.e., $C_{l=1}/C_{l=0}<20\%$ \citep{2023ApJ...956L...3A}, however, in the high-frequency range, this limit is more likely to be exceeded, a manifestation that aligns with the results reported in  \cite{2024ApJ...965..164G}.
}

\subsubsection{GW Sources in Shells}
\label{subsec:shell}
We are interested in whether the distribution of SMBBHs can reproduce the large-scale structure of the universe, and thus be used for cosmological purpose like galaxies. Given that the GW sources are situated at different distances from us, it is difficult to determine the physical scale corresponding to specific angular scales. 
To better study the clustering patterns exhibited in the spatial distribution of the GW sources from angular power spectrum, 
{  we chose one realization from the most optimal case with $\tau_{H}=0.1\,\mathrm{Gyr}$ and selected} GW sources located in different spherical thin shells with radius from $R-dR$ to $R+dR$, where we choose $dR=10$\,Mpc and radius $R$=382, 764 and 1528\,Mpc, which corresponds to redshift $z=0.09$, $0.19$ and $0.41$ respectively.

\begin{figure}[ht!]
\centering 
\includegraphics[width=0.6\textwidth]{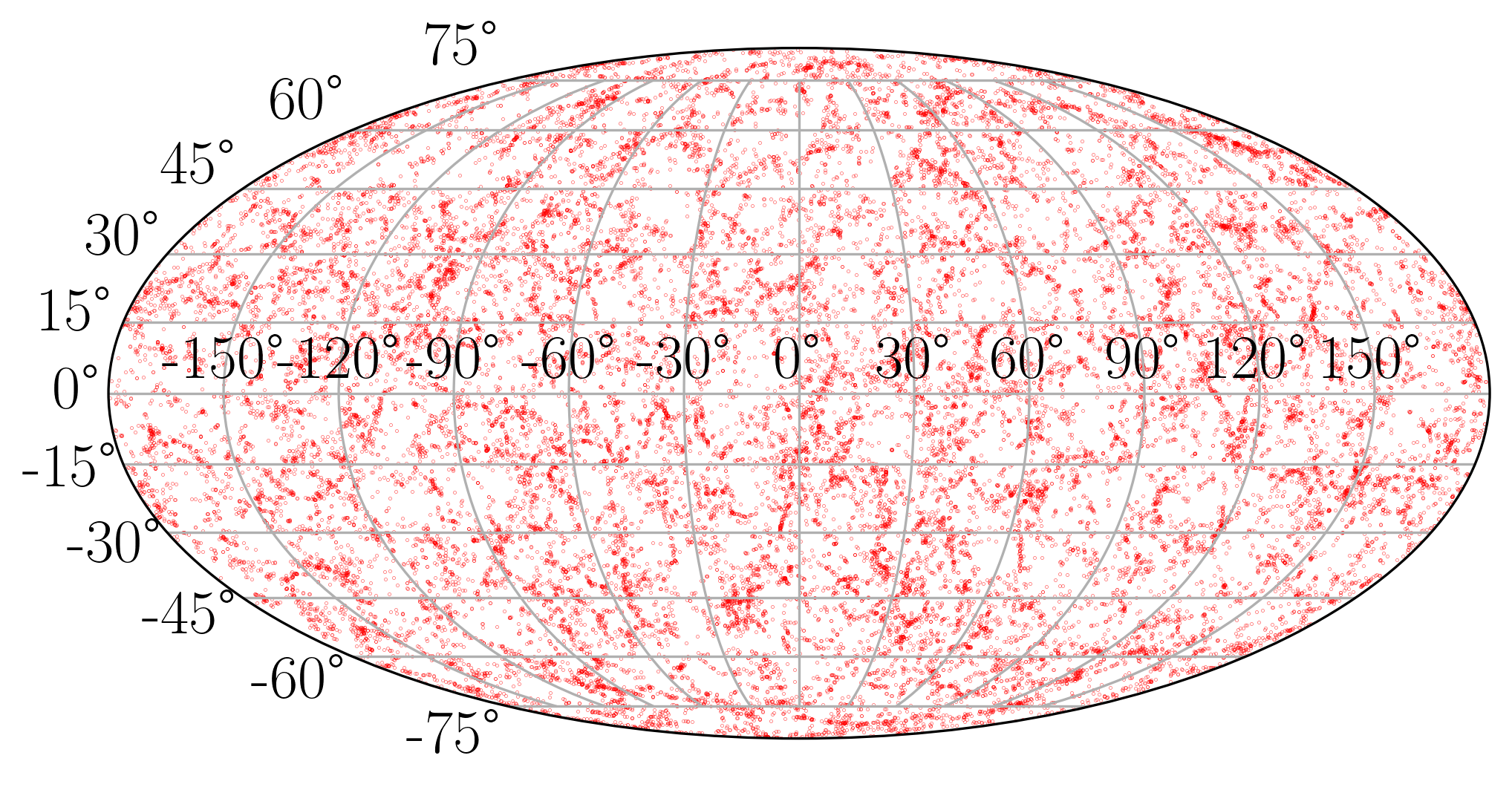}
\includegraphics[width=0.6\textwidth]{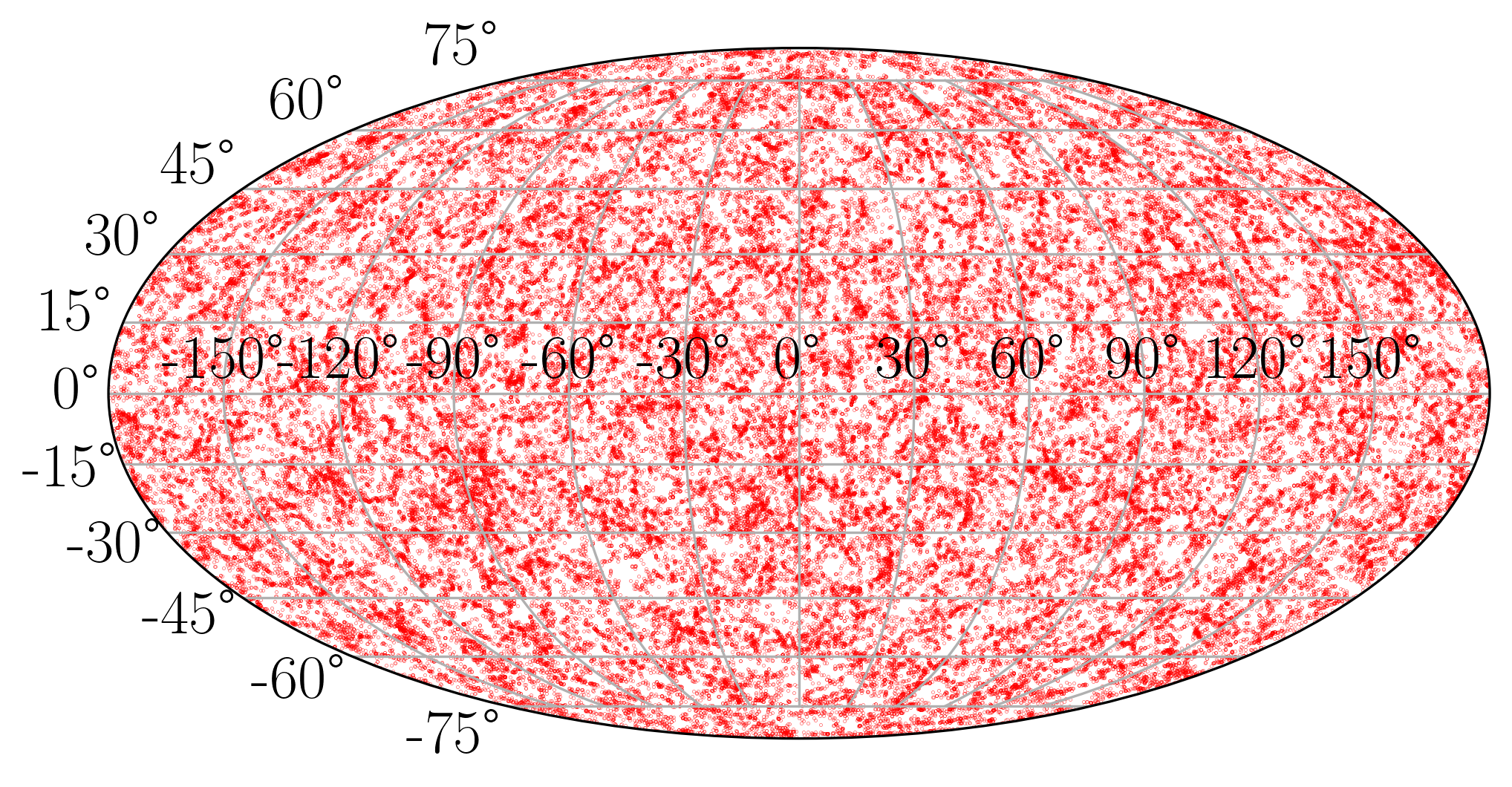}
\includegraphics[width=0.6\textwidth]{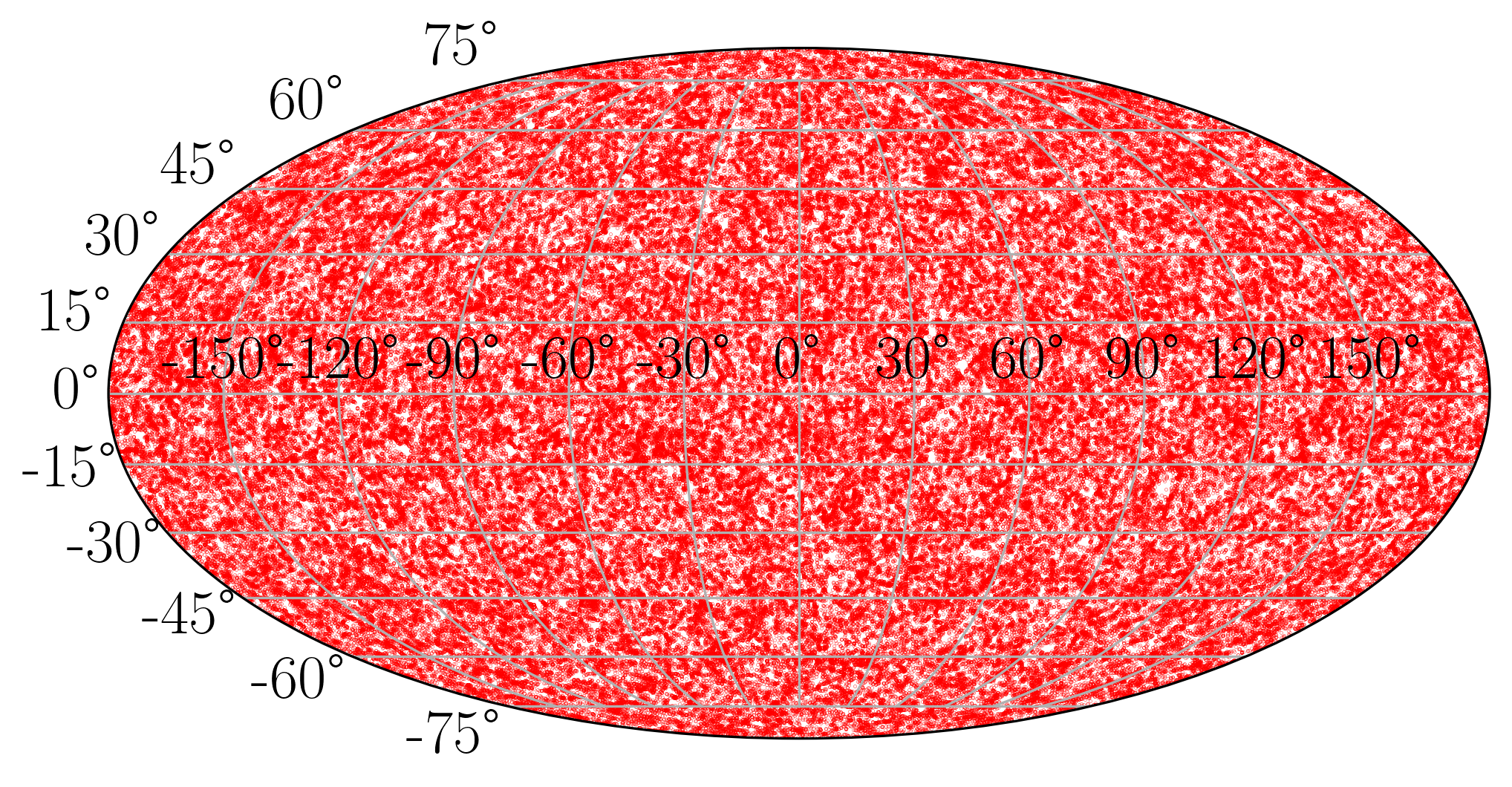} 
\caption{The angular distribution sky map of GW sources in spherical shells with radius $R=$ 382, 764 and 1528\,Mpc, respectively (from top to bottom) { from one representative realization with $\tau_H=0.1\,\mathrm{Gyr}$}, where red point represents the position of each GW source.}
\label{fig:GWshell}
\end{figure}

\begin{figure}[ht!]
\centering 
\includegraphics[width=0.8\textwidth]{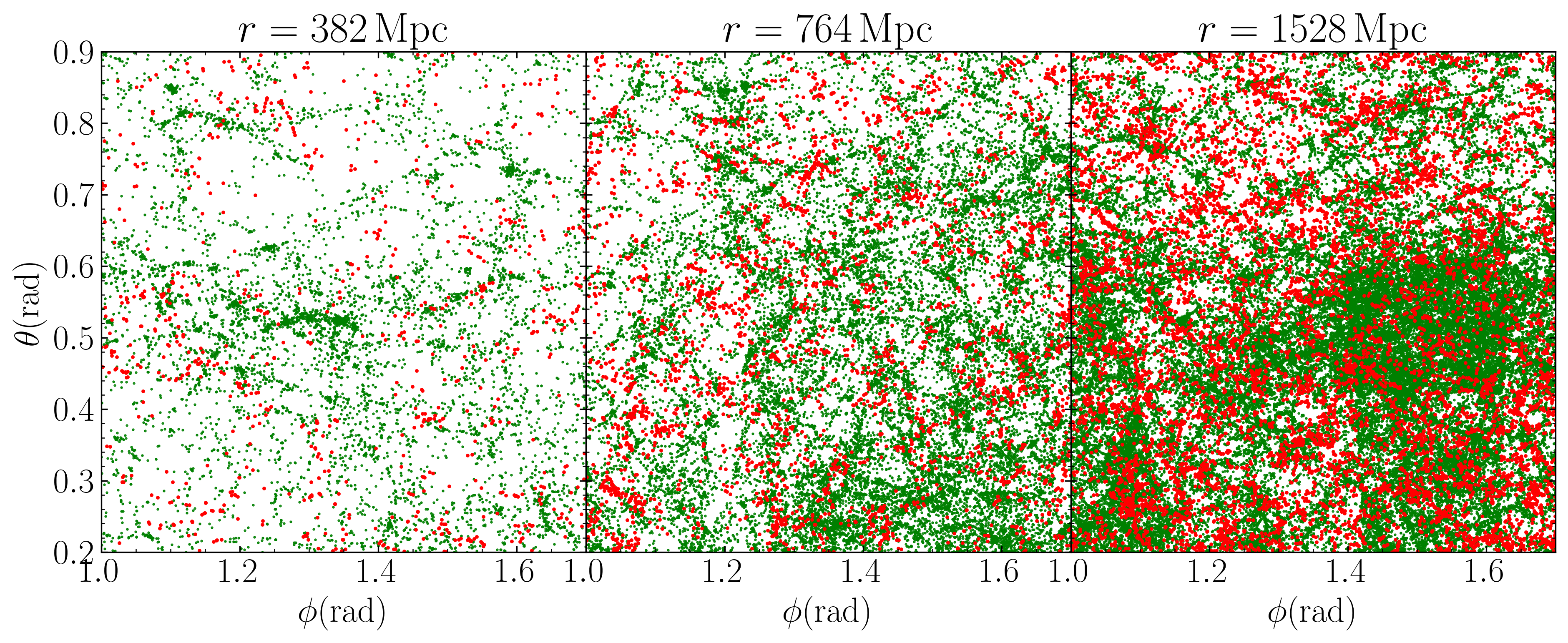}
\caption{A small portion ($\phi\in[0.7,1.7]$\,rad, $\theta\in[0.2,0.9]$\,rad) of the sky map with both GW sources and galaxies in spherical shells with radius $R=$382, 764 and 1528\,Mpc, respectively (from left to right). The red points represent the positions of SMBBHs, while the green points represent the positions of galaxies.}
\label{fig:GWandGal}
\end{figure}

Figure~\ref{fig:GWshell} shows the angular position distribution sky map of these GW sources in the three different spherical shells (from top to bottom). As the radius of the spherical shell increases, more and more SMBBHs gather within the range of the shell, thus appearing increasingly crowded. In Figure~\ref{fig:GWandGal} we also plotted a small patch of the celestial sphere, where we simultaneously displayed the distribution of SMBBHs and background galaxies. 
The distribution of SMBBHs is sparser than that of galaxies. It appears that SMBBHs alone may only provide incomplete information about the large-scale structure of the universe.

\begin{figure}[ht!]
\centering 
\includegraphics[width=0.6\textwidth]{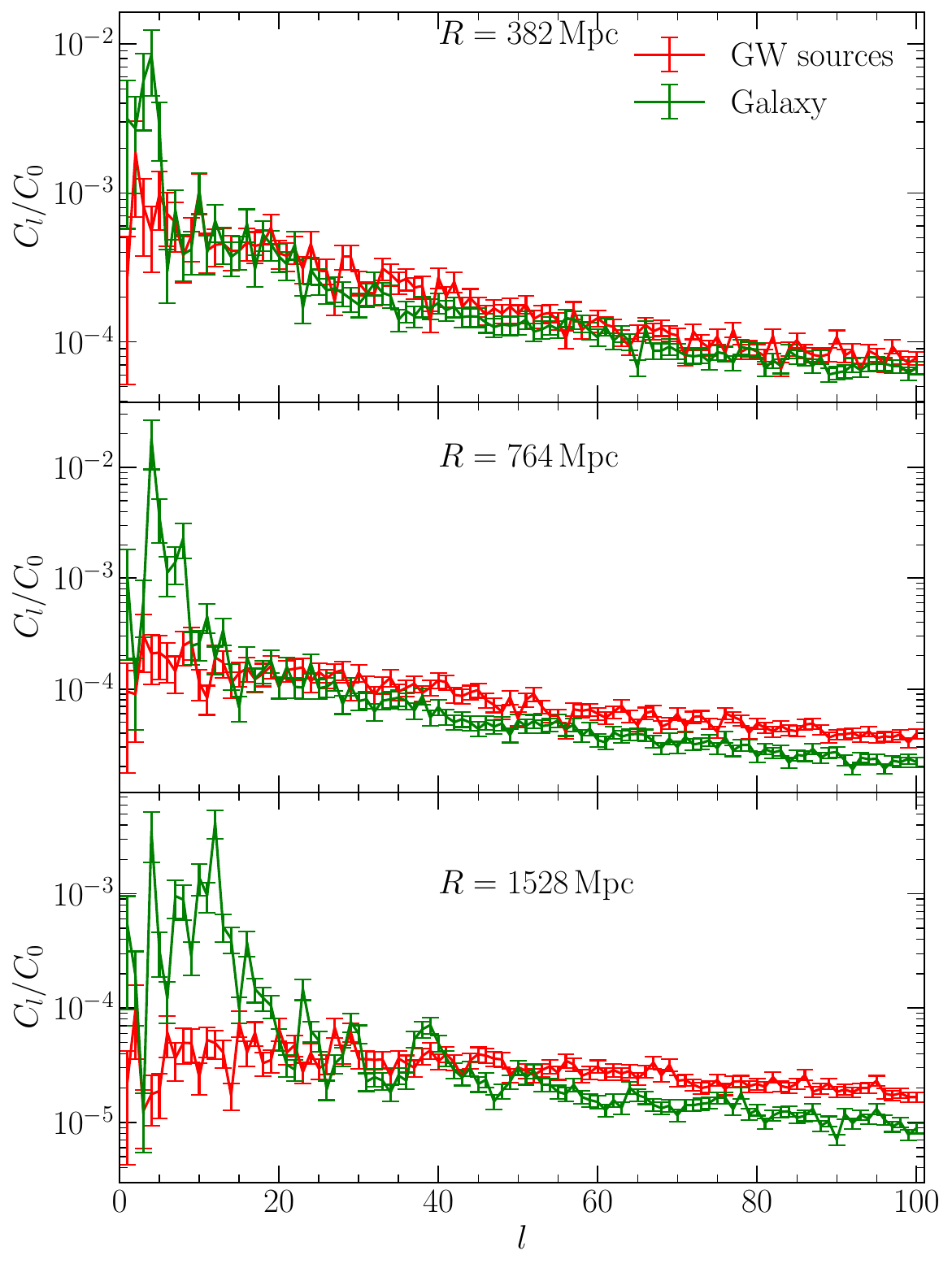} 
\caption{Angular power spectrum coefficients $C_l/C_0$ as a function of $l$ for the number density of GW sources (red lines) and galaxies (green lines) in different spherical shells with radius $R=382$, $764$ and $1528$ Mpc respectively (from up to down) { from one representative realization with $\tau_H=0.1\,\mathrm{Gyr}$}. }
\label{fig:shellcompare}
\end{figure}

In order to make a quantitative estimation about to what extent could the distribution of SMBBHs trace the underling matter distribution of the universe, we calculated the $C_l$ coefficients { with \texttt{Healpix}} for the number density of both GW sources and galaxies in these spherical shells and show the results in Figure~\ref{fig:shellcompare}. 
Different $l$ corresponds to different angular scale $\alpha$ based on the relationship $\alpha=\frac{\pi}{l}$.
When the radius $R$ of the spherical shells is fixed, a certain angular scale $\alpha$ corresponds to a specific physical scale $R\alpha$. 
We observe that, compared to SMBBHs, the $C_l/C_0$ values of galaxy distribution exhibit more significant fluctuations when $l$ is relatively small, which indicates the stronger clustering of galaxies on large scales. In contrast, $C_l/C_0$ of SMBBH distribution show less variation within the corresponding range. { On the other hand, the behavior of the $C_l/C_0$ coefficients of SMBBHs and galaxies show more agreement at higher multipole moments (large $l$). The values of $C_l/C_0$ of SMBBHs in this range are slightly larger than those of galaxies, indicating marginally stronger clustering of SMBBHs on smaller scales. Moreover, the discrepancy between the $C_l/C_0$ values decreases with smaller shell radius, while they become nearly identical within the smallest shell. This indicates that only on relatively smaller physical scales, the distribution of SMBBHs may exhibit similar patterns to that of galaxies. Our results are basically consistent with the behavior of cross correlations between GWB and galaxies shown in \cite{2024arXiv241100532S}, where larger $l$ (smaller scale) leads to stronger correlations between the two and higher sigma-level detectability.}

\begin{figure}[ht!]
\centering 
\includegraphics[width=0.9\textwidth]{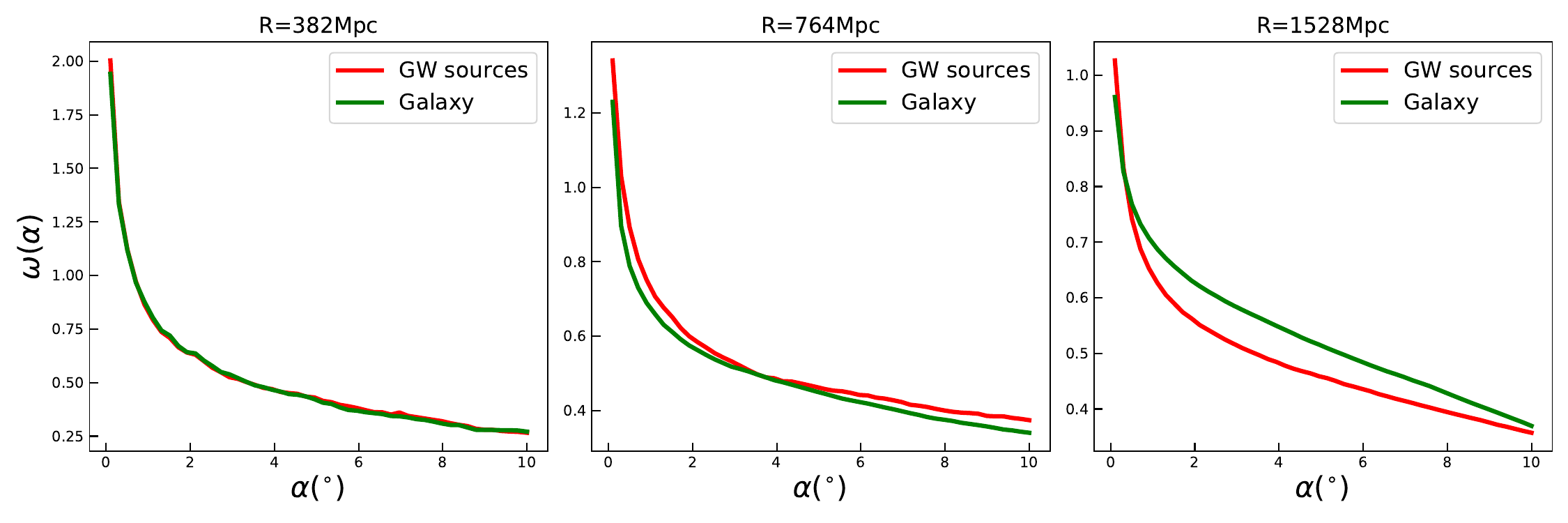} 
\caption{The two-point correlation function $\omega$ of galaxies and SMBBHs as a function of angle $\alpha(^{\circ})$ for spherical shells with radius $R=382$, $764$ and $1528$ Mpc respectively (from left to right) { from one representative realization with $\tau_H=0.1$ Gyr}. }
\label{fig:correlationshell}
\end{figure}

We further utilized \texttt{Corrfunc} to { calculate} the angular two point correlation function $\omega(\alpha)$ in these three shells as defined by the following formula:
\begin{equation}
\omega(\alpha)=\frac{DD(\alpha)-2DR(\alpha)+RR(\alpha)}{+RR(\alpha)}.
\label{eq:correlation}
\end{equation}
where $DD(\alpha)$ and $RR(\alpha)$ correspond to the number of SMBBH or galaxy pairs with angular separation $\alpha$ in data-data and random-random catalogs, respectively, whereas $DR(\alpha)$ corresponds to the number of
pairs with separation $\alpha$ calculated between data and random catalogs. The results are shown in Figure \ref{fig:correlationshell}.
In the figure, we observe that the angular correlation functions exhibit similar trends to those of the angular power spectra. { For smallest spherical shells, the correlation functions of SMBBHs and galaxies are nearly identical to each other. However, for larger spherical shells or larger angles, the correlation functions of the two show more deviations. The angular correlation of SMBBHs is slightly stronger than that of galaxies on smaller angular scales, while the angular correlation of galaxies is stronger than that of SMBBHs on larger angular scales.} The deviation behavior on large scales is understandable, because as can be clearly seen from Figure \ref{fig:GWshell} and Figure \ref{fig:GWandGal}, the distribution of SMBBHs is more homogeneous than that of galaxies. Therefore, they can only reproduce the clustering pattern of galaxies on small scales, while the more densely populated regions of galaxies on large scales are not reflected in the distribution of SMBBHs.

Of course, these are ideal theoretical discussions within the simulated universe, but these results may offer insights for exploring the correlations between the anisotropy of the GWB signal and
the distribution of galaxies. We will offer more discussion on this point in the discussion section.

\subsection{Individual Source Detectabilities}
\label{sec:detect}

\begin{table}
\caption{Parameter settings for several PTAs. Columns from left to right denote the notation of the PTA, total number of pulsars $N_{\rm p}$ observed by the PTA, the timing precision $\sigma_{\rm t}$, the total observational period $T_{\rm obs}$ and the mean cadence $\Delta t$ \citep{2022ApJ...939...55G,2023ApJ...955..132C,2025ApJ...978..104G}.
}
\begin{center}
\begin{tabular}{ccccc}
\hline \hline
PTAs & $N_{\rm p}$ & $\sigma_{\rm t}$\,(ns) & $T_{\rm obs}$\,(yr) & $\Delta t$\,(yr)\\ \hline

CPTA & 100 & 20 & 20  & 0.04 \\
SKA-PTA & $10^3$ & 10 & 20 & 0.04 \\ \hline \hline
\end{tabular}
\end{center}
\label{tab:para}
\end{table}
{ In the previous sections, we subtracted resolvable bright individual sources for current IPTA when analysing the characteristics of GWB signal in order to better reflect the current observation status.}
It is expected that more individual GW sources can be detected by more powerful PTAs with enough SNR $\varrho$ ($\varrho>8$) in the future. So these individual GW sources can actually be resolvable and separated from background. 
For future CPTA and SKA-PTA featuring greater detection capability and higher accuracy (The parameters expected to achieve for their experimental configurations are presented in Table~\ref{tab:para}), assuming a 20-years' observation period   and considering unresolvable GWB as noise, we calculated the SNR of our sampling SMBBHs based on { the following equation \citep{2022ApJ...939...55G, 2025ApJ...978..104G}:
\begin{equation}
\varrho^2 \approx N_{\rm p}\frac{\chi^2h_{\rm c}^2}{h_{\rm n}^2},
\end{equation}
where 
\begin{equation}
h_{\rm n}(f)=\sqrt{fS_{\rm n,s}+h_{\rm b}^2},
\label{eq:h_nB}
\end{equation}
and $S_{\rm n,s}(f)=8\pi^2\sigma_{\rm t}^2 f^2\Delta t$ is the power spectral density contributed by shot noise, and $h_{\rm b}$ is the unresolvable GWB signal we produced.}

\begin{figure}[htbp]
    \centering
    \begin{minipage}[t]{0.3\textwidth}
        \centering
        \includegraphics[width=\linewidth]{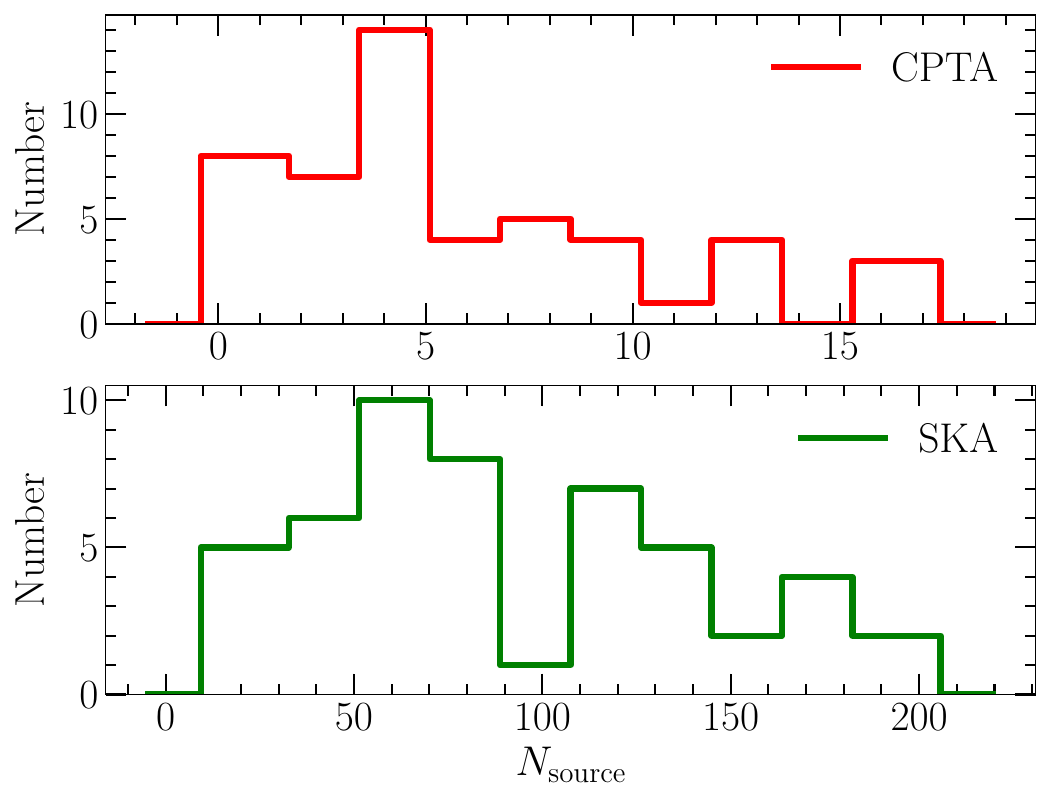}
    \end{minipage}
    \begin{minipage}[t]{0.3\textwidth}
        \centering
        \includegraphics[width=\linewidth]{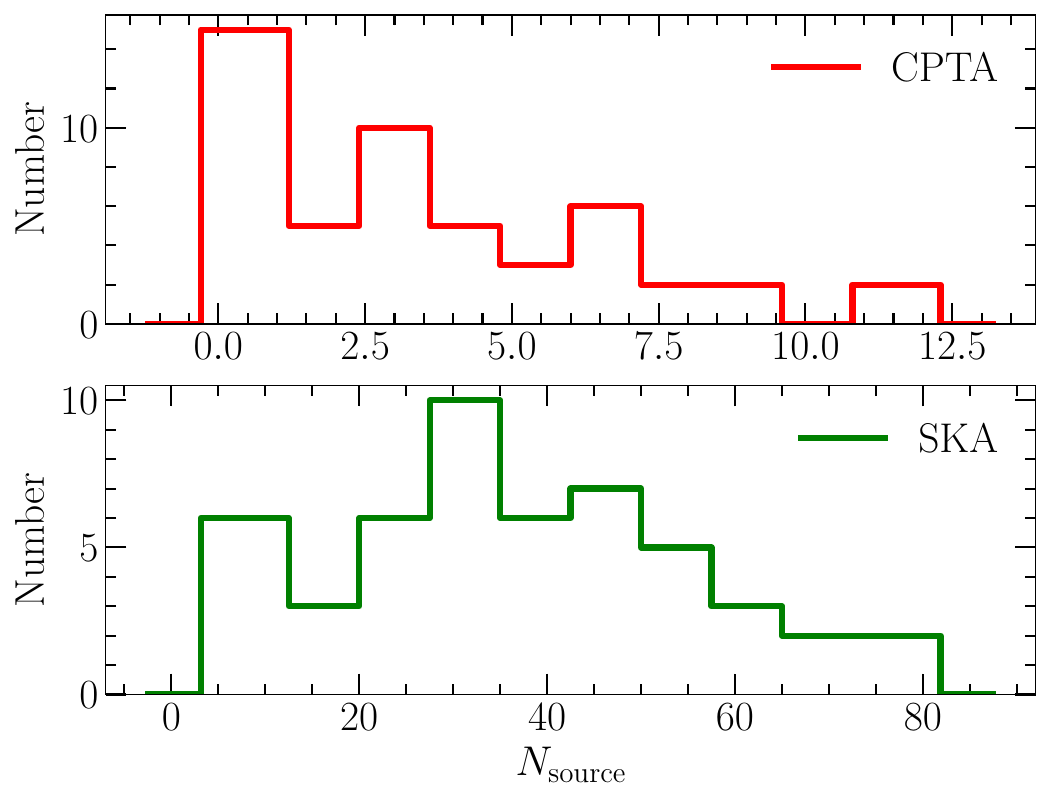}
    \end{minipage}
    \begin{minipage}[t]{0.3\textwidth}
        \centering
        \includegraphics[width=\linewidth]{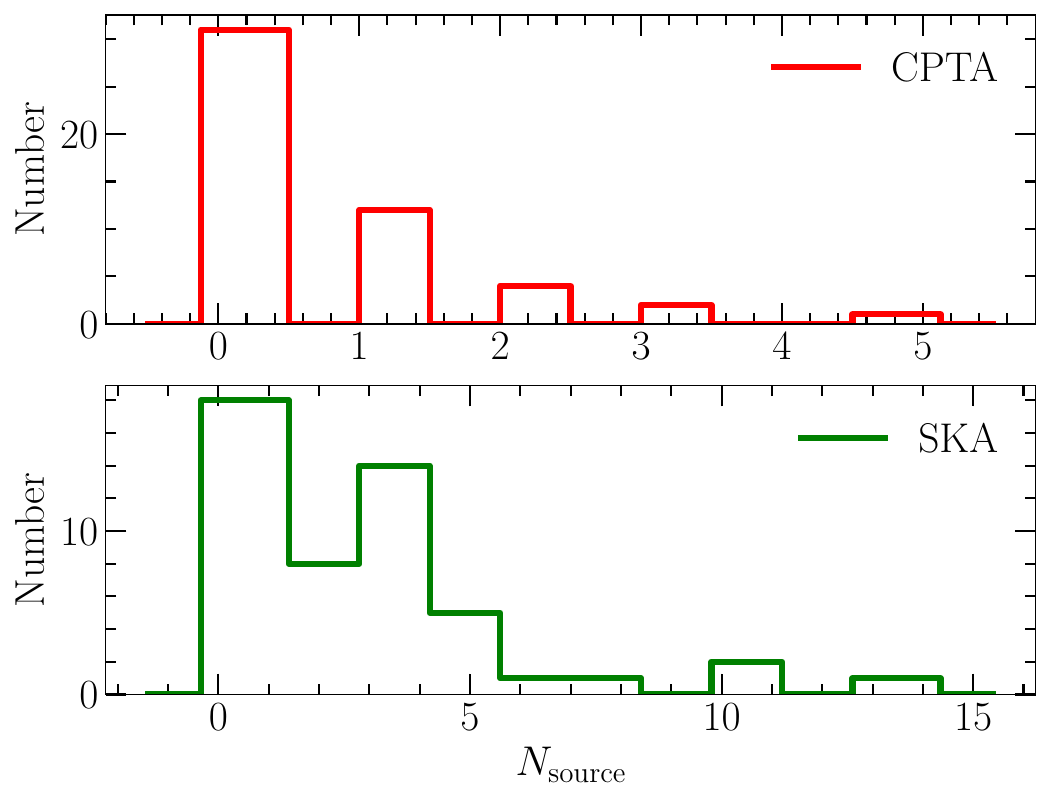}
    \end{minipage}
\caption{  The number of realizations versus the number count of GW sources detectable (SNR$>$8) by CPTA (red) and SKA-PTA (green) for $\tau_{H}=0.1$, $5$, $10\,\mathrm{Gyr}$ respectively (from left to right).}
\label{fig:Ndis}
\end{figure}

\begin{figure}[ht!]
\centering 
\begin{minipage}[t]{0.4\textwidth}
        \centering
        \includegraphics[width=\linewidth]{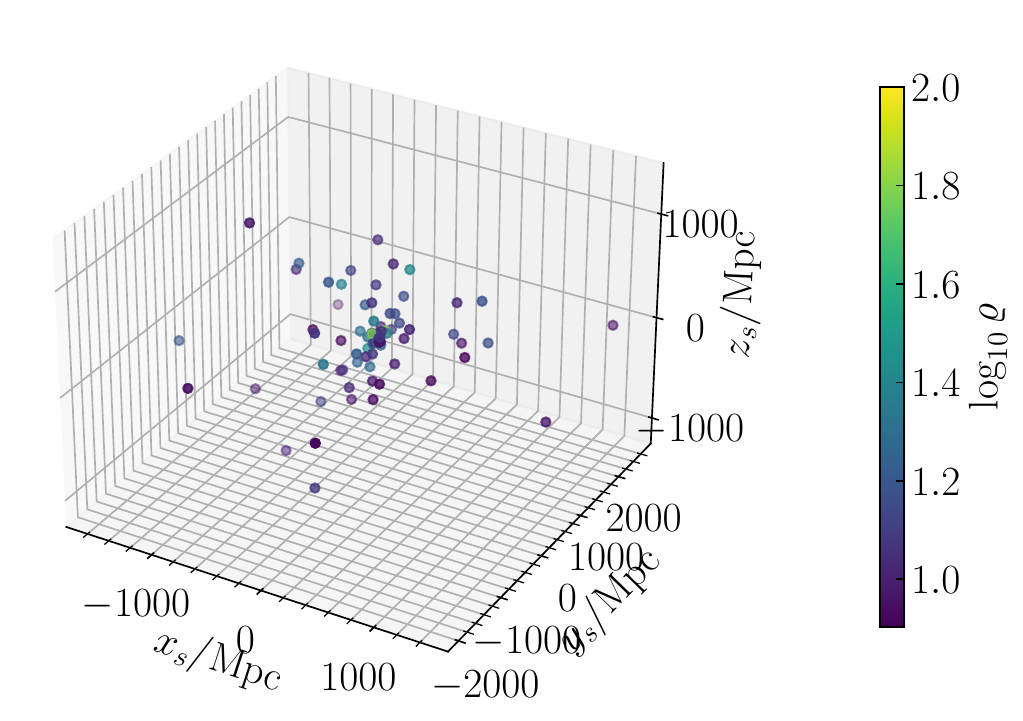}
    \end{minipage}
    \hfill  
    \begin{minipage}[t]{0.48\textwidth}
        \centering
        \includegraphics[width=\linewidth]{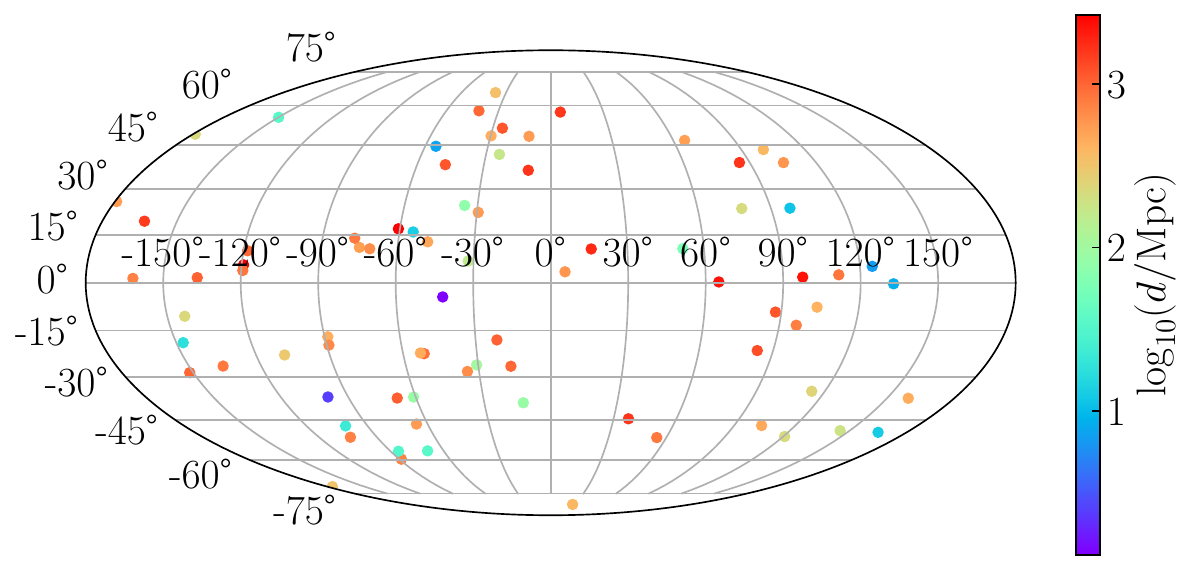}
    \end{minipage}
\caption{ The three dimensional spatial distribution (left) and the angular position distribution sky map (right) of individual GW sources detectable by SKA-PTA from one representative realization with $\tau_H=0.1\,\mathrm{Gyr}$, where colors represent the SNR and comoving distance of each source respectively. }
\label{fig:indGWmapC}
\end{figure}

\begin{figure}[htbp]
    \centering
    \begin{minipage}[t]{0.316\textwidth}
        \centering
        \includegraphics[width=\linewidth]{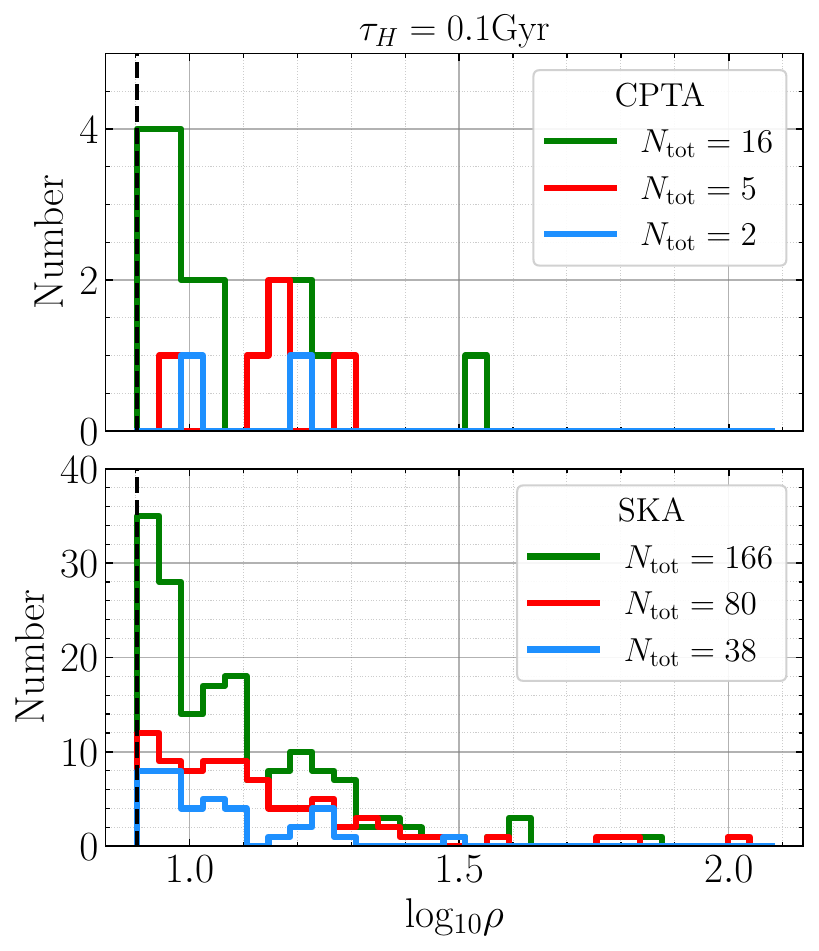}
    \end{minipage}
    \begin{minipage}[t]{0.3\textwidth}
        \centering
        \includegraphics[width=\linewidth]{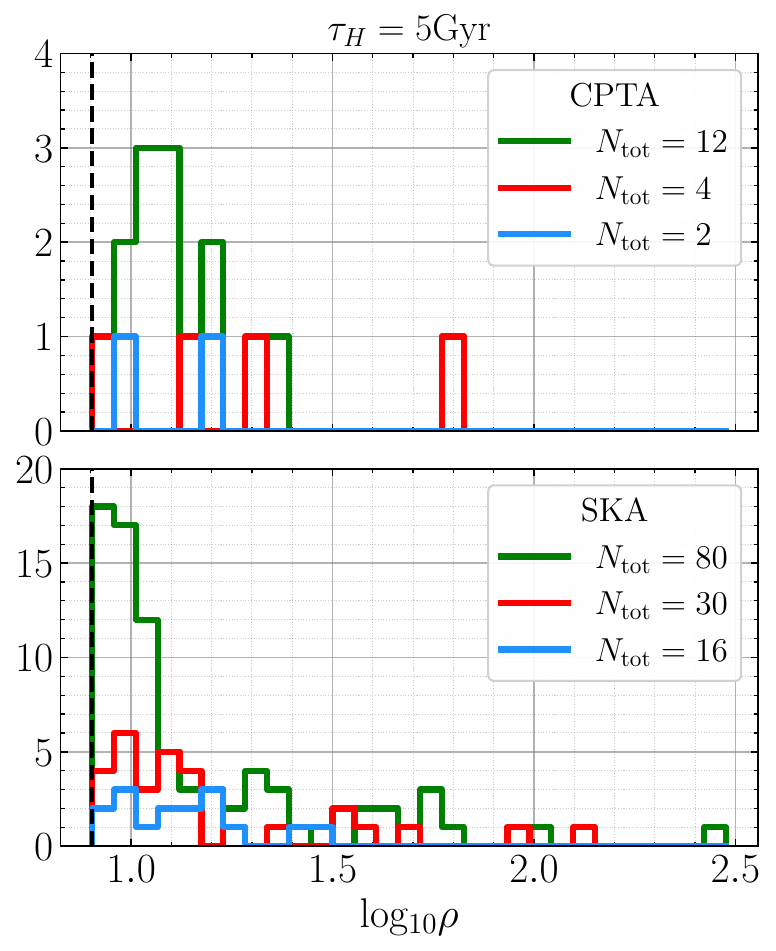}
    \end{minipage}
    \begin{minipage}[t]{0.293\textwidth}
        \centering
        \includegraphics[width=\linewidth]{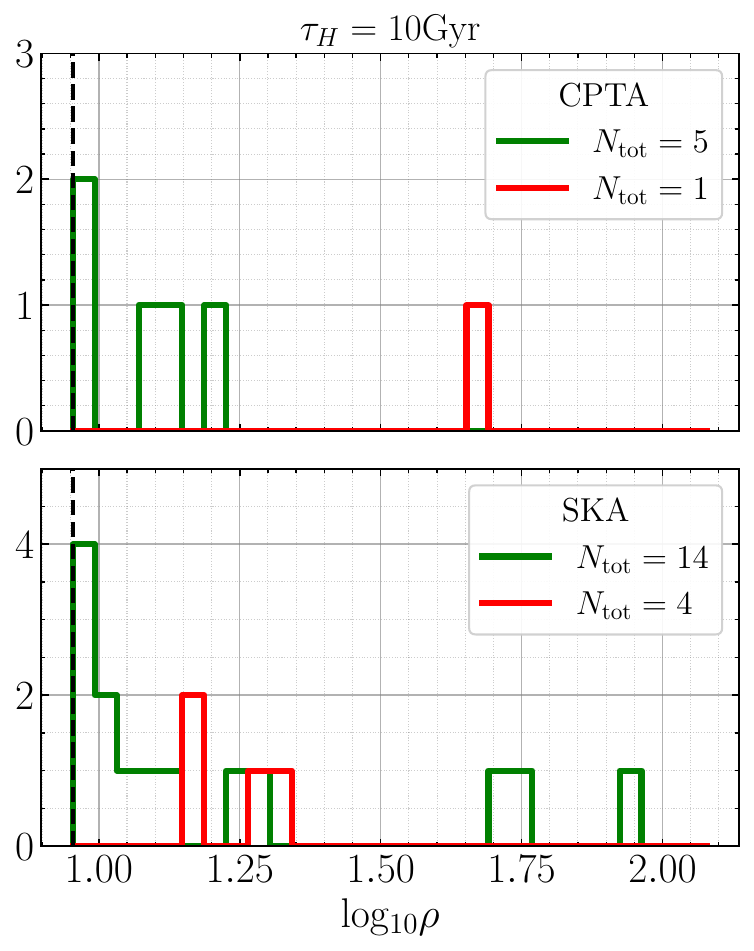}
    \end{minipage}
\caption{ The SNR distribution of resolvable individual GW sources for CPTA (upper panels) and SKA-PTA (lower panels) for $\tau_H=0.1, 5, 10\,\mathrm{Gyr}$ respectively (from left to right). Each panel presents results from three representative realizations with abundant, moderate, and sparse source counts, where $N_{\rm tot}$ denotes the total number of individual sources in each realization. The vertical black dashed lines indicate the SNR threshold at $\rho=8$.} 
\label{fig:SNRdis}
\end{figure}

\begin{figure}[htbp]
    \centering
    \begin{minipage}[t]{0.316\textwidth}
        \centering
        \includegraphics[width=\linewidth]{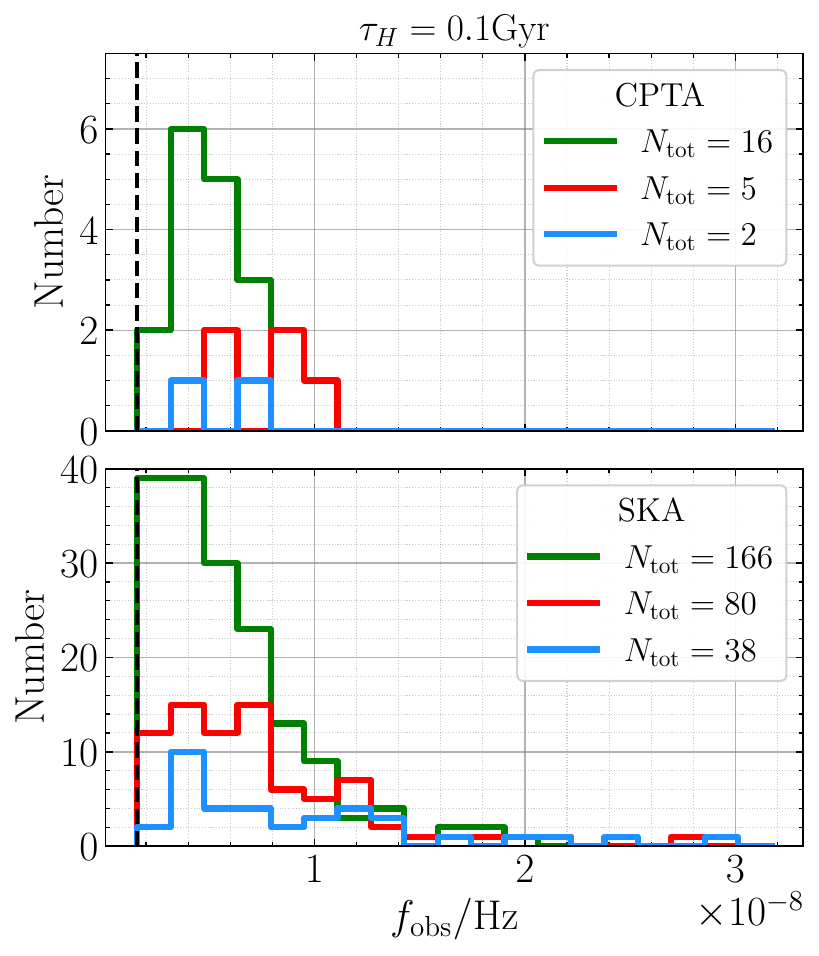}
    \end{minipage}
    \begin{minipage}[t]{0.3\textwidth}
        \centering
        \includegraphics[width=\linewidth]{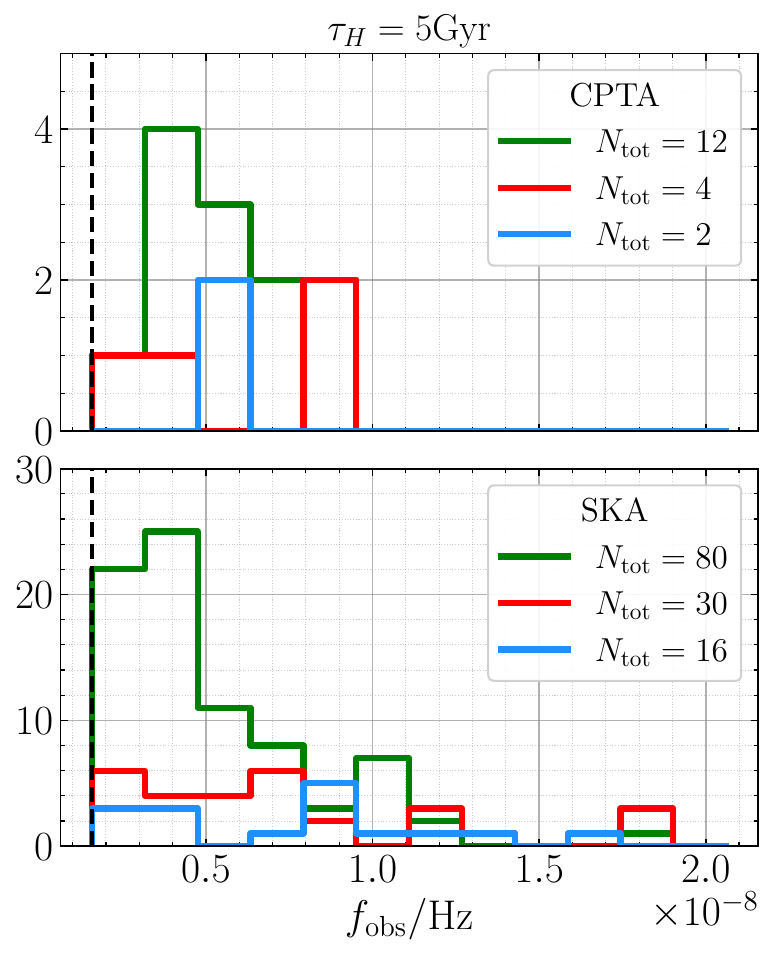}
    \end{minipage}
    \begin{minipage}[t]{0.293\textwidth}
        \centering
        \includegraphics[width=\linewidth]{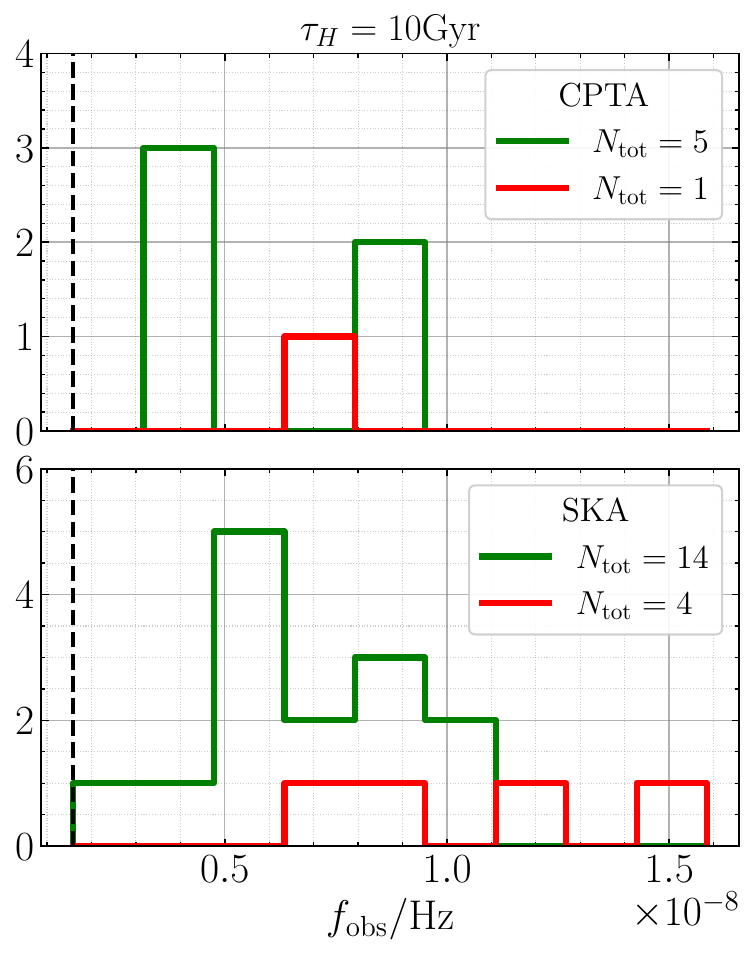}
    \end{minipage}
\caption{ The frequency distribution of resolvable individual GW sources for CPTA (upper panels) and SKA-PTA (lower panels) for $\tau_H=0.1, 5, 10\,\mathrm{Gyr}$ respectively (from left to right). Each panel presents results from three representative realizations with abundant, moderate, and sparse source counts, where $N_{\rm tot}$ denotes the total number of individual sources in each realization. The vertical black dashed lines indicate the lower frequency limit when total observation duration is 20 years.} 
\label{fig:freqdis}
\end{figure}

{Our results show that, future SKA is projected to resolve a median of 76, 32, and 2 individual GW sources with SNR $\varrho>8$ during 20 years’ observation time for $\tau_H=0.1$, $5$, $10\,\mathrm{Gyr}$ repectively (See also \citealt{2023ApJ...955..132C,2025A&A...694A.282T}), while CPTA is expected to identify a median of 5, 3, and 0 individual GW sources with the same observation duration, substantially lower than SKA. We have also confirmed that no realization exceeds the upper bound for the number of resolvable sources in each frequency bin that was proposed in \cite{2012PhRvD..86l4028B}.}
The distribution of the number of realizations versus the detectable individual GW source number counts by CPTA and SKA-PTA for all the three hardening timescales are shown in Figure~\ref{fig:Ndis}, and the results for the median values and variance of the number of individual sources are summarised in Table~\ref{tab:Ndis}. We also presented the three dimensional spatial distributions and two dimensional angular distribution of one representative realization for the most optimistic case with $\tau_{H}=0.1\,\mathrm{Gyr}$ in Figure~\ref{fig:indGWmapC} for SKA-PTA, where we use colors to represent the SNR and comoving distance of each source respectively. 
{ The statistical distribution of SNRs/frequencies of all resolvable individual GW sources by CPTA and SKA-PTA from three representative realizations with abundant, moderate, and sparse source counts are shown in Figure~\ref{fig:SNRdis}/Figure~\ref{fig:freqdis}.}
From the figures, one can observe the highest achievable SNR in different cases. { SKA-PTA, with its advanced sensitivity, is expected to detect GW sources with SNR values reaching around 100, whereas CPTA achieves notably lower SNRs, rarely exceeding 30. At the same time, SKA detects sources at frequencies greater than $10^{-8}$Hz in all scenarios, while CPTA typically detects sources within the $10^{-9}$Hz range.} 

\begin{table}
\caption{  The number of detectable (SNR$>$8) individual sources by CPTA and SKA-PTA for $\tau_{H}=0.1, 5, 10$\,Gyr. They are expressed as the median values plus or minus the values representing  1$\sigma$ confident interval (from 16\% to 84\%). 
}
\begin{center}
\begin{tabular}{c|cc}
\hline \hline
 & \multicolumn{2}{c}{SNR$>$8} \\ \hline
$\tau_{H}$/Gyr & CPTA & SKA-PTA \\\hline
0.1 & $5_{-4}^{+5}$ & $76_{-38}^{+61}$ \\ 
5 & $3_{-2}^{+4}$  &  $32_{-16}^{+21}$  \\ 
10 & $0_{-0}^{+1}$ & $2_{-2}^{+3}$ \\ \hline \hline %
\end{tabular}
%
\end{center}
\label{tab:Ndis}
\end{table}

\section{Lensing Effect}
\label{sec:lensing}
\begin{figure}[htbp]
    \centering
    \begin{minipage}[t]{0.35\textwidth}
        \centering
        \includegraphics[width=\linewidth]{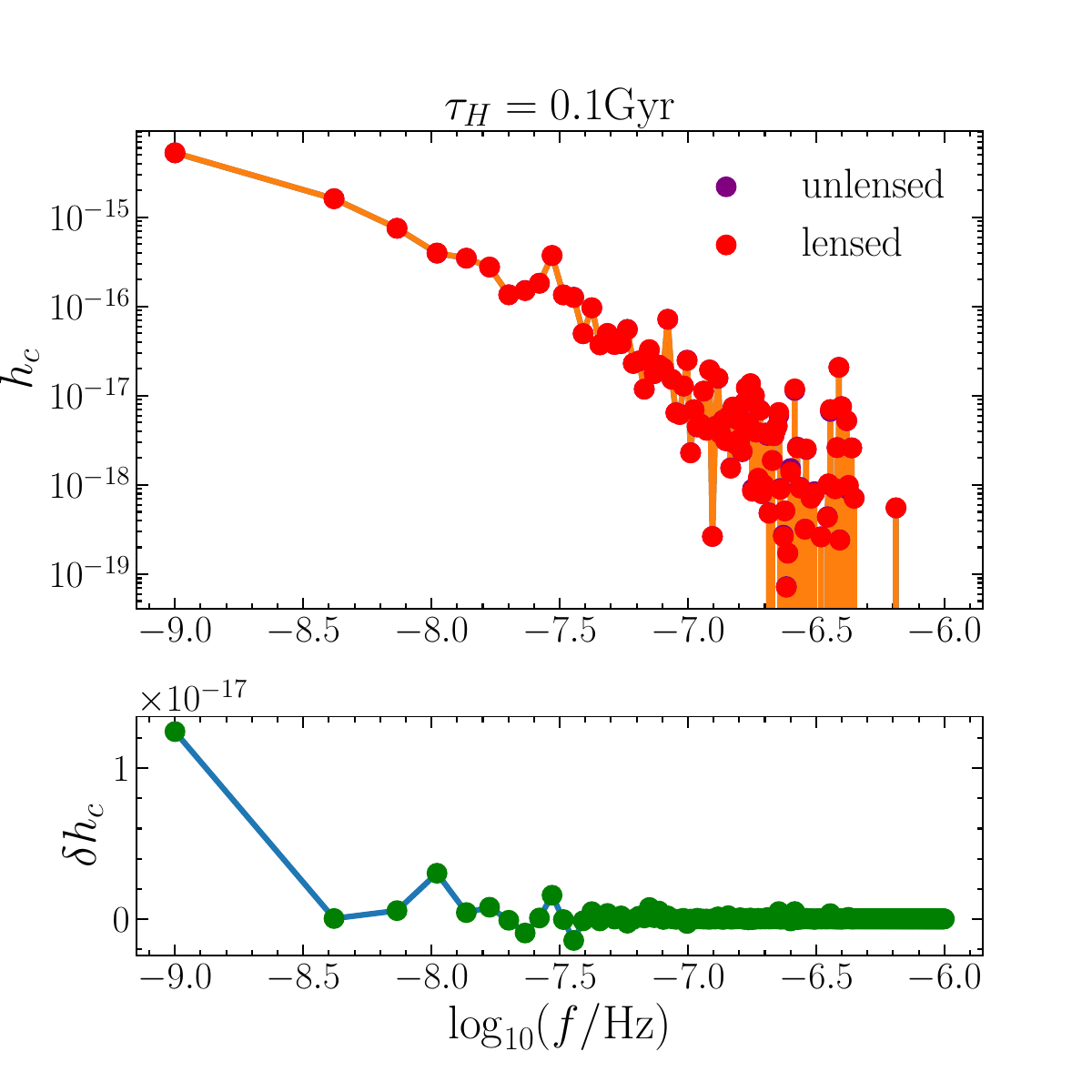}
    \end{minipage}
    \hspace{-0.83cm}
    \begin{minipage}[t]{0.35\textwidth}
        \centering
        \includegraphics[width=\linewidth]{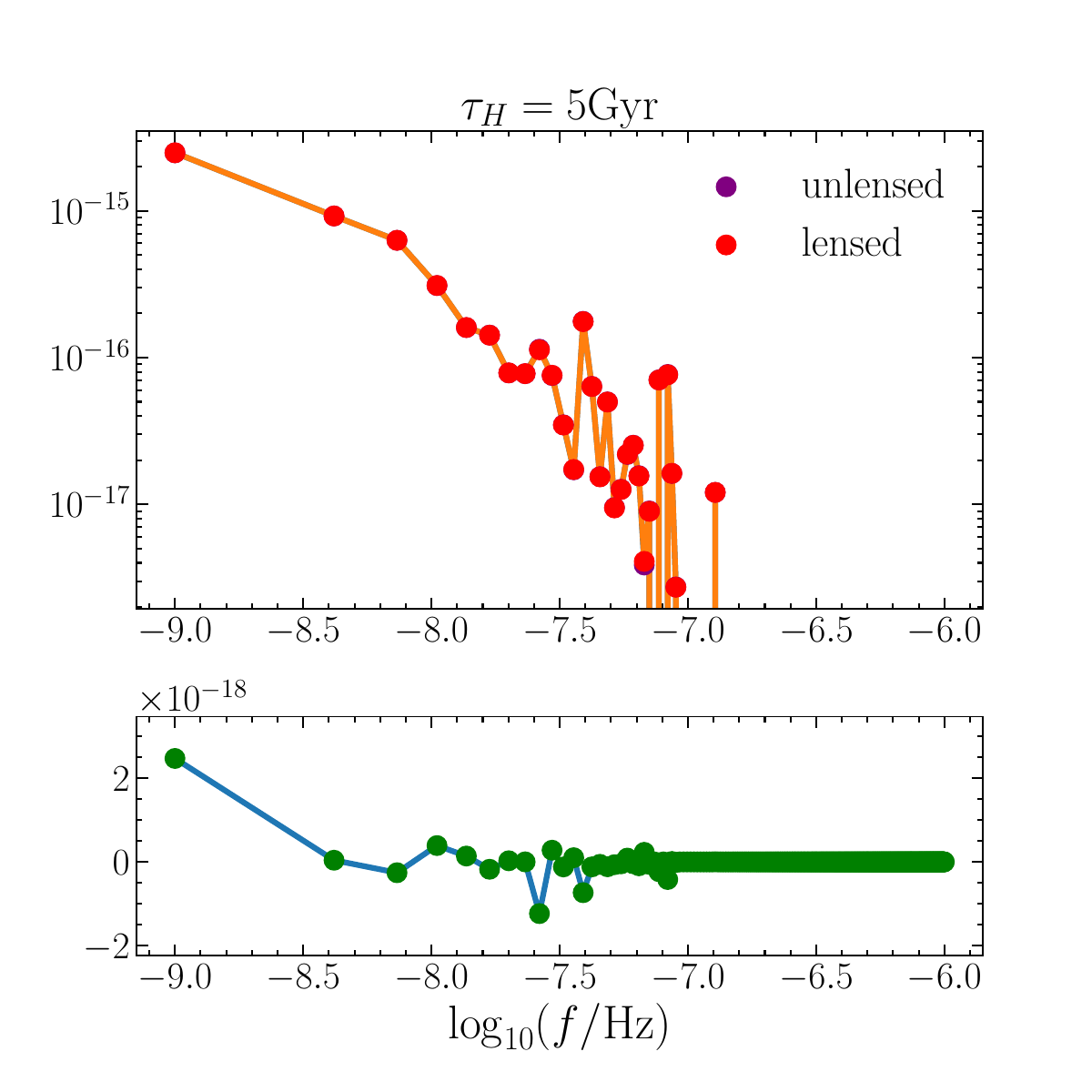}
    \end{minipage}
    \hspace{-0.83cm}
    \begin{minipage}[t]{0.35\textwidth}
        \centering
        \includegraphics[width=\linewidth]{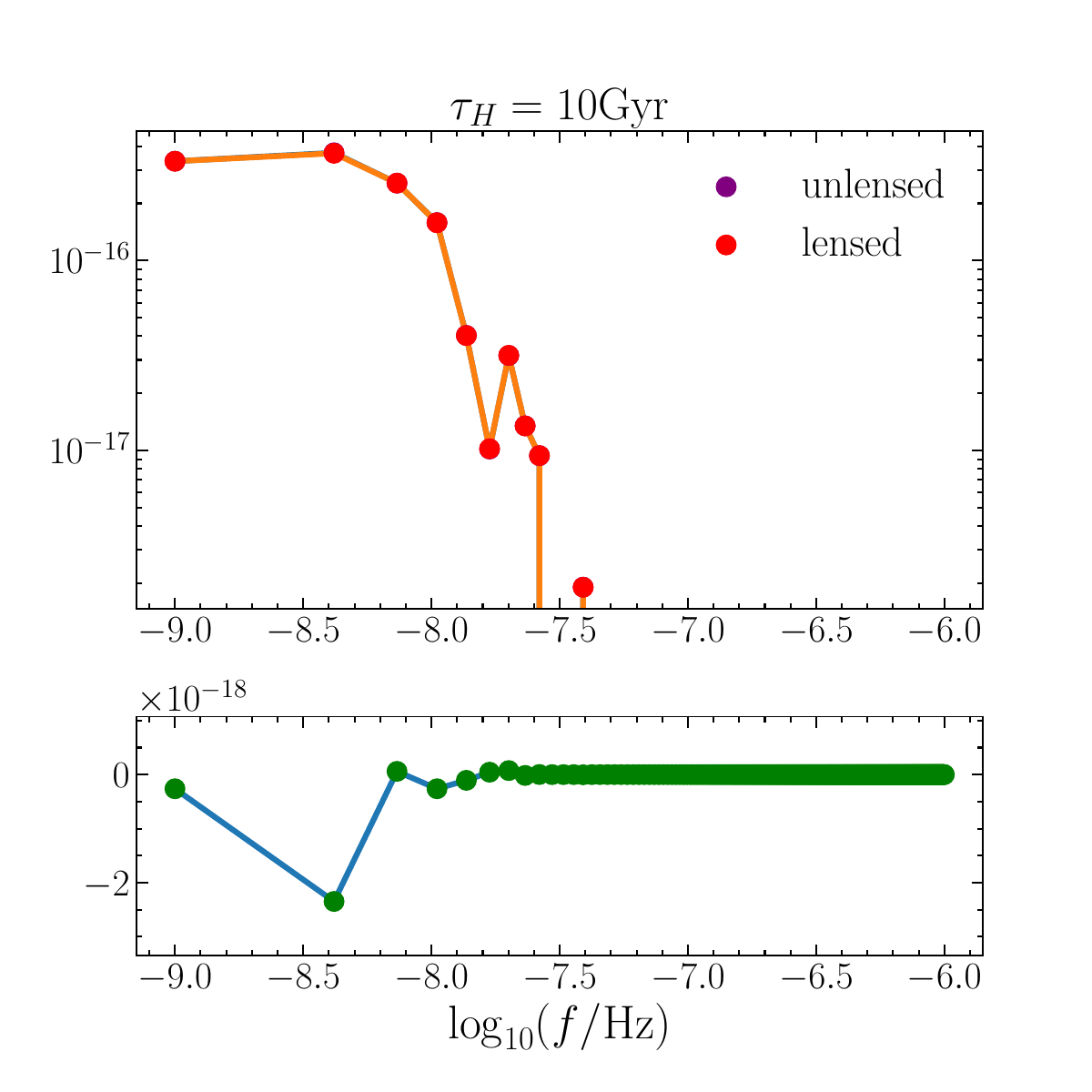}
    \end{minipage}
\caption{A comparison of the GWB characteristic amplitude with and without considering gravitational lensing effects { from one representative realization} for $\tau_H=0.1\,\mathrm{Gyr}$ (left), $5\,\mathrm{Gyr}$ (middle) and $10\,\mathrm{Gyr}$ (right) respectively. The lower panels show the differences in the characteristic amplitude between the two cases. Lensing effects induced variation is generally two to three orders of magnitude smaller than the original value.}
\label{fig:iso&rhocompare}
\end{figure}

\begin{figure}[htbp]
    \centering
    \begin{minipage}[t]{0.327\textwidth}
        \centering
        \includegraphics[width=\linewidth]{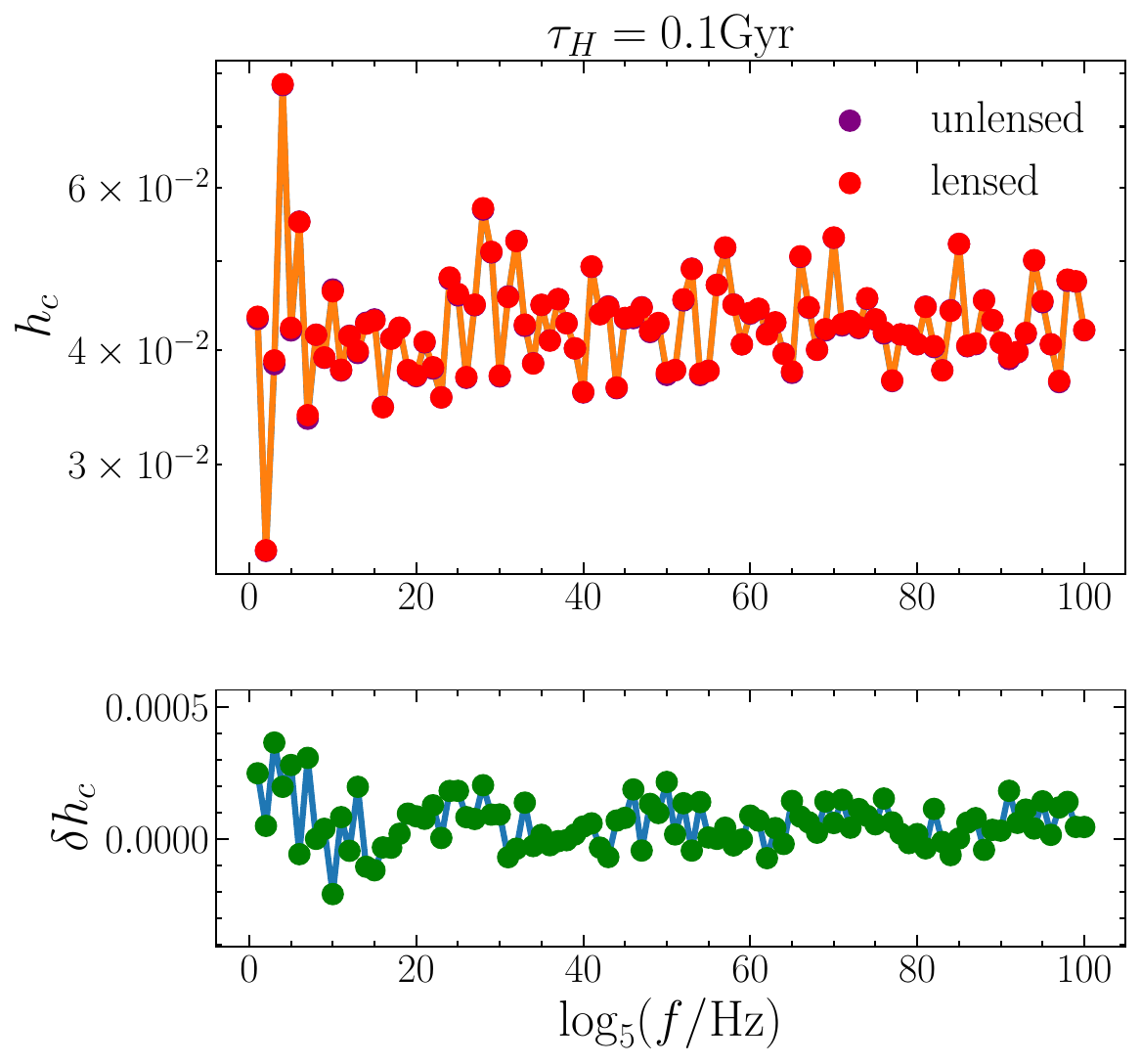}
    \end{minipage}
    \begin{minipage}[t]{0.31\textwidth}
        \centering
        \includegraphics[width=\linewidth]{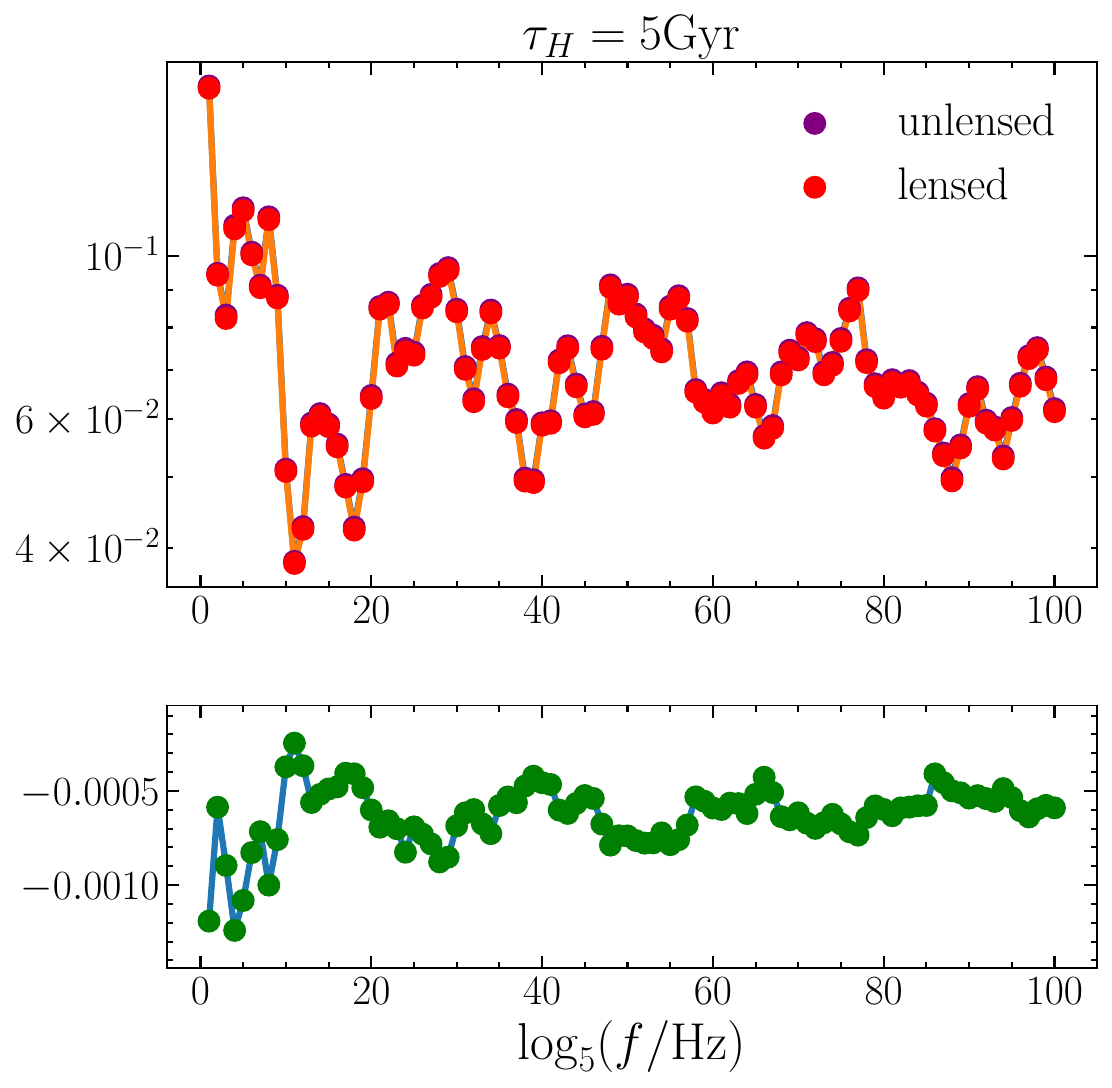}
    \end{minipage}
    \begin{minipage}[t]{0.31\textwidth}
        \centering
        \includegraphics[width=\linewidth]{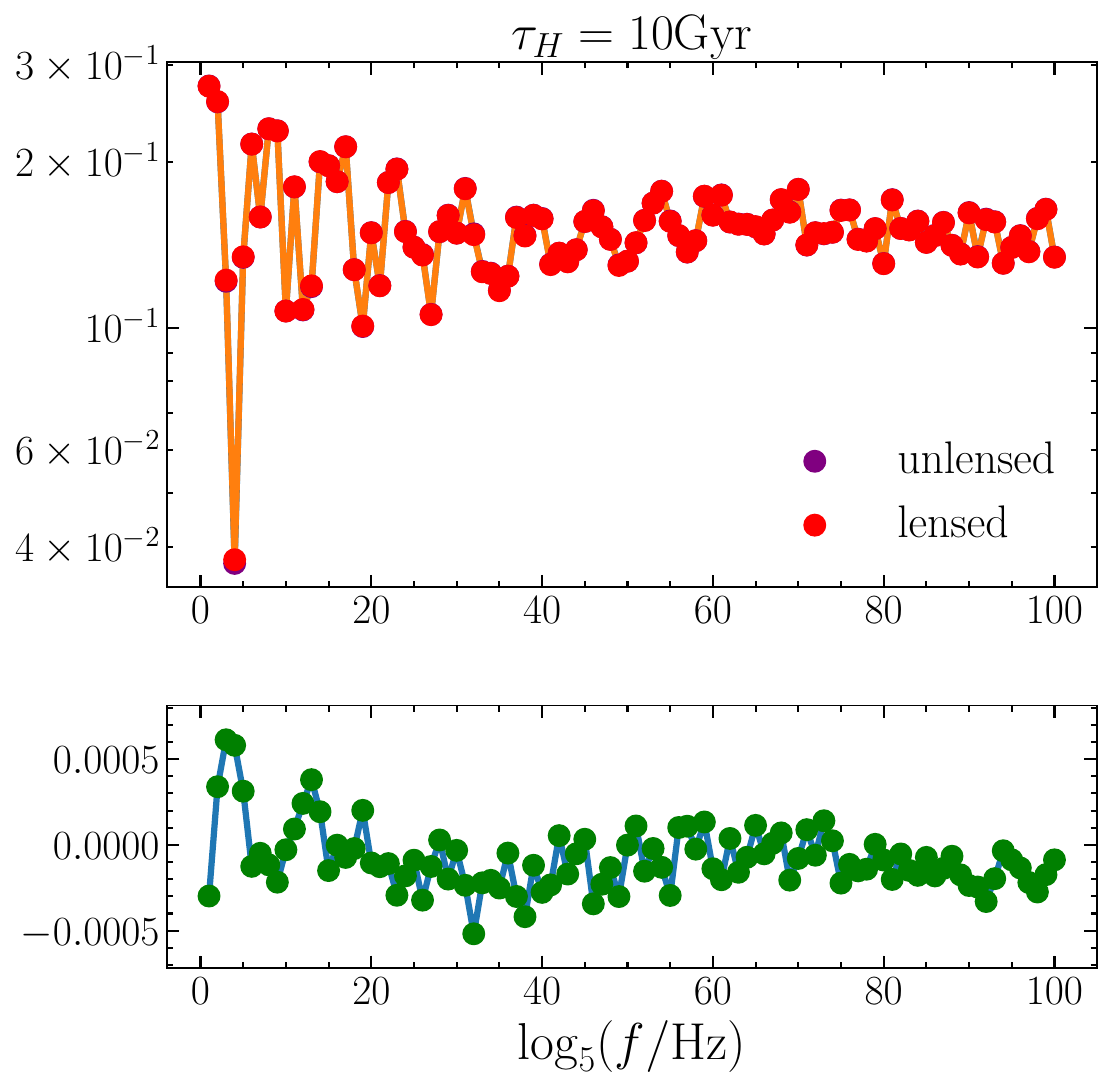}
    \end{minipage}
\caption{A comparison of the angular power spectrum of the GWB total energy density with and without considering gravitational lensing effects { from one representative realization} for $\tau_H=0.1$ Gyr (left), $5$ Gyr (middle) and $10$ Gyr (right) respectively. The lower panels show the differences in the $C_l/C_0$ coefficients between the two cases. Lensing effects induced variation is generally two to three orders of magnitude smaller than the original value.} 
\label{fig:clcompare}
\end{figure}

The gravitational wave signals from distant SMBBHs may be affected by the lensing effects of foreground galaxies or dark matter halos as they propagate towards Earth. In this section, we will discuss whether the gravitational lensing effect will have a significant impact on the various results related to GWB and individual sources that we have derived in previous sections. We adopt the lensing model used in \cite{Oguri:2018muv}, where they developed a hybrid framework to compute the magnification probability distribution function (PDF) that takes into account both weak and strong gravitational lensing effect for wide ranges of source redshifts and magnifications. We randomly assigned a magnification factor $\sqrt{\mu}$ to each gravitational wave event amplitude $h_i$ according to the PDF under this model. We recalculated the GWB isotropic and anisotropic properties with lensing effect, and compared the derived results with previous results in Figure~\ref{fig:iso&rhocompare} and Figure~\ref{fig:clcompare}.

From Figure \ref{fig:iso&rhocompare} we can see that, for the characteristic amplitude of the GWB, upon incorporating the lensing effect into our calculations, we observed a subtle change in the characteristic amplitude of GWB. Specifically, the amplitude exhibited a slight increase at the order of { $10^{-17}$ or $10^{-18}$}, which is two to three orders of magnitude smaller than the value of the amplitude itself. This increase is sufficiently small that it does not significantly alter the overall behavior of the GWB signal. For the anisotropy property of the GWB signal, { lensing effects induced variation of the angular power spectrum of total energy density is shown Figure~\ref{fig:clcompare}. We can see that, lensing induced variation on the $C_l$ coefficients is roughly at the order of $10^{-4}$, which is also two to three orders of magnitude smaller than the original value, the effect of gravitational lensing is not significant either. For the detectability of individual sources, the lensing effect can slightly increase the number of individual sources with SNR above 8 in some realizations. However, since the main limitation on the number of resolvable individual sources is the upper limit in each frequency bins, the overall impact of lensing on the individual source number count is negligible across all cases.}

\section{Conclusions and Discussion}
\label{sec:concl}

Utilizing data from numerical cosmological simulation, {  we generated multiple realizations of SMBBHs distributed in the observable universe as GW sources of PTA. While maintaining the features of cosmic large-scale structure in the mock universe, we studied the isotropic and anisotropic characteristics of SMBBH sources and GWB signal accounting for cosmic variance through different realizations, predicted the properties of resolvable individual GW sources by CPTA and SKA-PTA, and also examined the influence of gravitational lensing effects.}

{ For GWB, we calculated its characteristic amplitude and spectrum, and found that { the characteristic amplitude $h_c$ at frequency $f = 1~\mathrm{yr}^{-1}$ is $4.08 \times 10^{-16}$, $3.38 \times 10^{-16}$, and $2.19 \times 10^{-18}$ for $\tau_H = 0.1\,\mathrm{Gyr}$, $5\,\mathrm{Gyr}$, and $10\,\mathrm{Gyr}$, respectively; while at $f = 0.1\,\mathrm{yr}^{-1}$, $h_c=1.96 \times 10^{-15}$, $1.42 \times 10^{-15}$, and $3.13 \times 10^{-16}$ for the same $\tau_H$ values in corresponding order.}
We demonstrated the anisotropy properties in the distribution of simulated GW sources and the GWB signal using sky maps, and derived the angular power spectrum coefficients $C_l$ as a function of $l$ for both the total energy density of GWB and energy density in different frequency bins. In the total energy case, our results show that the directional dependence of frequency distribution represents a more dominant source of anisotropy compared to the contribution from bright individual sources. We predict that for most of the realizations in the cases of hardening timescales of $0.1, 5,$ and $10\,\mathrm{Gyr}$, the distribution of $C_1/C_0$ falls within the range of { $[3.0\times10^{-2}, 4.3\times10^{-1}]$, $[5.3\times10^{-2}, 6.8\times10^{-1}]$, $[7.7\times10^{-2}, 6.2\times10^{-1}]$ respectively (16\%-84\% distribution range)}, and are insensitive to the substraction of additional bright sources. { Regarding the} angular power spectrum of energy density in different frequency bins, $C_1/C_0$ shows great fluctuations at higher frequencies, while the lowest-frequency bin demonstrates a lower level of anisotropy compared to the total energy density case, and the differences in the $C_l/C_0$ distributions under the three hardening timescales become more pronounced. In the lowest frequency bin, our results show that $C_1/C_0$ falls mostly within the range of {$[2.2\times10^{-3}, 3.5\times10^{-2}$], $[4.0\times10^{-3}, 6.0\times10^{-2}]$, and  $[2.1\times10^{-2}, 2.9\times10^{-1}]$ for $\tau_H=0.1$, $5$, and $10\,\mathrm{Gyr}$ respectively (16\%-84\% distribution range)}, and the substraction of additional bright sources will further lower the level of anisotropy in the $\tau_H=0.1$ and $5\,\mathrm{Gyr}$ cases to {$[1.2\times10^{-3}, 9.1\times10^{-3}$] and $[2.7\times10^{-3}, 2.0\times10^{-2}]$}. Furthermore, We investigated the clustering properties of GW sources in spherical shells at various redshifts and discovered that, on relatively smaller scales, the clustering patterns of SMBBHs are more similar to that of galaxies. 
Since the GWB should be predominantly contributed by relatively nearby sources, this distribution pattern of GW sources may also be reflected in the GWB signal characteristics.
We also studied the detection capability of the future CPTA and SKA-PTA for individual sources. {From our calculation results, SKA-PTA is expected to identify a median of 76, 32, and 2 individual GW sources with SNR $\varrho>8$ during 20 years' observation time for $\tau_H=0.1$, $5$, and $10\,\mathrm{Gyr}$ respectively, while future CPTA is projected to resolve a median of 5, 3, and 0 individual GW sources with the same observation duration.} These mock individual sources can provide more realistic samples for further studies.
Finally, we took gravitational lensing effects into account and found that lensing effects on the characteristics of the GWB signal and the number of detectable individual sources are rather limited.}

Below, we offer some further discussions. As was mentioned at the end of Section~\ref{subsec:isotropic}, { the average value of the GWB characteristic amplitude from our work, even in the most optimistic case with $\tau_H=0.1$ Gyr, lies below the current experimental data points.} This discrepancy may stem from several factors. One such factor is that SAM inevitable has its own uncertainties in its model construction. { If we consider other versions of SAM that have slightly higher merger rates, e.g., Henriques 2015 or Izquierdo-Villalba 2019 \citep{2019MNRAS.483..503Y,David:stab3239}, it is still possible that SAM may offer results that is more compatible with observational data points. Nevertheless,  some realizations can still yield 
GWB signals that reach detection levels in specific frequency bins.}
Furthermore, it is also possible that the GWB signal detected is a complex superposition of signals from multiple sources across the universe. While SMBBHs are known to be significant contributors to this background, they are not the sole constituents. { Other physical processes with cosmological origins, such as cosmic string or cosmic phase transition, may also imprint their signatures on the GWB}, thereby complicating the interpretation of the observed signals and may also contribute to the discrepancy. However, it is important to emphasize that, despite these complexities, our current projected value of the GWB amplitude falls well within the theoretical range of [$1 \times 10^{-16}$, $5 \times 10^{-15}$] given by existing literature { in the $\tau_H=0.1$ and $5\,\mathrm{Gyr}$ cases.}

{The general physical scenario we considered in constructing light-cone SMBBH events is that we assume SMBBHs will evolve to a GW-dominated regime at 1\,nHz following the mergers of their progenitor galaxies through hardening processes characterized by a fixed hardening timescale $\tau_H$. Since current $N$-body simulations combined with semi-analytical models cannot provide accurate information about the frequencies and orbits of SMBBH systems, we also made the simplifying assumption of circular orbital evolution from 1\,nHz onward driven exclusively by GW emission. Although this is a rather simplified representation of the actual SMBBH evolution process, we've employed this relatively simple model to efficiently construct multiple realizations of the light-cone SMBBH dataset. In the future, we'll consider refine our light-cone construction method by incorporating non-circular orbits and more detailed evolution modeling. }


{ We have theoretically explored the possibility of reconstructing the large-scale structure of the universe using SMBBH events. These results may offer insights for exploring the correlations between the anisotropy of the GWB signal and the number distribution of galaxies, as is done in \cite{2024arXiv241100532S}. But it is important to recognize that since the GWB signal arises from the superposition of GW signals across different redshifts, the projection effect may weaken the correlation between GWB anisotropy and galaxy number distribution compared to the scenarios involving GW sources confined to single redshift shells considered here. In \cite{2024arXiv241100532S}, only SMBBHs within a single redshift bin were considered. It will be interesting to explore the connection between galaxy number distributions and more realistic GWB signals constructed using our complete light-cone data, as well as investigating its potential to constrain relative model parameters. We leave this to future work.}

It is also interesting to consider that the standard Hellings-Downs curve is derived on the assumption of an isotropic GWB \citep{1983ApJ...265L..39H}, the presence of anisotropic components in the GWB could affect the angular correlation, causing it to deviate from the standard curve. { This has been done in \cite{2024PhRvD.110l3507A,2025JCAP...03..011G} with theoretical framework}. It is possible that, based on our results, we can provide a numerical prediction on HD curve's configuration under the influence of anisotropy. We leave this also to future work.

{
\appendix 
\section{Mass evolution} \label{App:massevo}
For computational efficiency, we didn't consider black hole mass evolution after the merger of galaxies in the main text. In reality, masses of the SMBBHs may continue to grow within the hardening timescale. This appendix quantifies the resulting biases introduced by this simplification. Given that longer hardening timescales permit greater mass growth, we focus solely on the maximum hardening timescale $\tau_H=10\,\mathrm{Gyr}$ case to represent the potential upper limit of the impact of mass evolution on our results. We evolve the SMBBHs from the merger of their progenitor galaxies to the GW driven phase with the mass growth rate (including both radio mode and quasar mode) given by the SMBH growth model adopted by SAM (\cite{2011MNRAS.413..101G}, see also \cite{Croton:2005hbr}). 
We exclude direct mass growth from subsequent black hole mergers to avoid double counting, since as noted in the main text, such events are treated as independent SMBBH sources. For simplicity, we distribute any added mass between the two black holes of a binary pair according to their original mass ratio, thus preserving the mass ratio of the binary system. By tracking the mass evolution process, we derive updated mass parameters for all the SMBBHs. These masses can then be used to compute the characteristic amplitudes, anisotropy and individual source statistics for comparison with results presented in the main text. The results are presented in Table \ref{tab:compare} below.
\setcounter{figure}{0}
\setcounter{table}{0}
\renewcommand{\thefigure}{A\arabic{figure}}
\renewcommand{\thetable}{A\arabic{table}}
\begin{table}[H]
\caption{ Comparison of GWB characteristic amplitude, anisotropy, and resolvable source counts with versus without black hole mass evolution. Here, ME denotes the methodology incorporating black hole mass evolution after galaxy mergers, while NE refers to the approach presented in the main text that excludes such evolutionary effects, and $\rho_{\rm tot}$ is the total energy density, while $\rho_{f_{\rm min}}$ is the energy density in the lowest frequency bin starting from $f_{\rm min}$.
}
\begin{center}
\begin{tabular}{c|cc|cc|cc}
\hline \hline
 & \multicolumn{2}{c|}{Characteristic amplitude} & \multicolumn{2}{c|}{Anisotropy (16\%-84\% distribution range)} &  \multicolumn{2}{c}{Individual source} \\ \hline
Models & $h_c/\rm 0.1yr^{-1}$ & $h_c/\rm yr^{-1}$ & $C_1/C_0$ of $\rho_{\rm tot}$ & $C_1/C_0$ of $\rho_{\rm f_{\rm min}}$ & CPTA & SKA-PTA \\\hline
ME & $7.04 \times 10^{-16}$  &  $4.61\times 10^{-18}$ & $[7.2\times10^{-2}, 6.5\times10^{-1}]$ & $[1.9\times10^{-2}, 2.7\times10^{-1}]$ & $1_{-1}^{+1}$ & $8_{-4}^{+7}$ \\ 
NE & $3.13 \times 10^{-16}$  &  $2.19 \times 10^{-18}$ & $[7.7\times10^{-2}, 6.2\times10^{-1}]$  &  $[2.1\times10^{-2}, 2.9\times10^{-1}]$ & $0_{-0}^{+1}$  &  $2_{-2}^{+3}$  \\\hline \hline %
\end{tabular}
\end{center}
\label{tab:compare}
\end{table}
As shown in the table, the characteristic amplitudes increase when considering black hole mass evolution after galaxy mergers, yet they still remain at the same order of magnitude as the previous results. The number of individual sources also shows a slight increase, with a more significant rise seen in the number detectable by SKA-PTA. The variation in the distribution range (16th to 84th percentiles) of anisotropic power spectrum coefficients may stem from multiple factors. On one hand, mass enhancement could amplify the amplitudes of comparatively luminous individual sources, potentially increasing anisotropy. Conversely, growth opportunities for lower-mass systems may narrow mass disparities among sources, potentially reducing anisotropy. The 
table results demonstrate a broadening of the distribution range for total-energy anisotropy, and a minor overall reduction in anisotropy of energy density within the lowest-frequency bin. The results of both cases are, however, still generally within the same range as those previously reported.
Variations in the results of the  $\tau_H=5\,\mathrm{Gyr}$ case can be inferred from results shown above. Incorporates mass evolution will certainly shift both the characteristic amplitude and resolvable source counts toward values more comparable to the results in the $\tau_H=0.1\,\mathrm{Gyr}$ case. Given the shorter hardening timescale relative to the 10 Gyr case, the overall magnitude of variation should be smaller than that shown above. It is interesting to note that, the disparity in GWB properties—characteristic amplitude, anisotropy, and resolvable source counts—between the $\tau_H=0.1\,\mathrm{Gyr}$ and $\tau_H=5$Gyr cases are significantly smaller than their order-of-magnitude difference in hardening timescales.}

\section*{}
\noindent
We thank Prof. Youjun Lu, Qi Guo, Chengliang Wei for helpful suggestions. Xiao Guo is supported by the Postdoctoral Fellowship Program and China Postdoctoral Science Foundation under Grant Number BX20230104. We acknowledge the Beijing Super Cloud Computing Center for providing HPC resources that have contributed to the research results reported within this paper (URL: \url{http://www.blsc.cn/}).



\bibliographystyle{aasjournal}
\bibliography{main.bib}

\begin{thebibliography}{}
\expandafter\ifx\csname natexlab\endcsname\relax\def\natexlab#1{#1}\fi
\providecommand{\url}[1]{\href{#1}{#1}}
\providecommand{\dodoi}[1]{doi:~\href{http://doi.org/#1}{\nolinkurl{#1}}}
\providecommand{\doeprint}[1]{\href{http://ascl.net/#1}{\nolinkurl{http://ascl.net/#1}}}
\providecommand{\doarXiv}[1]{\href{https://arxiv.org/abs/#1}{\nolinkurl{https://arxiv.org/abs/#1}}}

\bibitem[{{Agazie} {et~al.}(2023{\natexlab{a}}){Agazie}, {Anumarlapudi}, {Archibald}, {Arzoumanian}, {Baker}, {B{\'e}csy}, {Blecha}, {Brazier}, {Brook}, {Burke-Spolaor}, {Burnette}, {Case}, {Charisi}, {Chatterjee}, {Chatziioannou}, {Cheeseboro}, {Chen}, {Cohen}, {Cordes}, {Cornish}, {Crawford}, {Cromartie}, {Crowter}, {Cutler}, {Decesar}, {Degan}, {Demorest}, {Deng}, {Dolch}, {Drachler}, {Ellis}, {Ferrara}, {Fiore}, {Fonseca}, {Freedman}, {Garver-Daniels}, {Gentile}, {Gersbach}, {Glaser}, {Good}, {G{\"u}ltekin}, {Hazboun}, {Hourihane}, {Islo}, {Jennings}, {Johnson}, {Jones}, {Kaiser}, {Kaplan}, {Kelley}, {Kerr}, {Key}, {Klein}, {Laal}, {Lam}, {Lamb}, {Lazio}, {Lewandowska}, {Littenberg}, {Liu}, {Lommen}, {Lorimer}, {Luo}, {Lynch}, {Ma}, {Madison}, {Mattson}, {McEwen}, {McKee}, {McLaughlin}, {McMann}, {Meyers}, {Meyers}, {Mingarelli}, {Mitridate}, {Natarajan}, {Ng}, {Nice}, {Ocker}, {Olum}, {Pennucci}, {Perera}, {Petrov}, {Pol}, {Radovan}, {Ransom}, {Ray}, {Romano}, {Sardesai}, {Schmiedekamp}, {Schmiedekamp},
  {Schmitz}, {Schult}, {Shapiro-Albert}, {Siemens}, {Simon}, {Siwek}, {Stairs}, {Stinebring}, {Stovall}, {Sun}, {Susobhanan}, {Swiggum}, {Taylor}, {Taylor}, {Turner}, {Unal}, {Vallisneri}, {van Haasteren}, {Vigeland}, {Wahl}, {Wang}, {Witt}, \& {Young}}]{2023ApJ...951L...8A}
{Agazie}, G., {Anumarlapudi}, A., {Archibald}, A.~M., {et~al.} 2023{\natexlab{a}}, \apjl, 951, L8, \dodoi{10.3847/2041-8213/acdac6}

\bibitem[{{Agazie} {et~al.}(2023{\natexlab{b}}){Agazie}, {Anumarlapudi}, {Archibald}, {Baker}, {B{\'e}csy}, {Blecha}, {Bonilla}, {Brazier}, {Brook}, {Burke-Spolaor}, {Burnette}, {Case}, {Casey-Clyde}, {Charisi}, {Chatterjee}, {Chatziioannou}, {Cheeseboro}, {Chen}, {Cohen}, {Cordes}, {Cornish}, {Crawford}, {Cromartie}, {Crowter}, {Cutler}, {D'Orazio}, {Decesar}, {Degan}, {Demorest}, {Deng}, {Dolch}, {Drachler}, {Ferrara}, {Fiore}, {Fonseca}, {Freedman}, {Gardiner}, {Garver-Daniels}, {Gentile}, {Gersbach}, {Glaser}, {Good}, {G{\"u}ltekin}, {Hazboun}, {Hourihane}, {Islo}, {Jennings}, {Johnson}, {Jones}, {Kaiser}, {Kaplan}, {Kelley}, {Kerr}, {Key}, {Laal}, {Lam}, {Lamb}, {Lazio}, {Lewandowska}, {Littenberg}, {Liu}, {Luo}, {Lynch}, {Ma}, {Madison}, {McEwen}, {McKee}, {McLaughlin}, {McMann}, {Meyers}, {Meyers}, {Mingarelli}, {Mitridate}, {Natarajan}, {Ng}, {Nice}, {Ocker}, {Olum}, {Pennucci}, {Perera}, {Petrov}, {Pol}, {Radovan}, {Ransom}, {Ray}, {Romano}, {Runnoe}, {Sardesai}, {Schmiedekamp}, {Schmiedekamp},
  {Schmitz}, {Schult}, {Shapiro-Albert}, {Siemens}, {Simon}, {Siwek}, {Stairs}, {Stinebring}, {Stovall}, {Sun}, {Susobhanan}, {Swiggum}, {Taylor}, {Taylor}, {Turner}, {Unal}, {Vallisneri}, {Vigeland}, {Wachter}, {Wahl}, {Wang}, {Witt}, {Wright}, \& {Young}}]{2023ApJ...952L..37A}
---. 2023{\natexlab{b}}, \apjl, 952, L37, \dodoi{10.3847/2041-8213/ace18b}

\bibitem[{{Agazie} {et~al.}(2023{\natexlab{c}}){Agazie}, {Anumarlapudi}, {Archibald}, {Arzoumanian}, {Baker}, {B{\'e}csy}, {Blecha}, {Brazier}, {Brook}, {Burke-Spolaor}, {Casey-Clyde}, {Charisi}, {Chatterjee}, {Cohen}, {Cordes}, {Cornish}, {Crawford}, {Cromartie}, {Crowter}, {DeCesar}, {Demorest}, {Dolch}, {Drachler}, {Ferrara}, {Fiore}, {Fonseca}, {Freedman}, {Gardiner}, {Garver-Daniels}, {Gentile}, {Glaser}, {Good}, {G{\"u}ltekin}, {Hazboun}, {Jennings}, {Johnson}, {Jones}, {Kaiser}, {Kaplan}, {Kelley}, {Kerr}, {Key}, {Laal}, {Lam}, {Lamb}, {Lazio}, {Lewandowska}, {Liu}, {Lorimer}, {Luo}, {Lynch}, {Ma}, {Madison}, {McEwen}, {McKee}, {McLaughlin}, {McMann}, {Meyers}, {Mingarelli}, {Mitridate}, {Ng}, {Nice}, {Ocker}, {Olum}, {Pennucci}, {Perera}, {Pol}, {Radovan}, {Ransom}, {Ray}, {Romano}, {Sardesai}, {Schmiedekamp}, {Schmiedekamp}, {Schmitz}, {Schult}, {Shapiro-Albert}, {Siemens}, {Simon}, {Siwek}, {Stairs}, {Stinebring}, {Stovall}, {Susobhanan}, {Swiggum}, {Taylor}, {Turner}, {Unal}, {Vallisneri},
  {Vigeland}, {Wahl}, {Witt}, \& {Young}}]{2023ApJ...956L...3A}
---. 2023{\natexlab{c}}, \apjl, 956, L3, \dodoi{10.3847/2041-8213/acf4fd}

\bibitem[{{Allen} {et~al.}(2024){Allen}, {Agarwal}, {Romano}, \& {Valtolina}}]{2024PhRvD.110l3507A}
{Allen}, B., {Agarwal}, D., {Romano}, J.~D., \& {Valtolina}, S. 2024, \prd, 110, 123507, \dodoi{10.1103/PhysRevD.110.123507}

\bibitem[{{Antoniadis} {et~al.}(2023){Antoniadis}, {Arumugam}, {Arumugam}, {Auclair}, {Babak}, {Bagchi}, {Bak Nielsen}, {Barausse}, {Bassa}, {Bathula}, {Berthereau}, {Bonetti}, {Bortolas}, {Brook}, {Burgay}, {Caballero}, {Caprini}, {Chalumeau}, {Champion}, {Chanlaridis}, {Chen}, {Cognard}, {Crisostomi}, {Dandapat}, {Deb}, {Desai}, {Desvignes}, {Dhanda-Batra}, {Dwivedi}, {Falxa}, {Fastidio}, {Ferdman}, {Franchini}, {Gair}, {Goncharov}, {Gopakumar}, {Graikou}, {Grie{\ss}meier}, {Gualandris}, {Guillemot}, {Guo}, {Gupta}, {Hisano}, {Hu}, {Iraci}, {Izquierdo-Villalba}, {Jang}, {Jawor}, {Janssen}, {Jessner}, {Joshi}, {Kareem}, {Karuppusamy}, {Keane}, {Keith}, {Kharbanda}, {Khizriev}, {Kikunaga}, {Kolhe}, {Kramer}, {Krishnakumar}, {Lackeos}, {Lee}, {Liu}, {Liu}, {Lyne}, {McKee}, {Maan}, {Main}, {Mickaliger}, {Middleton}, {Neronov}, {Nitu}, {Nobleson}, {Paladi}, {Parthasarathy}, {Perera}, {Perrodin}, {Petiteau}, {Porayko}, {Possenti}, {Prabu}, {Postnov}, {Quelquejay Leclere}, {Rana}, {Roper Pol}, {Samajdar},
  {Sanidas}, {Semikoz}, {Sesana}, {Shaifullah}, {Singha}, {Smarra}, {Speri}, {Spiewak}, {Srivastava}, {Stappers}, {Steer}, {Surnis}, {Susarla}, {Susobhanan}, {Takahashi}, {Tarafdar}, {Theureau}, {Tiburzi}, {Truant}, {van der Wateren}, {Valtolina}, {Vecchio}, {Venkatraman Krishnan}, {Verbiest}, {Wang}, {Wang}, \& {Wu}}]{2023arXiv230616227A}
{Antoniadis}, J., {Arumugam}, P., {Arumugam}, S., {et~al.} 2023, arXiv e-prints, arXiv:2306.16227, \dodoi{10.48550/arXiv.2306.16227}

\bibitem[{{Arzoumanian} {et~al.}(2020){Arzoumanian}, {Baker}, {Blumer}, {B{\'e}csy}, {Brazier}, {Brook}, {Burke-Spolaor}, {Chatterjee}, {Chen}, {Cordes}, {Cornish}, {Crawford}, {Cromartie}, {Decesar}, {Demorest}, {Dolch}, {Ellis}, {Ferrara}, {Fiore}, {Fonseca}, {Garver-Daniels}, {Gentile}, {Good}, {Hazboun}, {Holgado}, {Islo}, {Jennings}, {Jones}, {Kaiser}, {Kaplan}, {Kelley}, {Key}, {Laal}, {Lam}, {Lazio}, {Lorimer}, {Luo}, {Lynch}, {Madison}, {McLaughlin}, {Mingarelli}, {Ng}, {Nice}, {Pennucci}, {Pol}, {Ransom}, {Ray}, {Shapiro-Albert}, {Siemens}, {Simon}, {Spiewak}, {Stairs}, {Stinebring}, {Stovall}, {Sun}, {Swiggum}, {Taylor}, {Turner}, {Vallisneri}, {Vigeland}, \& {Witt}}]{2020ApJ...905L..34A}
{Arzoumanian}, Z., {Baker}, P.~T., {Blumer}, H., {et~al.} 2020, \apjl, 905, L34, \dodoi{10.3847/2041-8213/abd401}

\bibitem[{{Barnes} \& {Hernquist}(1996)}]{1996ApJ...471..115B}
{Barnes}, J.~E., \& {Hernquist}, L. 1996, \apj, 471, 115, \dodoi{10.1086/177957}

\bibitem[{{Barnes} \& {Hernquist}(1991)}]{1991ApJ...370L..65B}
{Barnes}, J.~E., \& {Hernquist}, L.~E. 1991, \apjl, 370, L65, \dodoi{10.1086/185978}

\bibitem[{{Blair} {et~al.}(2015){Blair}, {Ju}, {Zhao}, {Wen}, {Chu}, {Fang}, {Cai}, {Gao}, {Lin}, {Liu}, {Wu}, {Zhu}, {Reitze}, {Arai}, {Zhang}, {Flaminio}, {Zhu}, {Hobbs}, {Manchester}, {Shannon}, {Baccigalupi}, {Gao}, {Xu}, {Bian}, {Cao}, {Chang}, {Dong}, {Gong}, {Huang}, {Ju}, {Luo}, {Qiang}, {Tang}, {Wan}, {Wang}, {Xu}, {Zang}, {Zhang}, {Lau}, \& {Ni}}]{2015SCPMA..58.5748B}
{Blair}, D., {Ju}, L., {Zhao}, C., {et~al.} 2015, Science China Physics, Mechanics, and Astronomy, 58, 5748, \dodoi{10.1007/s11433-015-5748-6}

\bibitem[{{Blandford} {et~al.}(1984){Blandford}, {Narayan}, \& {Romani}}]{1984JApA....5..369B}
{Blandford}, R., {Narayan}, R., \& {Romani}, R.~W. 1984, Journal of Astrophysics and Astronomy, 5, 369, \dodoi{10.1007/BF02714466}

\bibitem[{{Boyle} \& {Terlevich}(1998)}]{1998MNRAS.293L..49B}
{Boyle}, B.~J., \& {Terlevich}, R.~J. 1998, \mnras, 293, L49, \dodoi{10.1046/j.1365-8711.1998.01264.x}

\bibitem[{{Boyle} \& {Pen}(2012)}]{2012PhRvD..86l4028B}
{Boyle}, L., \& {Pen}, U.-L. 2012, \prd, 86, 124028, \dodoi{10.1103/PhysRevD.86.124028}

\bibitem[{Brazier {et~al.}(2016)Brazier, Lassus, Petiteau, Possenti, Sesana, Vecchio, Lyne, Christy, Perera, Stappers, Tiburzi, Bassa, Mingarelli, Perrodin, Reardon, Champion, Nice, Madison, Stinebring, Fonseca, Graikou, Desvignes, Hobbs, Jones, Shaifullah, Theureau, Janssen, Cognard, Stairs, Mckee, Simon, Ellis, Swiggum, Cordes, Gair, Hessels, Wang, Liu, Stovall, Lee, Guillemot, Lentati, Levin, Toomey, Wen, Burgay, Kerr, Kramer, Vallisneri, McLaughlin, Gonzalez, Keith, Lam, Bhat, Garver-Daniels, Palliyaguru, Brem, Gentile, Lazarus, Rosado, Demorest, Lasky, Karuppusamy, Smits, Spiewak, van Haasteren, Ferdman, Shannon, Caballero, Manchester, Lynch, Babak, Burke-Spolaor, Chatterjee, Dai, Os?owski, Sanidas, Chamberlin, Ransom, Taylor, McWilliams, Dolch, Lazio, Pennucci, van Straten, Zhu, Siemens, You, Zhu, Wang, Arzoumanian, \& Verbiest}]{10.1093/mnras/stw347}
Brazier, A., Lassus, A., Petiteau, A., {et~al.} 2016, Monthly Notices of the Royal Astronomical Society, 458, 1267, \dodoi{10.1093/mnras/stw347}

\bibitem[{{Chen} {et~al.}(2020){Chen}, {Yu}, \& {Lu}}]{2020ApJ...897...86C}
{Chen}, Y., {Yu}, Q., \& {Lu}, Y. 2020, \apj, 897, 86, \dodoi{10.3847/1538-4357/ab9594}

\bibitem[{{Chen} {et~al.}(2023){Chen}, {Yu}, \& {Lu}}]{2023ApJ...955..132C}
---. 2023, \apj, 955, 132, \dodoi{10.3847/1538-4357/ace59f}

\bibitem[{{Cornish} \& {Sesana}(2013)}]{2013CQGra..30v4005C}
{Cornish}, N.~J., \& {Sesana}, A. 2013, Classical and Quantum Gravity, 30, 224005, \dodoi{10.1088/0264-9381/30/22/224005}

\bibitem[{{Cornish} \& {van Haasteren}(2014)}]{2014arXiv1406.4511C}
{Cornish}, N.~J., \& {van Haasteren}, R. 2014, arXiv e-prints, arXiv:1406.4511, \dodoi{10.48550/arXiv.1406.4511}

\bibitem[{{Creighton} \& {Anderson}(2011)}]{2011gwpa.book.....C}
{Creighton}, J., \& {Anderson}, W. 2011, {Gravitational-Wave Physics and Astronomy: An Introduction to Theory, Experiment and Data Analysis.} (Wiley -VCH Verlag GmbH \& Co. KGaA)

\bibitem[{Croton {et~al.}(2006)Croton, Springel, White, De~Lucia, Frenk, Gao, Jenkins, Kauffmann, Navarro, \& Yoshida}]{Croton:2005hbr}
Croton, D.~J., Springel, V., White, S. D.~M., {et~al.} 2006, Mon. Not. Roy. Astron. Soc., 365, 11, \dodoi{10.1111/j.1365-2966.2006.09994.x}

\bibitem[{{Detweiler}(1979)}]{1979ApJ...234.1100D}
{Detweiler}, S. 1979, \apj, 234, 1100, \dodoi{10.1086/157593}

\bibitem[{{Di Matteo} {et~al.}(2005){Di Matteo}, {Springel}, \& {Hernquist}}]{2005Natur.433..604D}
{Di Matteo}, T., {Springel}, V., \& {Hernquist}, L. 2005, \nat, 433, 604, \dodoi{10.1038/nature03335}

\bibitem[{{EPTA Collaboration} {et~al.}(2023){EPTA Collaboration}, {InPTA Collaboration}, {Antoniadis}, {Arumugam}, {Arumugam}, {Babak}, {Bagchi}, {Bak Nielsen}, {Bassa}, {Bathula}, {Berthereau}, {Bonetti}, {Bortolas}, {Brook}, {Burgay}, {Caballero}, {Chalumeau}, {Champion}, {Chanlaridis}, {Chen}, {Cognard}, {Dandapat}, {Deb}, {Desai}, {Desvignes}, {Dhanda-Batra}, {Dwivedi}, {Falxa}, {Ferdman}, {Franchini}, {Gair}, {Goncharov}, {Gopakumar}, {Graikou}, {Grie{\ss}meier}, {Guillemot}, {Guo}, {Gupta}, {Hisano}, {Hu}, {Iraci}, {Izquierdo-Villalba}, {Jang}, {Jawor}, {Janssen}, {Jessner}, {Joshi}, {Kareem}, {Karuppusamy}, {Keane}, {Keith}, {Kharbanda}, {Kikunaga}, {Kolhe}, {Kramer}, {Krishnakumar}, {Lackeos}, {Lee}, {Liu}, {Liu}, {Lyne}, {McKee}, {Maan}, {Main}, {Mickaliger}, {Ni{\c{t}}u}, {Nobleson}, {Paladi}, {Parthasarathy}, {Perera}, {Perrodin}, {Petiteau}, {Porayko}, {Possenti}, {Prabu}, {Quelquejay Leclere}, {Rana}, {Samajdar}, {Sanidas}, {Sesana}, {Shaifullah}, {Singha}, {Speri}, {Spiewak}, {Srivastava},
  {Stappers}, {Surnis}, {Susarla}, {Susobhanan}, {Takahashi}, {Tarafdar}, {Theureau}, {Tiburzi}, {van der Wateren}, {Vecchio}, {Venkatraman Krishnan}, {Verbiest}, {Wang}, {Wang}, \& {Wu}}]{2023A&A...678A..50E}
{EPTA Collaboration}, {InPTA Collaboration}, {Antoniadis}, J., {et~al.} 2023, \aap, 678, A50, \dodoi{10.1051/0004-6361/202346844}

\bibitem[{{Ferrarese} \& {Merritt}(2000)}]{2000ApJ...539L...9F}
{Ferrarese}, L., \& {Merritt}, D. 2000, \apjl, 539, L9, \dodoi{10.1086/312838}

\bibitem[{{Foster} \& {Backer}(1990)}]{1990ApJ...361..300F}
{Foster}, R.~S., \& {Backer}, D.~C. 1990, \apj, 361, 300, \dodoi{10.1086/169195}

\bibitem[{{Gardiner} {et~al.}(2024){Gardiner}, {Kelley}, {Lemke}, \& {Mitridate}}]{2024ApJ...965..164G}
{Gardiner}, E.~C., {Kelley}, L.~Z., {Lemke}, A.-M., \& {Mitridate}, A. 2024, \apj, 965, 164, \dodoi{10.3847/1538-4357/ad2be8}

\bibitem[{{Grimm} {et~al.}(2025){Grimm}, {Pijnenburg}, {Cusin}, \& {Bonvin}}]{2025JCAP...03..011G}
{Grimm}, N., {Pijnenburg}, M., {Cusin}, G., \& {Bonvin}, C. 2025, \jcap, 2025, 011, \dodoi{10.1088/1475-7516/2025/03/011}

\bibitem[{{Guo} {et~al.}(2013){Guo}, {White}, {Angulo}, {Henriques}, {Lemson}, {Boylan-Kolchin}, {Thomas}, \& {Short}}]{2013MNRAS.428.1351G}
{Guo}, Q., {White}, S., {Angulo}, R.~E., {et~al.} 2013, \mnras, 428, 1351, \dodoi{10.1093/mnras/sts115}

\bibitem[{{Guo} {et~al.}(2011){Guo}, {White}, {Boylan-Kolchin}, {De Lucia}, {Kauffmann}, {Lemson}, {Li}, {Springel}, \& {Weinmann}}]{2011MNRAS.413..101G}
{Guo}, Q., {White}, S., {Boylan-Kolchin}, M., {et~al.} 2011, \mnras, 413, 101, \dodoi{10.1111/j.1365-2966.2010.18114.x}

\bibitem[{{Guo} {et~al.}(2022){Guo}, {Lu}, \& {Yu}}]{2022ApJ...939...55G}
{Guo}, X., {Lu}, Y., \& {Yu}, Q. 2022, \apj, 939, 55, \dodoi{10.3847/1538-4357/ac9131}

\bibitem[{{Guo} {et~al.}(2025){Guo}, {Yu}, \& {Lu}}]{2025ApJ...978..104G}
{Guo}, X., {Yu}, Q., \& {Lu}, Y. 2025, \apj, 978, 104, \dodoi{10.3847/1538-4357/ad94ec}

\bibitem[{{Hellings} \& {Downs}(1983)}]{1983ApJ...265L..39H}
{Hellings}, R.~W., \& {Downs}, G.~S. 1983, \apjl, 265, L39, \dodoi{10.1086/183954}

\bibitem[{{Henriques} {et~al.}(2015){Henriques}, {White}, {Thomas}, {Angulo}, {Guo}, {Lemson}, {Springel}, \& {Overzier}}]{2015MNRAS.451.2663H}
{Henriques}, B. M.~B., {White}, S. D.~M., {Thomas}, P.~A., {et~al.} 2015, \mnras, 451, 2663, \dodoi{10.1093/mnras/stv705}

\bibitem[{{Hernquist}(1989)}]{1989Natur.340..687H}
{Hernquist}, L. 1989, \nat, 340, 687, \dodoi{10.1038/340687a0}

\bibitem[{{Hopkins} {et~al.}(2008){Hopkins}, {Hernquist}, {Cox}, \& {Kere{\v{s}}}}]{2008ApJS..175..356H}
{Hopkins}, P.~F., {Hernquist}, L., {Cox}, T.~J., \& {Kere{\v{s}}}, D. 2008, \apjs, 175, 356, \dodoi{10.1086/524362}

\bibitem[{{Hotinli} {et~al.}(2019){Hotinli}, {Kamionkowski}, \& {Jaffe}}]{2019OJAp....2E...8H}
{Hotinli}, S.~C., {Kamionkowski}, M., \& {Jaffe}, A.~H. 2019, The Open Journal of Astrophysics, 2, 8, \dodoi{10.21105/astro.1904.05348}

\bibitem[{Izquierdo-Villalba {et~al.}(2021)Izquierdo-Villalba, Sesana, Bonoli, \& Colpi}]{David:stab3239}
Izquierdo-Villalba, D., Sesana, A., Bonoli, S., \& Colpi, M. 2021, Monthly Notices of the Royal Astronomical Society, 509, 3488, \dodoi{10.1093/mnras/stab3239}

\bibitem[{{Jaffe} \& {Backer}(2003)}]{2003ApJ...583..616J}
{Jaffe}, A.~H., \& {Backer}, D.~C. 2003, \apj, 583, 616, \dodoi{10.1086/345443}

\bibitem[{{Joshi} {et~al.}(2018){Joshi}, {Arumugasamy}, {Bagchi}, {Bandyopadhyay}, {Basu}, {Dhanda Batra}, {Bethapudi}, {Choudhary}, {De}, {Dey}, {Gopakumar}, {Gupta}, {Krishnakumar}, {Maan}, {Manoharan}, {Naidu}, {Nandi}, {Pathak}, {Surnis}, \& {Susobhanan}}]{2018JApA...39...51J}
{Joshi}, B.~C., {Arumugasamy}, P., {Bagchi}, M., {et~al.} 2018, Journal of Astrophysics and Astronomy, 39, 51, \dodoi{10.1007/s12036-018-9549-y}

\bibitem[{{Kelley} {et~al.}(2017){Kelley}, {Blecha}, \& {Hernquist}}]{2017MNRAS.464.3131K}
{Kelley}, L.~Z., {Blecha}, L., \& {Hernquist}, L. 2017, \mnras, 464, 3131, \dodoi{10.1093/mnras/stw2452}

\bibitem[{{Kormendy} \& {Ho}(2013)}]{2013ARA&A..51..511K}
{Kormendy}, J., \& {Ho}, L.~C. 2013, \araa, 51, 511, \dodoi{10.1146/annurev-astro-082708-101811}

\bibitem[{{Kormendy} \& {Richstone}(1995)}]{1995ARA&A..33..581K}
{Kormendy}, J., \& {Richstone}, D. 1995, \araa, 33, 581, \dodoi{10.1146/annurev.aa.33.090195.003053}

\bibitem[{{Kramer} \& {Champion}(2013)}]{2013CQGra..30v4009K}
{Kramer}, M., \& {Champion}, D.~J. 2013, Classical and Quantum Gravity, 30, 224009, \dodoi{10.1088/0264-9381/30/22/224009}

\bibitem[{Lazio(2013)}]{Lazio2013SKA}
Lazio, T. J.~W. 2013, Classical and Quantum Gravity, 30, 224011.
\newblock \url{http://stacks.iop.org/0264-9381/30/i=22/a=224011}

\bibitem[{{Lee} {et~al.}(2011){Lee}, {Wex}, {Kramer}, {Stappers}, {Bassa}, {Janssen}, {Karuppusamy}, \& {Smits}}]{2011MNRAS.414.3251L}
{Lee}, K.~J., {Wex}, N., {Kramer}, M., {et~al.} 2011, \mnras, 414, 3251, \dodoi{10.1111/j.1365-2966.2011.18622.x}

\bibitem[{{Madau} {et~al.}(1996){Madau}, {Ferguson}, {Dickinson}, {Giavalisco}, {Steidel}, \& {Fruchter}}]{1996MNRAS.283.1388M}
{Madau}, P., {Ferguson}, H.~C., {Dickinson}, M.~E., {et~al.} 1996, \mnras, 283, 1388, \dodoi{10.1093/mnras/283.4.1388}

\bibitem[{Maggiore(2008)}]{maggiore2008gravitational}
Maggiore, M. 2008, Gravitational waves: Volume 1: Theory and experiments (Oxford: Oxford University Press)

\bibitem[{{Magorrian} {et~al.}(1998){Magorrian}, {Tremaine}, {Richstone}, {Bender}, {Bower}, {Dressler}, {Faber}, {Gebhardt}, {Green}, {Grillmair}, {Kormendy}, \& {Lauer}}]{1998AJ....115.2285M}
{Magorrian}, J., {Tremaine}, S., {Richstone}, D., {et~al.} 1998, \aj, 115, 2285, \dodoi{10.1086/300353}

\bibitem[{Manchester(2013)}]{manchester2013pulsar}
Manchester, R.~N. 2013, International Journal of Modern Physics D, 22, 1341007, \dodoi{10.1142/S0218271813410071}

\bibitem[{{Manchester} \& {IPTA}(2013)}]{2013CQGra..30v4010M}
{Manchester}, R.~N., \& {IPTA}. 2013, Classical and Quantum Gravity, 30, 224010, \dodoi{10.1088/0264-9381/30/22/224010}

\bibitem[{{McLaughlin}(2013)}]{2013CQGra..30v4008M}
{McLaughlin}, M.~A. 2013, Classical and Quantum Gravity, 30, 224008, \dodoi{10.1088/0264-9381/30/22/224008}

\bibitem[{{Mihos} \& {Hernquist}(1994)}]{1994ApJ...431L...9M}
{Mihos}, J.~C., \& {Hernquist}, L. 1994, \apjl, 431, L9, \dodoi{10.1086/187460}

\bibitem[{{Mihos} \& {Hernquist}(1996)}]{1996ApJ...464..641M}
---. 1996, \apj, 464, 641, \dodoi{10.1086/177353}

\bibitem[{{Miles} {et~al.}(2023){Miles}, {Shannon}, {Bailes}, {Reardon}, {Keith}, {Cameron}, {Parthasarathy}, {Shamohammadi}, {Spiewak}, {van Straten}, {Buchner}, {Camilo}, {Geyer}, {Karastergiou}, {Kramer}, {Serylak}, {Theureau}, \& {Venkatraman Krishnan}}]{2023MNRAS.519.3976M}
{Miles}, M.~T., {Shannon}, R.~M., {Bailes}, M., {et~al.} 2023, \mnras, 519, 3976, \dodoi{10.1093/mnras/stac3644}

\bibitem[{Mingarelli(2015)}]{mingarelli2015gravitational}
Mingarelli, C.~M. 2015, Gravitational wave astrophysics with pulsar timing arrays (Springer)

\bibitem[{{Mingarelli} {et~al.}(2013){Mingarelli}, {Sidery}, {Mandel}, \& {Vecchio}}]{2013PhRvD..88f2005M}
{Mingarelli}, C.~M.~F., {Sidery}, T., {Mandel}, I., \& {Vecchio}, A. 2013, \prd, 88, 062005, \dodoi{10.1103/PhysRevD.88.062005}

\bibitem[{{Nan} {et~al.}(2011){Nan}, {Li}, {Jin}, {Wang}, {Zhu}, {Zhu}, {Zhang}, {Yue}, \& {Qian}}]{2011IJMPD..20..989N}
{Nan}, R., {Li}, D., {Jin}, C., {et~al.} 2011, International Journal of Modern Physics D, 20, 989, \dodoi{10.1142/S0218271811019335}

\bibitem[{Oguri(2018)}]{Oguri:2018muv}
Oguri, M. 2018, Mon. Not. Roy. Astron. Soc., 480, 3842, \dodoi{10.1093/mnras/sty2145}

\bibitem[{{Perera} {et~al.}(2019){Perera}, {DeCesar}, {Demorest}, {Kerr}, {Lentati}, {Nice}, {Os{\l}owski}, {Ransom}, {Keith}, {Arzoumanian}, {Bailes}, {Baker}, {Bassa}, {Bhat}, {Brazier}, {Burgay}, {Burke-Spolaor}, {Caballero}, {Champion}, {Chatterjee}, {Chen}, {Cognard}, {Cordes}, {Crowter}, {Dai}, {Desvignes}, {Dolch}, {Ferdman}, {Ferrara}, {Fonseca}, {Goldstein}, {Graikou}, {Guillemot}, {Hazboun}, {Hobbs}, {Hu}, {Islo}, {Janssen}, {Karuppusamy}, {Kramer}, {Lam}, {Lee}, {Liu}, {Luo}, {Lyne}, {Manchester}, {McKee}, {McLaughlin}, {Mingarelli}, {Parthasarathy}, {Pennucci}, {Perrodin}, {Possenti}, {Reardon}, {Russell}, {Sanidas}, {Sesana}, {Shaifullah}, {Shannon}, {Siemens}, {Simon}, {Spiewak}, {Stairs}, {Stappers}, {Swiggum}, {Taylor}, {Theureau}, {Tiburzi}, {Vallisneri}, {Vecchio}, {Wang}, {Zhang}, {Zhang}, {Zhu}, \& {Zhu}}]{2019MNRAS.490.4666P}
{Perera}, B.~B.~P., {DeCesar}, M.~E., {Demorest}, P.~B., {et~al.} 2019, \mnras, 490, 4666, \dodoi{10.1093/mnras/stz2857}

\bibitem[{{Ransom} {et~al.}(2019){Ransom}, {Brazier}, {Chatterjee}, {Cohen}, {Cordes}, {DeCesar}, {Demorest}, {Hazboun}, {Lam}, {Lynch}, {McLaughlin}, {Ransom}, {Siemens}, {Taylor}, \& {Vigeland}}]{2019BAAS...51g.195R}
{Ransom}, S., {Brazier}, A., {Chatterjee}, S., {et~al.} 2019, in Bulletin of the American Astronomical Society, Vol.~51, 195.
\newblock \doarXiv{1908.05356}

\bibitem[{{Reardon} {et~al.}(2023){Reardon}, {Zic}, {Shannon}, {Hobbs}, {Bailes}, {Di Marco}, {Kapur}, {Rogers}, {Thrane}, {Askew}, {Bhat}, {Cameron}, {Cury{\l}o}, {Coles}, {Dai}, {Goncharov}, {Kerr}, {Kulkarni}, {Levin}, {Lower}, {Manchester}, {Mandow}, {Miles}, {Nathan}, {Os{\l}owski}, {Russell}, {Spiewak}, {Zhang}, \& {Zhu}}]{2023ApJ...951L...6R}
{Reardon}, D.~J., {Zic}, A., {Shannon}, R.~M., {et~al.} 2023, \apjl, 951, L6, \dodoi{10.3847/2041-8213/acdd02}

\bibitem[{{Sah} \& {Mukherjee}(2024)}]{2024arXiv240711669S}
{Sah}, M.~R., \& {Mukherjee}, S. 2024, arXiv e-prints, arXiv:2407.11669, \dodoi{10.48550/arXiv.2407.11669}

\bibitem[{{Sah} {et~al.}(2024){Sah}, {Mukherjee}, {Saeedzadeh}, {Babul}, {Tremmel}, \& {Quinn}}]{2024MNRAS.533.1568S}
{Sah}, M.~R., {Mukherjee}, S., {Saeedzadeh}, V., {et~al.} 2024, \mnras, 533, 1568, \dodoi{10.1093/mnras/stae1930}

\bibitem[{{Sanders} {et~al.}(1988){Sanders}, {Soifer}, {Elias}, {Madore}, {Matthews}, {Neugebauer}, \& {Scoville}}]{1988ApJ...325...74S}
{Sanders}, D.~B., {Soifer}, B.~T., {Elias}, J.~H., {et~al.} 1988, \apj, 325, 74, \dodoi{10.1086/165983}

\bibitem[{{Sato-Polito} \& {Kamionkowski}(2024)}]{2024PhRvD.109l3544S}
{Sato-Polito}, G., \& {Kamionkowski}, M. 2024, \prd, 109, 123544, \dodoi{10.1103/PhysRevD.109.123544}

\bibitem[{{Sazhin}(1978)}]{1978SvA....22...36S}
{Sazhin}, M.~V. 1978, \sovast, 22, 36

\bibitem[{{Schutz} \& {Ma}(2016)}]{2016MNRAS.459.1737S}
{Schutz}, K., \& {Ma}, C.-P. 2016, \mnras, 459, 1737, \dodoi{10.1093/mnras/stw768}

\bibitem[{{Semenzato} {et~al.}(2024){Semenzato}, {Casey-Clyde}, {Mingarelli}, {Raccanelli}, {Bellomo}, {Bartolo}, \& {Bertacca}}]{2024arXiv241100532S}
{Semenzato}, F., {Casey-Clyde}, J.~A., {Mingarelli}, C. M.~F., {et~al.} 2024, arXiv e-prints, arXiv:2411.00532, \dodoi{10.48550/arXiv.2411.00532}

\bibitem[{{Sesana}(2013{\natexlab{a}})}]{2013CQGra..30x4009S}
{Sesana}, A. 2013{\natexlab{a}}, Classical and Quantum Gravity, 30, 244009, \dodoi{10.1088/0264-9381/30/24/244009}

\bibitem[{{Sesana}(2013{\natexlab{b}})}]{2013MNRAS.433L...1S}
---. 2013{\natexlab{b}}, \mnras, 433, L1, \dodoi{10.1093/mnrasl/slt034}

\bibitem[{{Sesana}(2015)}]{2015ASSP...40..147S}
{Sesana}, A. 2015, in Astrophysics and Space Science Proceedings, Vol.~40, Gravitational Wave Astrophysics, 147, \dodoi{10.1007/978-3-319-10488-1_13}

\bibitem[{{Sesana} {et~al.}(2016){Sesana}, {Shankar}, {Bernardi}, \& {Sheth}}]{2016MNRAS.463L...6S}
{Sesana}, A., {Shankar}, F., {Bernardi}, M., \& {Sheth}, R.~K. 2016, \mnras, 463, L6, \dodoi{10.1093/mnrasl/slw139}

\bibitem[{{Sesana} \& {Vecchio}(2010)}]{2010CQGra..27h4016S}
{Sesana}, A., \& {Vecchio}, A. 2010, Classical and Quantum Gravity, 27, 084016, \dodoi{10.1088/0264-9381/27/8/084016}

\bibitem[{{Sesana} {et~al.}(2008){Sesana}, {Vecchio}, \& {Colacino}}]{2008MNRAS.390..192S}
{Sesana}, A., {Vecchio}, A., \& {Colacino}, C.~N. 2008, \mnras, 390, 192, \dodoi{10.1111/j.1365-2966.2008.13682.x}

\bibitem[{{Sesana} {et~al.}(2009){Sesana}, {Vecchio}, \& {Volonteri}}]{2009MNRAS.394.2255S}
{Sesana}, A., {Vecchio}, A., \& {Volonteri}, M. 2009, \mnras, 394, 2255, \dodoi{10.1111/j.1365-2966.2009.14499.x}

\bibitem[{{Smits} {et~al.}(2009){Smits}, {Lorimer}, {Kramer}, {Manchester}, {Stappers}, {Jin}, {Nan}, \& {Li}}]{2009A&A...505..919S}
{Smits}, R., {Lorimer}, D.~R., {Kramer}, M., {et~al.} 2009, \aap, 505, 919, \dodoi{10.1051/0004-6361/200911939}

\bibitem[{{Soltan}(1982)}]{1982MNRAS.200..115S}
{Soltan}, A. 1982, \mnras, 200, 115, \dodoi{10.1093/mnras/200.1.115}

\bibitem[{{Springel} {et~al.}(2005){Springel}, {White}, {Jenkins}, {Frenk}, {Yoshida}, {Gao}, {Navarro}, {Thacker}, {Croton}, {Helly}, {Peacock}, {Cole}, {Thomas}, {Couchman}, {Evrard}, {Colberg}, \& {Pearce}}]{2005Natur.435..629S}
{Springel}, V., {White}, S. D.~M., {Jenkins}, A., {et~al.} 2005, \nat, 435, 629, \dodoi{10.1038/nature03597}

\bibitem[{{Taylor} {et~al.}(2019){Taylor}, {Burke-Spolaor}, {Baker}, {Charisi}, {Islo}, {Kelley}, {Madison}, {Simon}, \& {Vigeland}}]{2019BAAS...51c.336T}
{Taylor}, S., {Burke-Spolaor}, S., {Baker}, P.~T., {et~al.} 2019, \baas, 51, 336.
\newblock \doarXiv{1903.08183}

\bibitem[{{Taylor} \& {Gair}(2013)}]{2013PhRvD..88h4001T}
{Taylor}, S.~R., \& {Gair}, J.~R. 2013, \prd, 88, 084001, \dodoi{10.1103/PhysRevD.88.084001}

\bibitem[{{Taylor} {et~al.}(2015){Taylor}, {Mingarelli}, {Gair}, {Sesana}, {Theureau}, {Babak}, {Bassa}, {Brem}, {Burgay}, {Caballero}, {Champion}, {Cognard}, {Desvignes}, {Guillemot}, {Hessels}, {Janssen}, {Karuppusamy}, {Kramer}, {Lassus}, {Lazarus}, {Lentati}, {Liu}, {Os{\l}owski}, {Perrodin}, {Petiteau}, {Possenti}, {Purver}, {Rosado}, {Sanidas}, {Smits}, {Stappers}, {Tiburzi}, {van Haasteren}, {Vecchio}, \& {Verbiest}}]{2015PhRvL.115d1101T}
{Taylor}, S.~R., {Mingarelli}, C.~M.~F., {Gair}, J.~R., {et~al.} 2015, \prl, 115, 041101, \dodoi{10.1103/PhysRevLett.115.041101}

\bibitem[{{Treister} {et~al.}(2012){Treister}, {Schawinski}, {Urry}, \& {Simmons}}]{2012ApJ...758L..39T}
{Treister}, E., {Schawinski}, K., {Urry}, C.~M., \& {Simmons}, B.~D. 2012, \apjl, 758, L39, \dodoi{10.1088/2041-8205/758/2/L39}

\bibitem[{{Truant} {et~al.}(2025){Truant}, {Izquierdo-Villalba}, {Sesana}, {Shaifullah}, \& {Bonetti}}]{2025A&A...694A.282T}
{Truant}, R.~J., {Izquierdo-Villalba}, D., {Sesana}, A., {Shaifullah}, G.~M., \& {Bonetti}, M. 2025, \aap, 694, A282, \dodoi{10.1051/0004-6361/202451556}

\bibitem[{{Ueda} {et~al.}(2003){Ueda}, {Akiyama}, {Ohta}, \& {Miyaji}}]{2003ApJ...598..886U}
{Ueda}, Y., {Akiyama}, M., {Ohta}, K., \& {Miyaji}, T. 2003, \apj, 598, 886, \dodoi{10.1086/378940}

\bibitem[{{van Haasteren}(2014)}]{2014gwdd.book.....V}
{van Haasteren}, R. 2014, {Gravitational Wave Detection and Data Analysis for Pulsar Timing Arrays} (Springer)

\bibitem[{{Wang} \& {Mohanty}(2017)}]{2017PhRvL.118o1104W}
{Wang}, Y., \& {Mohanty}, S.~D. 2017, Physical Review Letters, 118, 151104, \dodoi{10.1103/PhysRevLett.118.151104}

\bibitem[{{Xu} {et~al.}(2023){Xu}, {Chen}, {Guo}, {Jiang}, {Wang}, {Xu}, {Xue}, {Nicolas Caballero}, {Yuan}, {Xu}, {Wang}, {Hao}, {Luo}, {Lee}, {Han}, {Jiang}, {Shen}, {Wang}, {Wang}, {Xu}, {Wu}, {Manchester}, {Qian}, {Guan}, {Huang}, {Sun}, \& {Zhu}}]{2023RAA....23g5024X}
{Xu}, H., {Chen}, S., {Guo}, Y., {et~al.} 2023, Research in Astronomy and Astrophysics, 23, 075024, \dodoi{10.1088/1674-4527/acdfa5}

\bibitem[{{Yang} {et~al.}(2019){Yang}, {Hu}, \& {Li}}]{2019MNRAS.483..503Y}
{Yang}, Q., {Hu}, B., \& {Li}, X.-D. 2019, \mnras, 483, 503, \dodoi{10.1093/mnras/sty3126}

\bibitem[{{Yu} \& {Tremaine}(2002)}]{2002MNRAS.335..965Y}
{Yu}, Q., \& {Tremaine}, S. 2002, \mnras, 335, 965, \dodoi{10.1046/j.1365-8711.2002.05532.x}

\bibitem[{{Zheng} {et~al.}(2009){Zheng}, {Bell}, {Somerville}, {Rix}, {Jahnke}, {Fontanot}, {Rieke}, {Schiminovich}, \& {Meisenheimer}}]{2009ApJ...707.1566Z}
{Zheng}, X.~Z., {Bell}, E.~F., {Somerville}, R.~S., {et~al.} 2009, \apj, 707, 1566, \dodoi{10.1088/0004-637X/707/2/1566}

\bibitem[{{Zhu} {et~al.}(2014){Zhu}, {Hobbs}, {Wen}, {Coles}, {Wang}, {Shannon}, {Manchester}, {Bailes}, {Bhat}, {Burke-Spolaor}, {Dai}, {Keith}, {Kerr}, {Levin}, {Madison}, {Os{\l}owski}, {Ravi}, {Toomey}, \& {van Straten}}]{2014MNRAS.444.3709Z}
{Zhu}, X.~J., {Hobbs}, G., {Wen}, L., {et~al.} 2014, \mnras, 444, 3709, \dodoi{10.1093/mnras/stu1717}

\end{thebibliography}

\end{document}